\documentclass[fleqn,10pt]{book}
\usepackage[a4paper]{uutfphd}
\usepackage{epsfig}
\usepackage{cite}

\begin{document}
\newcommand{\beq}{\begin{equation}}
\newcommand{\eeq}{\end{equation}}
\newcommand{\bea}{\begin{eqnarray}}
\newcommand{\eea}{\end{eqnarray}}

\newcommand{\chii}{\raise.5ex\hbox{$\chi$}}
\newcommand{\R}{I \! \! R}
\newcommand{\N}{I \! \! N}

\newcommand{\noi}{\noindent}
\newcommand{\vs}{\vspace{5mm}}
\newcommand{\ie}{{\ensuremath{ i.e.\ }}}
\newcommand{\eg}{{\ensuremath{ e.g.\ }}}
\newcommand{\ea}{{\ensuremath{ et~al.\ }}}
\newcommand{\hf}{{\scriptstyle{1 \over 2}}}
\newcommand{\ih}{{\scriptstyle{i \over \hbar}}}
\newcommand{\hi}{{\scriptstyle{ \hbar \over i}}}
\newcommand{\dbrst}{\delta_{BRST}}

\newcommand{\ddr}[1]{{ 
 {\stackrel{\raise.1ex\hbox{$\leftarrow$}}{\delta^r}   } 
\over {   \delta {#1}}  }}
\newcommand{\ddl}[1]{{ 
 {\stackrel{\lower.3ex \hbox{$\rightarrow$}}{\delta^l}   }
 \over {   \delta {#1}}  }}
\newcommand{\dd}[1]{{  {\delta} \over {\delta {#1}}  }}
\newcommand{\pa}{\partial}

\newcommand{\lrpartial}{
 {\stackrel{\hbox{\small $\leftrightarrow$}}{\partial}}}

\newcommand{\lrpartialmu}{
 {\stackrel{\hbox{\small $\leftrightarrow$}}{\partial^\mu}}}

\newcommand{\eq}[1]{{(\ref{#1})}}
\newcommand{\mb}[1]{{\mbox{${#1}$}}}

\newtheorem{theorem}{Theorem}[chapter]
\newtheorem{proposition}[theorem]{Proposition}
\newtheorem{lemma}[theorem]{Lemma}
\newtheorem{claim}[theorem]{Claim}
\newtheorem{remark}[theorem]{Remark}
\newtheorem{example}[theorem]{Example}

\newcommand{\proofbox}{\begin{flushright}
${\,\lower0.9pt\vbox{\hrule \hbox{\vrule
height 0.2 cm \hskip 0.2 cm \vrule height 0.2 cm}\hrule}\,}$
\end{flushright}}




\def\lsim{\mathrel{\rlap{\lower4pt\hbox{\hskip1pt$\sim$}}
    \raise1pt\hbox{$<$}}}         
\def\gsim{\mathrel{\rlap{\lower4pt\hbox{\hskip1pt$\sim$}}
    \raise1pt\hbox{$>$}}}         
\def\esim{\mathrel{\rlap{\raise2pt\hbox{$\sim$}}
    \lower1pt\hbox{$-$}}}         
\newcommand{\cE}{\mathcal{E}}

\setlength{\marginparwidth}{3cm}
\newcommand{\comment}[1]{\marginpar{\sf #1}}

\renewcommand{\topfraction}{.9}
\renewcommand{\bottomfraction}{.9}
\renewcommand{\textfraction}{.1}
\renewcommand{\floatpagefraction}{.9}


\newcommand{\neu}{\tilde{\chi}}
    
\setlength{\unitlength}{1 cm}

\newcommand{\vertex}[5]
{\begin{equation}
  \begin{picture}(9,2.8)(0,0)
  \put(0.5,1.3){#2}
  \put(4.2,0.15){#3}
  \put(4.2,2.6){#4}
  \put(6.0,1.4){#5}
  \hspace{1.1cm}\epsfig{file=feyn/#1, height=2.8cm}
  \end{picture}
\end{equation}}

\newcommand{\vertexrev}[5]
{\begin{equation}
  \begin{picture}(9,2.8)(0,0)
  \put(0.5,0.15){#2}
  \put(0.5,2.6){#3}
  \put(4.2,1.3){#4}
  \put(6.0,1.4){#5}
  \hspace{1.1cm}\epsfig{file=feyn/#1, height=2.8cm}
  \end{picture}
\end{equation}}

\newcommand{\vertexmom}[7]
{\begin{equation}
  \begin{picture}(9,2.8)(0,0)
  \put(0.5,1.3){#2}
  \put(4.2,0.15){#3}
  \put(4.2,2.6){#4}
  \put(6.0,1.4){#5}
  \put(2.8,0.4){#6}
  \put(2.8,2.3){#7}
  \hspace{1.1cm}\epsfig{file=feyn/#1, height=2.8cm}
  \end{picture}
\end{equation}}

\newcommand{\vertexmome}[8]
{\begin{equation}
  \begin{picture}(9,2.8)(0,0)
  \put(0.5,1.3){#2}
  \put(4.2,0.15){#3}
  \put(4.2,2.6){#4}
  \put(6.0,1.7){#5}
  \put(6.2,1.1){#6}
  \put(2.8,0.4){#7}
  \put(2.8,2.3){#8}
  \hspace{1.1cm}\epsfig{file=feyn/#1, height=2.8cm}
  \end{picture}
\end{equation}}

\newcommand{\vertexgauge}[6]
{\begin{equation}
  \begin{picture}(9,2.8)(0,0)
  \put(0.5,1.3){#2}
  \put(4.2,0.15){#3}
  \put(4.2,2.6){#4}
  \put(1.8,1.7){$k_1$}
  \put(3.6,1.0){$k_2$}
  \put(3.6,1.8){$k_3$}
  \put(0.5,0.9){$\alpha$}
  \put(4.2,0.55){$\beta$}
  \put(4.2,2.2){$\gamma$}
  \put(6.0,1.7){#5}
  \put(6.2,1.1){#6}
  \hspace{1.1cm}\epsfig{file=feyn/#1, height=2.8cm}
  \end{picture}
\end{equation}}

\newcommand{\vertexind}[7]
{\begin{equation}
  \begin{picture}(9,2.8)(0,0)
  \put(0.5,1.3){#2}
  \put(4.2,0.15){#3}
  \put(4.2,2.6){#4}
  \put(6.0,1.4){#5}
  \put(3.55,0.15){#6}
  \put(3.55,2.6){#7}
  \hspace{1.1cm}\epsfig{file=feyn/#1, height=2.8cm}
  \end{picture}
\end{equation}}



\thispagestyle{empty}

\bigskip

\centerline{\Large \bf Aspects of Neutrino Detection}

\smallskip

\centerline{\Large \bf  of Neutralino Dark Matter}

\bigskip

\bigskip

\centerline{\large Joakim Edsj{\"o}\footnote{E-mail address: edsjo@teorfys.uu.se}}

\smallskip

\centerline{\em Department of Theoretical Physics, Uppsala
University,}
\centerline{\em Box 803, SE-751 08 Uppsala, Sweden}

\bigskip

\bigskip

\centerline{April, 1997}

\bigskip

\centerline{\em PhD Thesis}

\bigskip

\bigskip

\centerline{\bf Abstract}

\smallskip

Neutralino dark matter, and in particular different aspects of its 
detection at neutrino telescopes, has been studied within the Minimal 
Supersymmetric extension of the Standard Model, the MSSM.

The relic density of neutralinos has been calculated using
sophisticated routines for integrating the annihilation cross section
and the Boltzmann equation. As a new element, so called coannihilation
processes between the lightest neutralino and the heavier neutralinos
and charginos have also been included for any neutralino mass and
composition.

The detection rates at neutrino telescopes have been evaluated for
neutralino annihilation in both the Sun and the Earth using detailed
Monte Carlo simulations of the whole chain of processes from the
neutralino annihilation products in the core of the Sun or the Earth
to detectable muons at a neutrino telescope.

A comparison with other searches for supersymmetry at accelerators and
direct dark matter searches is also given.

The signal muon fluxes that current and future neutrino telescopes can
probe and the improvement in sensitivity that can be achieved with
angular and/or energy resolution of the neutrino-induced muons has
also been investigated.

The question of whether the neutralino mass can be extracted from the
width of the muon angular distribution, if a signal flux is observed,
has also been addressed.

\cleardoublepage

\thispagestyle{empty}

\vspace*{3cm}

\noindent \hspace{0.6\textwidth} {\large To}\\[2ex]

\noindent \hspace{0.6\textwidth} {\large my wife Lisa}

\noindent\hspace{0.6\textwidth} {\large and our dog Molly}

\clearpage

\thispagestyle{empty}

\noindent This thesis is based on the following papers

\bigskip

\begin{list}{-}{\renewcommand{\makelabel}[1]{\makebox[\labelwidth][l]{#1}}}

\item[I.]
J.~Edsj{\"o}, \emph{Neutrino-induced Muon Fluxes from Neutralino
  Annihilations in the Sun and in the Earth}, 
  Nucl.\ Phys.\ {\bf B} (Proc.\ Suppl.) {\bf 43} (1995) 265.

\item[II.]
J.~Edsj{\"o} and P.~Gondolo, \emph{WIMP mass determination with
  neutrino telescopes}, Phys.\ Lett.\ {\bf B357} (1995) 595.

\item[III.] 
L.~Bergstr{\"o}m, J.~Edsj{\"o} and P.~Gondolo, \emph{Indirect
  neutralino detection rates in neutrino telescopes},
Phys.\ Rev.\ {\bf D55} (1997) 1765.

\item[IV.] 
L.~Bergstr{\"o}m, J.~Edsj{\"o} and 
M.~Kamionkowski, \emph{Astrophysical-Neutrino Detection with Angular
  and Energy Resolution}, Astrop.\ Phys., in press.

\item[V.] 
J.~Edsj{\"o} and P.~Gondolo, \emph{Neutralino Relic Density including 
Coannihilations}, submitted to Phys.\ Rev.\ {\bf D}.

\end{list}

\bigskip


\cleardoublepage
\setcounter{page}{1}

\tableofcontents

\chapter{Introduction}
\label{Intro}

Many cosmological observations show a definite need of dark matter,
which can make up more than 95\% of the mass in the Universe. One
usually defines $\Omega=\rho/\rho_{\rm crit}$ where $\rho$ is the
density in the Universe and $\rho_{\rm crit}$ is the so called
critical density for which the Universe would be flat.  Rotation
curves of galaxies indicate that
\begin{equation}
  \Omega \gsim 0.1
\end{equation}
in contrast to the luminous mass density  
\begin{equation}
  \Omega_{\rm luminous} \lsim 0.01
\end{equation}
which clearly indicates the existence of dark matter. Moreover,
motions of galaxies in clusters and superclusters indicate that 
\cite{pecvel}
\begin{equation}
  \Omega \gsim 0.2\mbox{--}0.3
\end{equation}
or maybe even higher.  $\Omega$ is also bounded from above due to the 
age of the Universe being at least 10$^{10}$ years \cite{KolbTurner}, 
\begin{equation}
  \Omega h^2 \lsim 1
\end{equation}
where $h$ is the Hubble constant in units of 100 km Mpc$^{-1}$
s$^{-1}$. The value of $h$ is still a bit uncertain with different
experimental determinations ranging from 0.5 to 0.9. 

What can this dark matter be then? Studies of Big Bang nucleosynthesis
predict values of the abundances of $^2$H, $^3$He, $^4$He and $^7$Li
which, when compared with observed abundances, give \cite{primabund}
\begin{equation}
  0.008 \lsim \Omega_{\rm baryon} h^2 < 0.024.
\end{equation}
We thus see that the dark matter cannot be made up of baryons, but
some more 'exotic' relic from the big bang is needed to explain the
high values of $\Omega$ observed.  There exist several hypothetical
candidates of which a Weakly Interacting Massive Particle (WIMP) is a
major candidate. WIMPs will freeze out in the early Universe when they
are non-relativistic and their relic density is approximately given by
\cite{jkg}
\begin{equation} \label{eq:omegasimp}
  \Omega_{\rm WIMP} h^2 \simeq \frac{3 \times 10^{-27} \mbox{ cm$^3$
  s$^{-1}$}} {\langle \sigma_A v \rangle}
\end{equation}
where $\langle \sigma_A v \rangle$ is the thermally averaged
annihilation cross section. A weakly interacting particle is expected
to have an annihilation cross section of the order of $\langle
\sigma_A v \rangle \sim \alpha_{\rm ew}^2 / (100 \mbox{ GeV})^2 \sim
10^{-25}$ cm$^3$ s$^{-1}$ which gives an $\Omega$ of about the
magnitude wanted. Hence, if there are any WIMPs left from the big
bang, they are expected to have a relic density that can be enough to
explain the dark matter problem.

One of the leading WIMP candidates is the lightest supersymmetric
particle, the neutralino. Assuming that the dark matter is really
constituted by WIMPs (or more specifically by neutralinos), how can
one find them? First of all, if WIMPs constitute the dark matter, they
will have clumped together making up a halo of
the galaxy (containing most of our galaxy's mass). Hence they will be
all around us and the Earth (and the solar system itself) will move
through this halo during its motion through the galaxy.

There are in principle two different kinds of experiments proposed to
search for the dark matter in the halo of the galaxy: direct and
indirect searches. In direct experiments one looks for these WIMPs
passing by a detector and scattering off some nucleus. This scattering
can be detected and, if found, would be an evidence for WIMPs in the
galactic halo. In the indirect searches, one looks not for the WIMPs
directly, but for signals coming from annihilation of two WIMPs. For
example, their annihilations in the halo will result in a $\gamma$-
and $\bar{p}$-flux which can be searched for. These WIMPs can also get
elastically scattered while passing the Sun or the Earth and get
gravitationally trapped. They will then accumulate at the center of
the Sun and the Earth where their annihilation eventually will produce
neutrinos which can be detected. This last indirect way of searching
for WIMPs is the main topic of this thesis. Though most discussions
will be devoted to neutralinos, many of the results in this thesis
will be applicable to any WIMP.

In Chapter~\ref{MSSMdef} the Minimal Supersymmetric extension of the
Standard Model (MSSM), in which we work, will be defined, in
Chapter~\ref{ExpCon} the experimental constraints on the MSSM will be
reviewed, in Chapter~\ref{RelDens} a detailed calculation of the relic
density of neutralinos will be performed and in Chapter~\ref{Indirect}
the expected neutrino-induced muon fluxes at neutrino telescopes will
be evaluated.  We then close by some concluding remarks in
Chapter~\ref{Concl}.

\chapter{Definition of the MSSM}
\label{MSSMdef}

\section{Introduction}

We work in the Minimal Supersymmetric extension of the Standard Model
(MSSM) with $N=1$ supersymmetry generators and we will essentially
follow the notation of Ref.~\cite{HaberKane,GuHa86}. We will only give
a short introduction to supersymmetry phenomenology in this chapter
and the interested reader is referred to
Ref.~\cite{jkg,HaberKane,GuHa86,WessBagger} for more details.

Supersymmetry is a symmetry relating fermions to bosons such that for
each fermionic degree of freedom there is a bosonic degree of freedom.
This extends the particle content of the Standard Model (SM) such that
each particle in the SM has a corresponding superpartner (or
partners).  More specifically, the particle content in the MSSM is the
same as of the SM plus the superpartners and two Higgs doublets
(instead of one as in the SM). Two Higgs doublets are needed to give
mass to both up- and down-type quarks and will result in five physical
Higgs bosons.  If supersymmetry were unbroken, a SM particle and its
superpartner would have the same mass and quantum numbers (except for
spin). Since we haven't seen these particles, we can conclude that
supersymmetry is broken at the energies probed by present
accelerators.

In Table~\ref{tab:susyparticles} we list the `normal' particles and
their corresponding superpartners. Note that some `normal' particles
have more than one superpartner, e.g.\ each quark has two squarks,
$\tilde{q}_{L}$ and $\tilde{q}_{R}$ as superpartners, but the number
of degrees of freedom (2 for the quark (spin $\frac{1}{2}$) and 1 for
each squark (spin 0)) sums up to be the same for the normal particle
and its superpartner(s). The general notation is to have a tilde on
the symbol for the superpartners, but for the charginos and
neutralinos we will usually drop the tilde since there is no risk for
misinterpretations anyway.

\begin{table}
\footnotesize
\begin{center}
\begin{tabular}{lllllll} \hline 
  \multicolumn{2}{l}{Normal particles/fields} & \multicolumn{5}{l}{Supersymmetric partners} \\
  & & \multicolumn{3}{l}{Interaction eigenstates} & \multicolumn{2}{l}{Mass 
  eigenstates} \\
  Symbol & Name & Symbol & Name & & Symbol & Name \\ \hline
  $q=d,c,b,u,s,t$ & quark & $\tilde{q}_{L}$, $\tilde{q}_{R}$ & 
  squark & & $\tilde{q}_{1}$, $\tilde{q}_{2}$ & squark \\
  $l=e,\mu,\tau$ & lepton & $\tilde{l}_{L}$, $\tilde{l}_{R}$ & slepton & 
  & $\tilde{l}_{1}$, $\tilde{l}_{2}$ & slepton \\
  $\nu = \nu_{e}, \nu_{\mu}, \nu_{\tau}$ & neutrino & $\tilde{\nu}$ & 
  sneutrino & & $\tilde{\nu}$ & sneutrino \\
  $g$ & gluon & $\tilde{g}$ & gluino & & $\tilde{g}$ & gluino \\
  $W^\pm$ & $W$-boson & $\tilde{W}^\pm$ & wino & & & \\
  $H^-$ & Higgs boson & $\tilde{H}_{1}^-$ & higgsino & 
  \raisebox{-.25ex}[0ex][0ex]{$\left. \raisebox{0ex}[-3.3ex][3.3ex]{}
  \right\}$} &  $\tilde{\chi}_{1,2}^\pm$ & chargino \\
  $H^+$ & Higgs boson & $\tilde{H}_{2}^+$ & higgsino & & & \\
  $B$ & $B$-field & $\tilde{B}$ & bino & & & \\
  $W^3$ & $W^3$-field & $\tilde{W}^3$ & wino & & & \\
  $H_{1}^0$ & Higgs boson & 
  \raisebox{-1.75ex}[0ex][0ex]{$\tilde{H}_{1}^0$} & 
  \raisebox{-1.75ex}[0ex][0ex]{higgsino} & 
  \raisebox{.25ex}[0ex][0ex]{$\left. \raisebox{0ex}[-5.25ex][5.25ex]{}
  \right\}$} & \raisebox{0.5ex}[0ex][0ex]{$\tilde{\chi}_{1,2,3,4}^0$} & 
  \raisebox{.5ex}[0ex][0ex]{neutralino} \\[0.5ex]
  $H_{2}^0$ & Higgs boson & 
  \raisebox{-1.75ex}[0ex][0ex]{$\tilde{H}_{2}^0$} & 
  \raisebox{-1.75ex}[0ex][0ex]{higgsino} & & & \\[0.5ex]
  $H_{3}^0$ & Higgs boson & & & & & \\[0.5ex] \hline
\end{tabular}
\caption{The `normal' particles and their superpartners in the MSSM.}
\label{tab:susyparticles}
\end{center}
\end{table}

\section{The superpotential, supersymmetry breaking and $R$-parity}

To write down the Lagrangian for the MSSM, one should introduce the 
superfield formalism. This is not within the scope of this thesis and 
we will just write down the superpotential and the soft supersymmetry 
breaking potential for reference and the reader is referred to 
Ref.~\cite{jkg,HaberKane,GuHa86,WessBagger} for details.

The superpotential is given by
\begin{eqnarray}
  W = \epsilon_{ij} \left(
  - {\bf \hat{e}}_{R}^{*} {\bf Y}_E {\bf \hat{l}}^i_{L} {\hat H}^j_1 
  - {\bf \hat{d}}_{R}^{*} {\bf Y}_D {\bf \hat{q}}^i_{L} {\hat H}^j_1 
  + {\bf \hat{u}}_{R}^{*} {\bf Y}_U {\bf \hat{q}}^i_{L} {\hat H}^j_2 
  - \mu {\hat H}^i_1 {\hat H}^j_2 
  \right) 
  \label{eq:superpotential}
\end{eqnarray}
where $i$ and $j$ are SU(2) indices, the Yukawa couplings $\bf Y$ are
matrices in generation space and $\bf \hat{e}$, $\bf \hat{l}$, $\bf
\hat{u}$, $\bf \hat{d}$ and $\bf \hat{q}$ are the superfields of the
leptons and sleptons and of the quarks and squarks.  The lefthanded
components are SU(2) doublets and the righthanded are SU(2) singlets.

We then introduce all possible soft supersymmetry breaking terms 
(without violating gauge-invariance or breaking baryon or lepton number) 
in the potential
\begin{eqnarray}
  \label{eq:Vsoft}
  \lefteqn{V_{{\rm soft}} =  
  \epsilon_{ij} \Big(
   {\bf \tilde{e}}_{R}^{*} {\bf A}_E {\bf Y}_E {\bf \tilde{l}}^i_{L} H^j_1 
  + {\bf \tilde{d}}_{R}^{*} {\bf A}_D {\bf Y}_D {\bf \tilde{q}}^i_{L} 
  H^j_1} \nonumber \\
  & &
  - {\bf \tilde{u}}_{R}^{*} {\bf A}_U {\bf Y}_U {\bf \tilde{q}}^i_{L} H^j_2 
  - B \mu H^i_1 H^j_2 + {\rm h.c.} 
  \Big) \nonumber \\ &&
  + H^{i*}_1 m_1^2 H^i_1 + H^{i*}_2 m_2^2 H^i_2
  \nonumber \\ && +
  {\bf \tilde{q}}_{L}^{i*} {\bf M}_{Q}^{2} {\bf \tilde{q}}^i_{L} + 
  {\bf \tilde{l}}_{L}^{i*} {\bf M}_{L}^{2} {\bf \tilde{l}}^i_{L} + 
  {\bf \tilde{u}}_{R}^{*} {\bf M}_{U}^{2} {\bf \tilde{u}}_{R} + 
  {\bf \tilde{d}}_{R}^{*} {\bf M}_{D}^{2} {\bf \tilde{d}}_{R} + 
  {\bf \tilde{e}}_{R}^{*} {\bf M}_{E}^{2} {\bf \tilde{e}}_{R} 
  \nonumber \\ && +
  \frac{1}{2} M_1 \tilde{B} \tilde{B} +
  \frac{1}{2} M_2 \left( \tilde{W}^3 \tilde{W}^3 +
        2 \tilde{W}^+ \tilde{W}^- \right) +
  \frac{1}{2} M_3 \tilde{g} \tilde{g}
\end{eqnarray}
where the soft trilinear couplings $\bf A$ and the soft sfermion
masses $\bf M$ are matrices in generation space and the fields $\bf
\tilde{e}$, $\bf \tilde{l}$, $\bf \tilde{u}$, $\bf \tilde{d}$ and $\bf
\tilde{q}$ are the scalar components of the superfields corresponding
to the superpartners of the SM\@.  The $L$ and $R$ subscripts on the
sfermion fields refers to the chirality of the fermion they are
superpartners of. $\tilde{B}$, $\tilde{W}^3$ and $\tilde{W}^\pm$ are
the fermionic superpartners of the SU(2) gauge fields and $\tilde{g}$
is the gluino field. $\mu$ is the higgsino mass parameter and $M_1$,
$M_2$ and $M_3$ are the gaugino mass parameters.  $B$ is the soft
bilinear coupling and $m_{1,2}$ are Higgs mass parameters.  Notice
that we have now introduced many new parameters, but this is the price
to pay until we know how supersymmetry breaking occurs.

The superpotential and soft supersymmetry breaking potential we have
now introduced are not the most general ones, unless we assume that
the so called $R$-parity is conserved. $R$-parity is a discrete
symmetry being $1$ for `normal' particles and $-1$ for their
superpartners. $R$-parity has to be put in by hand and if conserved
automatically prevents baryon and lepton number violation which would
otherwise be allowed unsuppressed at tree level. $R$-parity
conservation also implies that the Lightest Supersymmetric Particle,
the LSP, is stable which is a very welcome consequence. In this thesis
we will assume throughout that $R$-parity is conserved.

\section{Electroweak symmetry breaking and Higgs bosons}
\label{sec:EWHiggs}

Electroweak symmetry breaking is caused by the fields $H_1$ and $H_2$
acquiring vacuum expectation values
\begin{equation}
  \langle H_1 \rangle = \left( \begin{array}{c}
  v_1 \\
  0 \end{array} \right) , \qquad
  \langle H_2 \rangle = \left( \begin{array}{c}
  0 \\
  v_2 \end{array} \right)
\end{equation}
where $v_1$ and $v_2$ can be chosen real and non-negative by using
appropriate phases for the Higgs fields. They are related to the $W$
boson mass by
\begin{equation}
  m_W^2 = \frac{1}{2} g^2 (v_1^2 + v_2^2) 
\end{equation}
and we also have the convenient expression for the $Z$ boson mass
\begin{equation}
  m_Z^2 = \frac{1}{2}\left( g^2 + g'^2 \right) \left( v_1^2 + v_2^2
  \right)
\end{equation}
where $g$ and $g'$ are the usual SU(2) and U(1) gauge coupling
constants.  In the unitary gauge we then replace the fields $H_i$ with
$H_i + \langle H_i \rangle$, $i=1,2$. We define the ratio of the
vacuum expectation values,
\begin{equation}
  \tan \beta = \frac{v_2}{v_1}.
\end{equation}

There are five physical Higgs bosons in the MSSM, $H_1^0$, $H_2^0$,
$H_3^0$ and $H^\pm$. In another frequently used notation, the neutral
Higgs bosons are denoted by $H$, $h$ and $A$ respectively. We will use
the notation $H_1^0$, $H_2^0$ and $A$ (and sometimes $H_3^0$) for the
Higgs bosons.  Of the neutral ones, $A$ is CP-odd and $H_1^0$ and
$H_2^0$ are CP-even. The CP-even Higgs bosons are generally mixtures
of the interaction eigenstates and the mixing angle is denoted by
$\alpha$ where $-\pi/2 \leq \alpha \leq 0$.

In Eq.~(\ref{eq:Vsoft}) there are three parameters in the Higgs
sector, $m_1$, $m_2$ and $B$. The constraints coming from minimizing
the Higgs potential removes one of these and we are left with two
independent parameters, see e.g.\ Ref.~\cite{HiggsHunter} for details.
We have already chosen $\tan \beta$ as one of them and it is
convenient to choose the mass of the CP-odd Higgs boson, $m_A$, as our
second free parameter.  The masses of the other Higgs bosons are then
at tree-level given by
\begin{eqnarray}
  m^2_{H_2^0,H_1^0} & = & \frac{1}{2} \left[ m_A^2 + m_Z^2 \mp
 \sqrt{(m_A^2+m_Z^2)^2-4 m_Z^2m_A^2 \cos^2 2 \beta} \right]
 \label{eq:mhH} \\
  m_{H^\pm}^2 & = & m_A^2 + m_W^2.
\end{eqnarray}
As seen by Eq.~(\ref{eq:mhH}), the mass of the lightest Higgs boson
$m_{H_2^0}$ is at tree level bounded from above,
\begin{equation}
  m_{H_2^0} \leq m_Z | \cos 2 \beta|.
\end{equation}

The Higgs boson masses do however get large radiative corrections and
we have used the renormalization group improved 2-loop leading log
corrections in Ref.~\cite{carena}. For other references on such
effective potential approaches, see Ref.~\cite{effpot}.  The upper
bound on the $H_2^0$ mass depends on the mass of the top quark, $m_t$.
For $m_t<200$ GeV, the upper bound is $m_{H_2^0}<150$ GeV and for
$m_t=175$ GeV, it is $m_{H_2^0}<130$ GeV \cite{HiggsLEP2}.

\section{Neutralinos}

The neutralinos are linear combinations of the superpartners of the 
gauge bosons and the Higgs bosons. In the basis 
$(\tilde{B},\tilde{W}_{3},\tilde{H}_{1}^0,\tilde{H}_{2}^0)$ their mass 
matrix is given by
\begin{equation} \label{eq:neumass}
  {\cal M}_{\tilde \chi^0} = 
  \left( \begin{array}{cccc}
  {M_1} & 0 & -\frac{g'v_1}{\sqrt{2}} & +\frac{g'v_2}{\sqrt{2}} \\
  0 & {M_2} & +\frac{gv_1}{\sqrt{2}} & -\frac{gv_2}{\sqrt{2}} \\
  -\frac{g'v_1}{\sqrt{2}} & +\frac{gv_1}{\sqrt{2}} & 0 & -\mu \\
  +\frac{g'v_2}{\sqrt{2}} & -\frac{gv_2}{\sqrt{2}} & -\mu & 0 \\
  \end{array} \right).
\end{equation}
The neutralino mass matrix can be diagonalized
analytically to give the four neutralinos,
\begin{equation} \label{eq:neulinear}
  \tilde{\chi}^0_i = 
  N_{i1} \tilde{B} + N_{i2} \tilde{W}^3 + 
  N_{i3} \tilde{H}^0_1 + N_{i4} \tilde{H}^0_2 ,
\end{equation}
the lightest of which, $\tilde{\chi}_1^0$, to be called \emph{the}
neutralino, $\chi$, is then the candidate for the dark matter in the
Universe. We have chosen to work with the convention where the matrix
$N_{ij}$ is complex and the mass eigenvalues are all positive.  The
gaugino fraction of neutralino $i$ is defined as
\begin{equation}
  Z_g^i = |N_{i1}|^2 + |N_{i2}|^2,
\end{equation}
and we will call the lightest neutralino higgsino-like when $Z_{g}<0.01$, 
mixed when $0.01 \leq Z_{g} \leq 0.99$ and gaugino-like when 
$Z_{g}>0.99$ where the shorthand notation $Z_{g} \equiv Z_{g}^1$ is 
used for the lightest neutralino gaugino fraction.

The neutralino mass matrix, Eq.~(\ref{eq:neumass}), is valid at tree
level but gets loop corrections due to dominantly quark-squark loops
\cite{NeuLoop1, NeuLoop2}. This effect is most important when
calculating the relic density of higgsino-like neutralinos (as we will
see in Chapter~\ref{RelDens}), where both the next-to-lightest
neutralino and the lightest chargino are close in mass to the lightest
neutralino and in this case even small corrections to the masses are
important. The most important one-loop corrections are corrections
to entries $(3,3)$ and $(4,4)$ in the neutralino mass matrix,
Eq.~(\ref{eq:neumass}), and they are given by \cite{NeuLoop1,NeuLoop2}
\begin{eqnarray}
  \delta_{33} & = & - \frac{3}{16\pi^2}Y_{b}^2 m_b \sin (2 
  \theta_{\tilde{b}}) {\rm Re} \left[
  B_{0}(Q,b,\tilde{b}_{1})-B_{0}(Q,b,\tilde{b}_{2}) \right] \\
  \delta_{44} & = & - \frac{3}{16\pi^2}Y_{t}^2 m_t \sin (2 
  \theta_{\tilde{t}}) {\rm Re} \left[
  B_{0}(Q,t,\tilde{t}_{1})-B_{0}(Q,t,\tilde{t}_{2}) \right]
\end{eqnarray}
where $m_{b}$ and $m_{t}$ are the masses of the $b$ and $t$ quarks,
\begin{equation}
  Y_{b} = \frac{gm_{b}}{\sqrt{2}m_{W}\cos \beta} 
  \quad \mbox{and} \quad 
  Y_{t}=\frac{gm_{t}}{\sqrt{2}m_{W} \sin \beta}
\end{equation}  
are the Yukawa couplings of the $b$ and $t$ quark,
$\theta_{\tilde{b}}$ and $\theta_{\tilde{t}}$ are the mixing angles of
the squark mass eigenstates ($\tilde{q}_{1} = \tilde{q}_{L} \cos
\theta_{\tilde{q}} + \tilde{q}_{R} \sin \theta_{\tilde{q}}$) and
$B_{0}$ is the two-point function for which we use the convention in
\cite{NeuLoop1,NeuLoop2}.  Expressions for $B_{0}$ can be found in
e.g.\ \cite{B-functions}.  For the momentum scale $Q$ we use $|\mu|$
as suggested in \cite{NeuLoop1}.  Note that the loop corrections
depend on the mixing angles of the squarks as given by the soft
supersymmetry breaking parameters ${\bf A}_{U,D}$ in the soft
supersymmetry breaking potential
Eq.~(\ref{eq:Vsoft}) (see also Section
\ref{sec:squarks} below).

\section{Charginos}

The chargino mass terms in the Lagrangian are given by
\begin{equation}
  (\tilde{W}^- \tilde{H}_1^-) {\cal M}_{\tilde{\chi}^\pm} \left(
  \begin{array}{c}
  \tilde{W}^+ \\
  \tilde{H}_2^+ \end{array} \right) + \mbox{h.c.}
\end{equation}
where the mass matrix,
\begin{equation} \label{eq:chamass}
  {\cal M}_{\tilde{\chi}^\pm} = 
  \left( \begin{array}{cc}
  {M_2} & {gv_2} \\
  {gv_1} & \mu \\
  \end{array} \right),
\end{equation}
is diagonalized by
\begin{eqnarray}
  \tilde{\chi}^-_i & = & U_{i1} \tilde{W}^- + U_{i2} \tilde{H}_1^- \\
  \tilde{\chi}^+_i & = & V_{i1} \tilde{W}^+ + V_{i2} \tilde{H}_2^+.
\end{eqnarray}
We choose det$(U)=1$ and $U^* {\cal M}_{\tilde{\chi}^\pm} V^\dagger =
\mbox{diag}(m_{\tilde{\chi}_1^\pm},m_{\tilde{\chi}_2^\pm})$ with
non-negative chargino masses. The chargino mass matrix also gets
one-loop corrections as the neutralino mass matrix, but these are
always negligible compared to the neutralino mass corrections
\cite{NeuLoop1} and can hence safely be neglected.

\section{Squarks and sleptons}
\label{sec:squarks}

For the squarks we choose a basis where the squarks are rotated in the
same way as the corresponding quarks in the standard model. We follow
the conventions of the Particle Data Group \cite{PDG} where the mixing
is put in the left-handed $d$-quark fields. 

The squark mass matrices then look like
\begin{eqnarray}
  {\cal M}_{\tilde u}^2 &\!\!\!\!=\!\!\!\!& \left( \begin{array}{cc}
  \mbox{\bf M}_Q^2 + \mbox{\bf m}_u^\dagger \mbox{\bf m}_u +
      D_{LL}^{u} \mbox{\bf 1} &
    \mbox{\bf m}_u^\dagger 
        ( {\bf A}_U^\dagger - \mu^* \cot\beta ) \\
   ( {\bf A}_U - \mu \cot\beta ) \mbox{\bf m}_u &
  \mbox{\bf M}_U^2 + \mbox{\bf m}_u \mbox{\bf m}_u^\dagger +
      D_{RR}^{u} \mbox{\bf 1} \\
  \end{array} \right)
  \label{mutilde} \\
  {\cal M}_{\tilde d}^2 &\!\!\!\!=\!\!\!\!& \left( \begin{array}{cc}
  {\mbox{\bf K}^\dagger \mbox{\bf M}_Q^2 \mbox{\bf K}+
  \mbox{\bf m}_d\mbox{\bf m}_d^\dagger+D_{LL}^{d}\mbox{\bf 1}}&
  {\mbox{\bf m}_d^\dagger ( {\bf A}_D^\dagger-\mu^*\tan\beta )}\\
  {( {\bf A}_D-\mu\tan\beta ) \mbox{\bf m}_d}&
  {\mbox{\bf M}_D^2+\mbox{\bf m}_d^\dagger\mbox{\bf m}_d+
      D_{RR}^{d}\mbox{\bf 1}}\\
  \end{array} \right). 
  \label{mdtilde}
\end{eqnarray}
For the sneutrinos and sleptons we in the same way get the mass matrices
\begin{eqnarray}
  {\cal M}^2_{\tilde\nu} & = & \mbox{\bf M}_L^2 + D^\nu_{LL} \mbox{\bf
  1} \\
  {\cal M}^2_{\tilde e} & = &\left( \begin{array}{cc}
  {\mbox{\bf M}_L^2+\mbox{\bf m}_e\mbox{\bf m}_e^\dagger+
       D_{LL}^{e}\mbox{\bf 1}}&
  {\mbox{\bf m}_e^\dagger ( {\bf A}_E^\dagger-\mu^*\tan\beta )}\\
  {( {\bf A}_E-\mu\tan\beta ) \mbox{\bf m}_e}&
  {\mbox{\bf M}_E^2+\mbox{\bf m}_e^\dagger\mbox{\bf m}_e+
       D_{RR}^{e}\mbox{\bf 1}}\\
  \end{array} \right)
  \label{metilde}
\end{eqnarray}
where
\begin{eqnarray}
  D^f_{LL} & = & m_Z^2\cos 2\beta(T_{3f}-e_f\sin^2\theta_W) \\
  D^f_{RR} & = & m_Z^2\cos 2\beta e_f\sin^2\theta_W
\end{eqnarray}
with $T_{3f}$ being the third component of the weak isospin and $e_f$
being the charge in units of the elementary charge $e$ ($e>0$).
In the basis we have chosen, we have
\begin{eqnarray}
  {\bf m}_u & = & \mbox{diag}(m_u,m_c,m_t) \\
  {\bf m}_d & = & \mbox{diag}(m_d,m_s,m_b) \\
  {\bf m}_e & = & \mbox{diag}(m_e,m_\mu,m_\tau)
\end{eqnarray}
where we have used $m_t=175$ GeV for the top quark mass.

We now need to find the mass eigenstates, $\tilde{f}_k$, that
diagonalize these mass matrices. The relations between the mass
eigenstates and the interaction 
eigenstates $\tilde{f}_L$  and $\tilde{f}_R$ are 
\begin{eqnarray}
  \tilde{f}_{La} & = & \sum_{k=1}^6 \tilde{f}_k {\bf 
\Gamma}_{FL}^{*ka} , \\
  \tilde{f}_{Ra} & = & \sum_{k=1}^6 \tilde{f}_k {\bf 
\Gamma}_{FR}^{*ka}
\end{eqnarray} 
where the mixing matrices $\bf \Gamma$ have dimension $6\times3$
for squarks and charged sleptons and dimension $3\times3$ for sneutrinos.

We now have to specify the soft supersymmetry breaking parameters
${\bf A}_U$, ${\bf A}_D$, ${\bf A}_L$, ${\bf M}_Q$, ${\bf M}_U$, ${\bf
  M}_D$, ${\bf M}_E$ and ${\bf M}_L$. To reduce the number of free
parameters we make the simple Ansatz
\begin{equation}
  \left\{ \begin{array}{lcl}
  {\bf A}_U & = & \mbox{\rm diag}(0,0,A_t) \\ 
  {\bf A}_D & = &  \mbox{\rm diag}(0,0,A_b) \\
  {\bf A}_E & = & {\bf 0} \\
  {\bf M}_Q & = &  {\bf M}_U = {\bf M}_D = {\bf M}_E =
   {\bf M}_L = m_0 {\bf 1}
  \end{array} \right. \label{eq:Aansatz}
\end{equation}
which, since the matrices are diagonal, does not introduce any
tree-level flavour changing neutral currents (FCNCs).

\section{GUT assumptions}
\label{sec:GUTassump}

To reduce the number of free parameters further we will make the usual
Grand Unified Theory (GUT) assumptions for the gaugino mass parameters 
$M_1$, $M_2$ and $M_3$,
\begin{eqnarray}
  M_1 & = & \frac{5}{3} \tan^2 \theta_W M_2 \simeq 0.5 M_2
  \label{eq:M1M2} \\
  M_2 &= & \frac{\alpha_{\rm ew}}{\sin^2 \theta_W \alpha_s} M_3 
  \simeq 0.3 M_3 \label{eq:M2M3}
\end{eqnarray}
where $\alpha_{\rm ew}$ is the fine-structure constant and 
$\alpha_{s}$ is the strong coupling constant.
These relations come from the assumption that the gaugino mass
parameters unify at the unification scale as given by the gauge coupling
unification. 

\begin{table}
\scriptsize
\begin{center}
\begin{tabular}{lrrrrrrr} \hline
Scan & \multicolumn{1}{l}{normal} & 
\multicolumn{1}{l}{light} & 
\multicolumn{1}{l}{generous} & 
\multicolumn{1}{l}{high} & 
\multicolumn{1}{l}{high} & 
\multicolumn{1}{l}{light} &
\multicolumn{1}{l}{heavy} \\
     & &
\multicolumn{1}{l}{Higgs}               &            & 
\multicolumn{1}{l}{mass 1} & 
\multicolumn{1}{l}{mass 2} &
\multicolumn{1}{l}{higgsinos} & 
\multicolumn{1}{l}{gauginos} \\ \hline
$\mu^{\rm min}$ [GeV]  &   $-5000$ &   $-5000$ &  $-10000$ &    1000 &  
$-30000$ & $-100$ & 1000 \\
$\mu^{\rm max}$ [GeV]  &    5000 &    5000 &   10000 &   30000 &    
$-1000$ & 100 & 30000 \\
$M_2^{\rm min}$ [GeV]  &   $-5000$ &   $-5000$ &  $-10000$ &    1000 &    
1000 & $-1000$ & 1.9$\mu$/$-1.9\mu$ \\
$M_2^{\rm max}$ [GeV]  &    5000 &    5000 &   10000 &   30000 &   
30000 & 1000 & 2.1$\mu$/$-2.1\mu$ \\
$\tan \beta^{\rm min}$ &     1.2 &     1.2 &     1.2 &     1.2 &     
1.2 & 1.2 & 1.2 \\
$\tan \beta^{\rm max}$ &      50 &      50 &      50 &      50 &      
50 & 2.1 & 50 \\
$m_A^{\rm min}$ [GeV]  &       0 &       0 &       0 &       0 
&       0 & 0 & 0\\
$m_A^{\rm max}$ [GeV]  &    1000 &     150 &    3000 &   10000 &   
10000 & 1000 & 10000 \\
$m_0^{\rm min}$ [GeV]  &     100 &     100 &     100 &    1000 &    
1000 & 100 & 1000 \\
$m_0^{\rm max}$ [GeV]  &    3000 &    3000 &    5000 &   30000 &   
30000 & 3000 & 30000 \\
$A_b^{\rm min}$        & $-3m_0$ & $-3m_0$ & $-3m_0$ & $-3m_0$ & 
$-3m_0$ & $-3m_0$ & $-3m_0$ \\
$A_b^{\rm max}$        &  $3m_0$ &  $3m_0$ &  $3m_0$ &  $3m_0$ &  
$3m_0$  & $3m_0$ & $3m_0$ \\
$A_t^{\rm min}$        & $-3m_0$ & $-3m_0$ & $-3m_0$ & $-3m_0$ & 
$-3m_0$ & $-3m_0$ & $-3m_0$ \\
$A_t^{\rm max}$        &  $3m_0$ &  $3m_0$ &  $3m_0$ &  $3m_0$ &  
$3m_0$ & $3m_0$ & $3m_0$ \\
No. of models          &    4655 &    3342 &    3938 &    1000 &    
999 & 177 & 250  \\ \hline
\end{tabular}
\end{center}
\caption{The span of MSSM parameters used for the different scans.
For $\mu$ and $M_2$ the scans are uniform in the logarithms of
the parameters and for the other parameters they are uniform in the 
parameters themselves. The number of models refers to the number
of models satisfying all experimental constraints given in 
Section~\protect\ref{sec:AccCon}.}
\label{tab:scans}
\end{table}

\section{Feynman rules}
\label{sec:feyn}

The Feynman rules for the MSSM are given in Appendix~\ref{FeynApp}.
The rules given there are basically a compilation of the rules found
in Ref.~\cite{HaberKane,GuHa86,Mandl} but slightly rewritten in a form
suitable for general analytical calculations as well as numerical
implementations.

\section{MSSM parameter scans}
\label{sec:scans}

The general MSSM contains 63 free parameters \cite{jkg}, but with the
assumptions made in the previous sections in this chapter, we have
reduced the number of parameters to the seven parameters $\mu$, $M_2$,
$\tan \beta$, $m_A$, $m_0$, $A_b$ and $A_t$. It is however a
non-trivial task to sample this seven-dimensional parameter space in a
complete way. In an attempt to do this we have performed several
different scans in the parameter space, some of which are quite
general and some of which are more specialized to find interesting
regions of the parameter space.  In Table~\ref{tab:scans} we list the
different scans we have used in the explicit calculations in the
subsequent chapters. 

Remember, though, that the actual look of our scatter plots in
Chapters~\ref{RelDens} and \ref{Indirect} might change if different scans
were used. One should especially not pay any attention to the density
of points in different regions: it is just an artifact of our
scanning.

One might argue that the highest values of the massive parameters are
unnatural and require fine-tuning. We have in this thesis
taken a more phenomenological approach allowing even these high
values.

\chapter{Experimental Constraints}
\label{ExpCon}

For each set of parameters in the MSSM we have a unique model with
given mass spectrum, particle properties etc. Supersymmetry is
searched for both at accelerators and in dark matter searches and some
of these models will already be excluded. In the sections below, both
accelerator searches and dark matter searches will be discussed
briefly. Note, however, that we only use the experimental constraints
coming from accelerator searches to rule out models. We will, however,
compare with direct dark matter searches later on.

\section{Accelerator searches}
\label{sec:AccCon}

Since supersymmetry introduces many new particles these can affect 
what is seen at accelerators, either directly by finding a new 
particle or indirectly by changing some measured width or branching 
ratio. Below is given the currently most relevant bounds on the 
MSSM coming from accelerator searches.

\subsection{Neutralinos and Charginos}

The most effective limit on the chargino mass comes
from LEP2, through the search for the process $e^+
e^- \rightarrow \chi^+ \chi^-$ where $\chi^+$ is the lightest
chargino. This essentially puts the constraint $m_{\chi^+} >
\frac{1}{2}\sqrt{s}$ (to within a few GeV \cite{ChargLEP2})\@. 
From LEP2 the present bound on the chargino mass is \cite{LEP2}
\begin{equation} \label{eq:LEPmcha}
  m_{\chi^+} > 85 \mbox{ GeV.}
\end{equation}

The neutralinos can also be produced at LEP and would contribute to
the invisible width of the $Z$ boson. It is however difficult to
relate this to a model-independent limit on the neutralino mass. In
case there is no sfermion mixing, $m_{\tilde{f}_L}=m_{\tilde{f}_R}$,
and all squarks are degenerate in mass (except for $\tilde{t}_L$ and
$\tilde{t}_R$), the limit from LEP would be $m_\chi \gsim 23$ GeV when
$\tan \beta>3$\@.  We have implemented this limit by calculating
the invisible width of the $Z$ boson directly for each MSSM model.  If
the width, to which neutrinos, sneutrinos and neutralinos contribute,
is above the experimental limit
\begin{equation}
  \Gamma_Z^{\rm invisible} < 0.5024 \mbox{ GeV}
\end{equation}
the model is excluded. 

Due to our GUT assumption, Eq.~(\ref{eq:M1M2}), the lightest chargino
is never heavier than twice the neutralino mass though, and hence
the chargino mass bound is more effective in constraining the
neutralino mass in our models than the invisible $Z$ width bound is.

\subsection{Higgs bosons}

In the MSSM, the lightest Higgs boson, $H_2^0$, has a tree level mass
$m_{H_2^0}<m_Z$, but loop corrections can increase the mass up to
about 150 GeV (as described in Section~\ref{sec:EWHiggs}). If no Higgs
boson is seen up this mass, this would mean that the MSSM is ruled
out. The lightest Higgs boson is searched for at LEP2, where the main
processes are $e^+ e^- \rightarrow H_2^0 Z^0$ and $e^+ e^- \rightarrow
H_2^0 A$. The cross sections for these production channels are
proportional to $\sin^2 (\beta-\alpha)$ and $\cos^2 (\beta-\alpha)$
respectively and are hence complementary to each other. More details
about Higgs searches at LEP2 can be found in e.g.\ 
Ref.~\cite{HiggsLEP2}, from which we will use results for detection
prospects of supersymmetry at LEP2 in Section~\ref{sec:MuonPred}.

The present LEP2 bound on the lightest Higgs boson mass
is approximately given by \cite{LEP2}
\begin{equation} \label{eq:H2bound}
  m_{H_2^0} > 62.5 \mbox{ GeV,}
\end{equation}
but can be made more stringent by making the bound dependent on
$\sin^2 (\beta - \alpha)$. The bound then gets about 10 GeV higher at high 
$\sin^2 (\beta - \alpha)$, but we have not included this
more stringent mass bound here.

\subsection{Squarks and gluinos}

Squarks and gluinos are primarily searched for at hadron colliders. 
When produced they will eventually decay to the lightest neutralino
which will escape the detector leading to a missing energy
event. Since the squark and gluino decays depend very
much on the neutralino sector these limits will be quite model
dependent. Assuming the GUT assumptions,
Eqs.~(\ref{eq:M1M2})--(\ref{eq:M2M3}), to hold, one can derive a limit
on the squark and gluino masses as \cite{PDG}
\begin{eqnarray}
  m_{\tilde{q}} & > & 176 \mbox{ GeV} \\
  m_{\tilde{g}} & > & 154 \mbox{ GeV.}
\end{eqnarray}
There is however a controversy if there is still a window of light
gluinos, $\sim$ 1--4 GeV open or not.

\subsection{Sleptons}

Charged sleptons are searched for at $e^+e^-$ colliders where
sleptons can be produced and eventually decay to the lightest
neutralino resulting in missing energy. As for the squarks and gluino
searches, the bounds will depend on details in the neutralino sector,
especially on the neutralino mass. The LEP limits on the slepton
masses are \cite{PDG}
\begin{eqnarray}
  m_{\tilde{\nu}} & > & 37.1 \mbox{ GeV} \\
  m_{\tilde{e}} & > & 45 \mbox{ GeV if $m_\chi<41$ GeV} \\
  m_{\tilde{\mu}} & > & 45 \mbox{ GeV if $m_\chi<41$ GeV} \\
  m_{\tilde{\tau}} & > & 45 \mbox{ GeV if $m_\chi<38$ GeV}
\end{eqnarray}

\subsection{Other searches}

Even though we have chosen the MSSM parameters to avoid tree level
FCNCs, these can occur as one-loop corrections. It turns out that the
$b \rightarrow s \gamma$ decay width as measured by the CLEO
experiment \cite{CLEO} is an important constraint on the MSSM since
squark loops (in case of squark mixing) can change this width.  We
have used the following constraint on the decay width $b \rightarrow s
\gamma$,
\begin{equation}
  1.0 \times 10^{-4} < BR(b \rightarrow s \gamma) < 4.0 \times 10^{-4}
\end{equation}
where the branching ratio $BR(b \rightarrow s \gamma)$ is calculated
with QCD corrections included using the method in 
Ref.~\cite{bsg,bsgLarsPaolo}.

\section{Dark matter searches}

If neutralinos make up the dark matter in the Universe, they can also
be searched for by different direct and indirect dark matter searches.
The direct searches look for neutralino scattering off nuclei in a
detector.  This scattering releases some energy in the detector which
can be measured. The indirect searches look for indications of
neutralino annihilation, e.g.\ in the galactic halo producing
antiprotons, positrons or gamma rays or in the center of the Sun and
Earth producing high energy neutrinos which can be detected by
neutrino telescopes as explained in detail in Chapter~\ref{Indirect}.

We have not used any of these dark matter searches to exclude models, 
but we will compare the indirect detection rates in neutrino 
telescopes with direct detection rates in Chapter~\ref{Indirect}.

\chapter{Relic Density Calculations}
\label{RelDens}

Since the neutralino is a WIMP its annihilation 
cross section is expected to be of about the right magnitude to give 
a relic density $\Omega_\chi h^2 \sim 1$\@. The neutralino is not 
invented to solve the dark matter problem but comes from particle 
physics considerations and it is very interesting that it turns out to 
have a relic density in the right regime to be able to make up the dark 
matter in the Universe.

The relic density of neutralinos has been calculated by several
authors during the years
\cite{NeuLoop1,GriestSeckel,relcalc,McDonald,MizutaYamaguchi,SWO,DreesNojiri}
and a simple, but approximate, way of calculating the relic density
can be found in e.g.\ Ref.~\cite{KolbTurner}. This is rather
approximate since it assumes the cross section to be a nice function
expandable in $v^2$ where $v$ is the relative velocity of the
annihilating particles.  This expansion is often very bad, e.g.\ when
there are thresholds and resonances. These problems have been treated
in a semi-analytical way in Ref.~\cite{GriestSeckel}.  Instead of
using these approximate expressions we use the full cross section and
solve the Boltzmann equation numerically with the method given in
Ref.~\cite{GondoloGelmini,JEpaper5}. This way we automatically take
care of thresholds and resonances.

When any other supersymmetric particles are close in mass to the
lightest neutralino they will also be present at the time when the
neutralino freezes out in the early Universe.  When this happens so
called coannihilations can take place between all these supersymmetric
particles present at freeze-out.  This was first noted by Griest and
Seckel \cite{GriestSeckel} who investigated this for the rather
accidental case where squarks are of about the same mass as the
lightest neutralino.  Later, coannihilations between the lightest
neutralino and the lightest chargino were investigated by Mizuta and
Yamaguchi \cite{MizutaYamaguchi} for higgsinos lighter than the $W$
boson.  Drees and Nojiri \cite{DreesNojiri} investigated
coannihilations between the lightest and the
next-to-lightest neutralino, which are not as important as the
chargino-neutralino coannihilations.  Recently, Drees et al.\ 
\cite{NeuLoop1} reinvestigated coannihilations for light higgsinos
taking one-loop corrections to the neutralino and chargino masses into
account.

We have performed a more general analysis and evaluated the relic 
density $\Omega_\chi h^2$ including coannihilation processes between 
all charginos and neutralinos lighter than $2.1m_{\chi}$ for a general 
neutralino with any mass, $m_\chi$, and composition, $Z_g$. We have 
however not included coannihilations with squarks which occurs more 
accidentally than the in many cases unavoidable mass degeneracy 
between the lightest neutralinos and the lightest chargino.

In the following sections, the method by which the relic density is
evaluated when coannihilations are included \cite{JEpaper5} will be
described and our results will be presented and discussed.

\section{The Boltzmann equation}

We want to generalize the formulas in Ref.~\cite{GondoloGelmini} to
include coannihilations. We will do that by starting from the
expressions in Ref.~\cite{GriestSeckel} which will then be rewritten
into a more convenient form.

Consider annihilation of $N$ supersymmetric particles with masses
$m_{i}$ and internal degrees of freedom $g_{i}$.  Order them such that
$m_{1} \leq m_{2} \leq \cdots \leq m_{N-1} \leq m_{N}$. For the
lightest neutralino, the notation $m_1$ and $m_\chi$ will be used
interchangeably.  The evolution of the number density of particle $i$
is given by
\begin{eqnarray} \label{eq:Boltzmann}
  \frac{dn_{i}}{dt} 
  &=& 
  -3 H n_{i} 
  - \sum_{j=1}^N \langle \sigma_{ij} v_{ij} \rangle 
    \left( n_{i} n_{j} - n_{i}^{\rm{eq}} n_{j}^{\rm{eq}} \right) 
  \nonumber \\ 
  & & 
  - \sum_{j\ne i} 
  \big[ \langle \sigma'_{Xij} v_{ij} \rangle 
        \left( n_i n_X - n_{i}^{\rm{eq}} n_{X}^{\rm{eq}} \right)
      - \langle \sigma'_{Xji} v_{ij} \rangle
        \left( n_j n_X - n_{j}^{\rm{eq}} n_{X}^{\rm{eq}} \right)
  \big]
  \nonumber \\ 
  & &
  - \sum_{j\ne i} 
  \big[ \Gamma_{ij} 
        \left( n_i - n_i^{\rm{eq}} \right) 
      - \Gamma_{ji} 
        \left( n_j - n_j^{\rm{eq}} \right) 
  \big]
\end{eqnarray}
where
\begin{eqnarray}
  \sigma_{ij}  & = & \sum_X \sigma (\chi_i \chi_j \rightarrow X) \\
  \sigma'_{Xij} & = & \sum_Y \sigma (\chi_i X \rightarrow \chi_j Y) \\
  \Gamma_{ij}  & = & \sum_X \Gamma (\chi_i \rightarrow \chi_j X)
\end{eqnarray}
are the total annihilation cross sections, the inclusive scattering
cross sections and the inclusive decay rates respectively and $X$ and $Y$
are (sets of) standard model particles involved in the
interactions. The 'relative velocity' is defined by
\begin{equation}
  v_{ij} = \frac{\sqrt{(p_{i} \cdot p_{j})^2-m_{i}^2 m_{j}^2}}{E_{i} E_{j}}
\end{equation}
with $p_i$ and $E_i$ being the four-momentum and energy of particle $i$.
$n_i$ are the number densities of the corresponding
particles given by
\begin{equation}
  n_{i}^{\rm{eq}} = \frac{g_{i}}{(2\pi)^3} \int d^3{\bf p}_{i}f_{i}
\end{equation}
with ${\bf p}_i$ being the three-momentum of particle $i$ and $f_i$
being the equilibrium distribution function which in
the Maxwell-Boltzmann approximation is given by
\begin{equation}
  f_{i} = e^{-E_{i}/T}
\end{equation}
where $T$ is the temperature.
Since we assume that $R$-parity holds, all supersymmetric particles will
eventually decay to the LSP and we thus only have to consider the
total number density of supersymmetric particles $ n= \sum_{i=1}^N n_{i}$.
By summing Eq.~(\ref{eq:Boltzmann}) over all SUSY particles $i$ we get
the evolution equation for $n$,
\begin{equation}
  \frac{dn}{dt} = -3Hn - \sum_{i,j=1}^N \langle \sigma_{ij} v_{ij} \rangle 
  \left( n_{i}n_{j} - n_{i}^{\rm{eq}}n_{j}^{\rm{eq}} \right)
\end{equation}
where the terms on the second and third lines in
Eq.~(\ref{eq:Boltzmann}) cancel in the sum. 
The scattering rate of supersymmetric particles off particles in the
thermal background is much faster than their annihilation rate,
because the scattering cross sections $\sigma'_{Xij}$ are of the same
order of magnitude as the annihilation cross sections $\sigma_{ij}$
but the background particle density $n_X$ is much larger than each of
the supersymmetric particle densities $n_i$ when the former are
relativistic and the latter are non-relativistic, and so suppressed by
a Boltzmann factor. In this case, the $\chi_i$ distributions remain in
thermal equilibrium, and in particular their ratios are equal to the
equilibrium values,
\begin{equation}
  \frac{n_{i}}{n} \simeq \frac{n_{i}^{\rm{eq}}}{n^{\rm{eq}}}.
\end{equation}
We then get
\begin{equation} \label{eq:Boltzmann2}
  \frac{dn}{dt} =
  -3Hn - \langle \sigma_{\rm{eff}} v \rangle 
  \left( n^2 - n_{\rm{eq}}^2 \right)
\end{equation}
where
\begin{equation} \label{eq:sigmaveffdef}
  \langle \sigma_{\rm{eff}} v \rangle = \sum_{ij} \langle
  \sigma_{ij}v_{ij} \rangle \frac{n_{i}^{\rm{eq}}}{n^{\rm{eq}}}
  \frac{n_{j}^{\rm{eq}}}{n^{\rm{eq}}}.
\end{equation}

\section{Thermal averaging}
\label{sec:thermav}

Now reformulate the thermal average, Eq.~(\ref{eq:sigmaveffdef}), into
more convenient expressions. 

First, using the Maxwell-Boltzmann approximation we get 
\cite{GondoloGelmini,JEpaper5} 
\begin{equation} \label{eq:neq}
  n_{\rm eq} = \sum_i n_i^{\rm eq} = \frac{T}{2\pi^2} \sum_i g_i m_{i}^2
  K_{2} \left( \frac{m_{i}}{T}\right)
\end{equation}
where $K_{2}$ is the modified Bessel function of the second kind of 
order 2. Then rewrite
Eq.~(\ref{eq:sigmaveffdef}) as
\begin{equation} \label{eq:sigmaveff}
  \langle \sigma_{\rm{eff}} v \rangle =
  \frac{A}{n_{\rm{eq}}^2}
\end{equation}
where
\begin{equation} 
  A = \sum_{ij} \int W_{ij} \frac{g_i f_i d^3p_i}{(2\pi)^3 2E_i}
  \frac{g_j f_j d^3p_j}{(2\pi)^3 2E_j} .
\label{eq:Aij2}
\end{equation}
is the total annihilation rate per unit volume at temperature $T$.
$W_{ij}$ is the annihilation rate and is related to the cross section
through\footnote{The quantity $w_{ij}$ in Ref.\ \cite{SWO} is
  $W_{ij}/4$.}
\begin{equation} \label{eq:Wijcross}
  W_{ij} = 4 p_{ij} \sqrt{s} \sigma_{ij} = 4 \sigma_{ij} \sqrt{(p_i
\cdot p_j)^2 - m_i^2 m_j^2} = 4 E_{i} E_{j} \sigma_{ij} v_{ij}
\end{equation}
where 
\begin{equation}
   p_{ij} = 
  \frac{\left[s-(m_i+m_j)^2\right]^{1/2}
  \left[s-(m_i-m_j)^2\right]^{1/2}}{2\sqrt{s}}.
\end{equation}

For a two-body final state, $W_{ij}$ is given by
\begin{equation} \label{eq:Wij2body}
  W^{\rm{2-body}}_{ij} = \frac{|\bf k|}{16\pi^2 g_i g_j S_f \sqrt{s}}
  \sum_{\rm{internal~d.o.f.}} \int \left| {\cal M} \right|^2
  d\Omega ,
\end{equation}
where $\bf k$ is the final center-of-mass momentum, $S_f$ is a symmetry
factor equal to 2 for identical final particles, and the integration
is over the outgoing directions of one of the final particles.
As usual, an average over initial internal degrees of freedom is 
performed.

Now consider annihilation of two particles, $i$ and $j$, with masses
$m_{i}$ and $m_{j}$ and statistical degrees of freedom $g_{i}$ and
$g_{j}$. If we use Boltzmann statistics (good for
$T\lsim m$) we can put Eq.~(\ref{eq:Aij2}) into the form
\begin{equation} \label{eq:Aij3}
  A = \sum_{ij}
  \int g_i g_j W_{ij} e^{-E_{i}/T} e^{-E_{j}/T} 
\frac{d^3p_i}{(2\pi)^3 2E_i}
  \frac{d^3p_j}{(2\pi)^3 2E_j},
\end{equation}
where ${\bf p}_{i}$ and ${\bf p}_{j}$ are the three-momenta and
$E_{i}$ and $E_{j}$ are the energies of the colliding particles.
We can now follow the procedure in Ref.~\cite{GondoloGelmini} as done
in Ref.~\cite{JEpaper5} and perform some of the integrations in
Eq.~(\ref{eq:Aij3}) to arrive at
\begin{equation} \label{eq:As}
  A =  \frac{T}{32
  \pi^4} \sum_{ij} \int_{(m_i+m_j)^2}^\infty ds g_ig_jp_{ij} W_{ij}
  K_{1} \left( \frac{\sqrt{s}}{T}\right)
\end{equation}
where $K_{1}$ is the modified Bessel function of the second kind of 
order 1.

Now we have what we need to perform the sum in
Eq.~(\ref{eq:Aij2}) to get $\langle \sigma_{\rm{eff}} v
\rangle$. Let
\begin{eqnarray} \label{eq:weff}
  W_{\rm{eff}} & = & \sum_{ij}\frac{p_{ij}}{p_{\rm{eff}}}
  \frac{g_ig_j}{g_1^2} W_{ij} \nonumber \\
  & = &
  \sum_{ij} \sqrt{\frac{[s-(m_{i}-m_{j})^2][s-(m_{i}+m_{j})^2]}
  {s(s-4m_1^2)}} \frac{g_ig_j}{g_1^2} W_{ij}
\end{eqnarray}
with
\begin{equation} \label{eq:peff}
   p_{\rm{eff}} = p_{11} = \frac{1}{2}
  \sqrt{s-4m_{1}^2}.
\end{equation}
Since $W_{ij}(s) = 0 $ for $s \le (m_i+m_j)^2$, the radicand in
Eq.~(\ref{eq:weff}) is never negative.  

Eq.~(\ref{eq:As}) can be written in a form more suitable for numerical
integration by using $p_{\rm{eff}}$ instead of $s$ as integration
variable.  From Eq.~(\ref{eq:peff}), $ ds = 8 p_{\rm{eff}}
dp_{\rm{eff}} $, and we have
\begin{equation}
\label{eq:Apeff}
  A = \frac{g_1^2 T}{4 \pi^4} \int_{0}^\infty dp_{\rm eff}
  p^2_{\rm eff} W_{\rm eff} K_{1} \left( \frac{\sqrt{s}}{T}\right).
\end{equation}

\begin{figure}
  \centerline{\epsfig{file=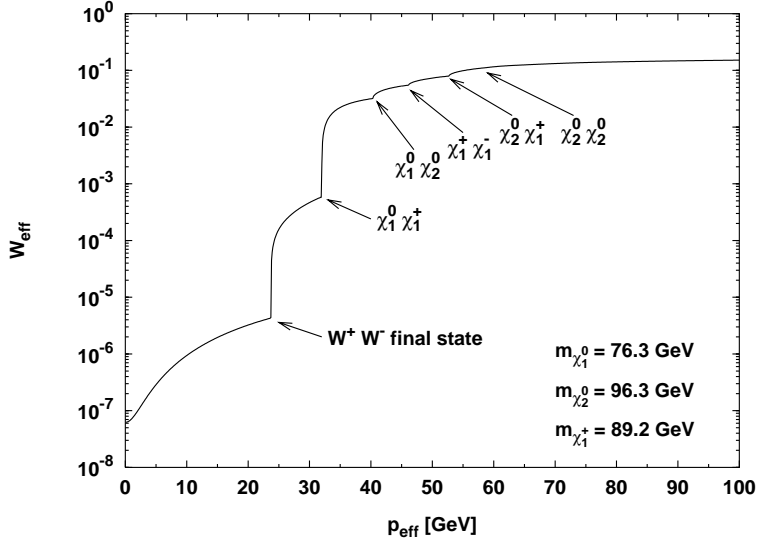,width=0.75\textwidth}}
  \caption{The effective invariant annhiliation rate, $W_{\rm eff}$, 
  as a function of $p_{\rm eff}$ for model 1 in
  Table~\protect\ref{tab:reprmod}. The final state threshold for annihilation 
  into $W^+ W^-$ and the coannihilation thresholds, as given by 
  Eq.~(\protect\ref{eq:weff}), are indicated.  The $\chi_2^0 
  \chi_2^0$ coannihilation threshold is too small to be seen.}
  \label{fig:effrate}
\end{figure}

We can then finally write Eq.~(\ref{eq:sigmaveff}) as
\begin{equation} \label{eq:sigmavefffin2}
  \langle \sigma_{\rm{eff}}v \rangle = \frac{\int_0^\infty
  dp_{\rm{eff}} p_{\rm{eff}}^2 W_{\rm{eff}} K_1 \left(
  \frac{\sqrt{s}}{T} \right) } { m_1^4 T \left[ \sum_i \frac{g_i}{g_1}
  \frac{m_i^2}{m_1^2} K_2 \left(\frac{m_i}{T}\right) \right]^2}\,.
\end{equation}
This expression is very similar to the case without coannihilations,
the difference being the denominator and the replacement of the
invariant rate with the effective invariant rate.

\begin{figure}
  \centerline{\epsfig{file=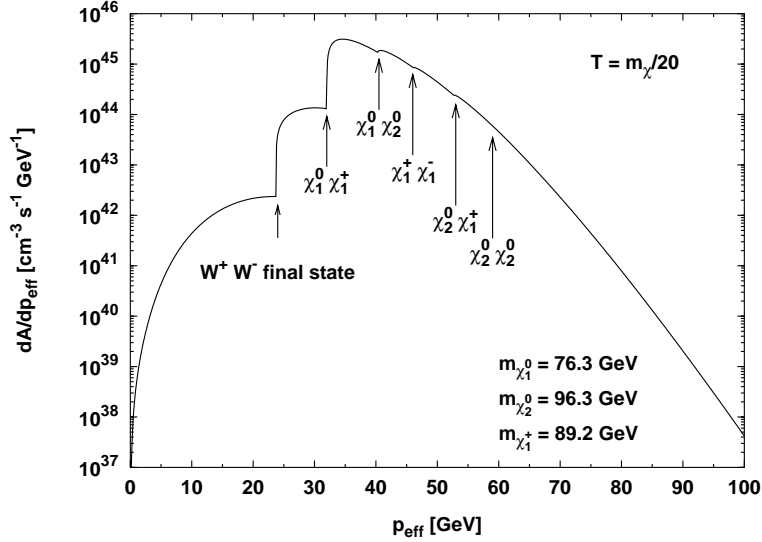,width=0.75\textwidth}}
  \caption{Total differential annihilation rate per unit volume, 
    $dA/dp_{\rm eff} = (T/\pi^4) p_{\rm eff}^2 K_1(p_{\rm eff},T)
  W_{\rm eff}$, for the same model as in
  Fig.~\protect\ref{fig:effrate}.  We have chosen to evaluate
  $dA/dp_{\rm eff}$ for $T=m_\chi/20$ which is a typical value at
  freeze-out. The Boltzmann suppression at higher $p_{\rm eff}$ should
  be evident.}
  \label{fig:k1effrate}
\end{figure}

In the effective annihilation rate, $W_{\rm eff}$,
coannihilations appear as thresholds at $\sqrt{s}$ equal to the sum of
the masses of the coannihilating particles.  We show an example in
Fig.~\ref{fig:effrate} where it is clearly seen that the
coannihilation thresholds appear in the effective invariant rate just
as final state thresholds do. 
In Fig.~\ref{fig:k1effrate} we show the
differential of $A$ with respect to $p_{\rm eff}$, $dA/dp_{\rm eff}$.
The Boltzmann suppression at higher $p_{\rm eff}$, contained in the
exponential decay of $K_{1}$, is clearly visible.  We have in
Fig.~\ref{fig:k1effrate} evaluated the modified Bessel function at the
temperature $T=m_{\chi}/20$ which is a typical freeze-out temperature.
When the temperature is higher, the peak will shift to the right and
when it is lower it will shift to the left.  For the particular model
shown in Figs.~\ref{fig:effrate}--\ref{fig:k1effrate}, the relic
density is evaluated to be $\Omega_\chi h^2=0.030$ when
coannihilations are included and $\Omega_\chi h^2=0.18$ when they are
not.

We end this section with a comment on the internal degrees of
freedom $g_i$.  A neutralino is a Majorana fermion and has two
internal degrees of freedom, $g_{\chi_{i}^0}=2$. A chargino can be
treated either as two separate species $\chi_{i}^+$ and
$\chi_{i}^-$, each with internal degrees of freedom
$g_{\chi^+}=g_{\chi^-}=2$, or, more simply, as a single species
$\chi_{i}^\pm$ with $g_{\chi_{i}^\pm}=4$ internal degrees of
freedom.  

\section{Reformulation of the Boltzmann equation}

We can now put Eq.~(\ref{eq:Boltzmann2}) into a more convenient form
by instead of the number density considering the ratio of the number
density to the entropy density,
\begin{equation} \label{eq:ydef}
  Y = \frac{n}{s}
\end{equation}
and instead of having the time $t$ as independent variable we choose
$x=m_1/T$ with $m_1$ being the LSP mass and $T$ being the temperature.
By following Ref.~\cite{GondoloGelmini} we can write the evolution
equation as
\begin{equation} \label{eq:Boltzmann4}
  \frac{dY}{dx} = - \sqrt{\frac{\pi}{45G}} \frac{g_{*}^{1/2}m_1}{x^2}
  \langle \sigma_{\rm{eff}} v \rangle \left( Y^2 -
  Y_{\rm{eq}}^2 \right) 
\end{equation}
where $Y_{\rm eq}$ is given by
\begin{equation}
  Y_{\rm{eq}} = \frac{n_{\rm{eq}}}{s} = 
  \frac{45 x^2}{4 \pi^4 h_{\rm{eff}}(T)} \sum_i g_i
  \left( \frac{m_i}{m_1} \right)^2 K_{2} \left( x 
\frac{m_{i}}{m_1}\right)
\end{equation}
with $G$ being the Gravitational constant, the parameter $g_{*}^{1/2}$
being defined as
\begin{equation}
  g_{*}^{1/2} = \frac{h_{\rm{eff}}}{\sqrt{g_{\rm{eff}}}}
  \left( 1+\frac{T}{3h_{\rm{eff}}} \frac{d h_{\rm{eff}}}{dT}
  \right)
\end{equation}
and $g_{\rm eff}$ and $h_{\rm eff}$ being the effective degrees of
freedom as given in the usual parameterizations of the energy and entropy
densities
\begin{equation} \label{eq:geffheff}
  \rho = g_{\rm{eff}}(T) \frac{\pi^2}{30} T^4
  \quad , \quad 
  s = h_{\rm{eff}}(T) \frac{2\pi^2}{45} T^3.
\end{equation}

We have evaluated $g_{\rm eff}$, $h_{\rm eff}$ and $g_*^{1/2}$ using
the methods given in Ref.~\cite{GondoloGelmini} assuming the QCD phase
transition to occur at 150 MeV\@. Our results are not sensitive to the
value of $T_{QCD}$ though, since the neutralino freeze-out temperature
is always much larger than $T_{QCD}$.

To obtain the relic density we should integrate
Eq.~(\ref{eq:Boltzmann4}) from $x=0$ to $x_0=m_\chi/T_0$ where $T_0$
is the photon temperature of the Universe today. The relic density
today in units of the critical density is then given by
\begin{equation}
  \Omega_\chi = \rho_\chi^0/\rho_{\rm
  crit}=m_\chi s_0 Y_0/\rho_{\rm crit}
\end{equation}
where $\rho_{\rm crit}=3 H^2/8 \pi G$ is the critical density, $s_0$
is the entropy density today and $Y_0$ is the solution of the
integration of Eq.~(\ref{eq:Boltzmann4}). With a
background radiation temperature of $T_0=2.726$ K we obtain
\begin{equation} \label{eq:omegah2}
  \Omega_\chi h^2 = 2.755\times 10^8 \frac{m_\chi}{\mbox{GeV}} Y_0.
\end{equation}

\section{Annihilation cross sections}
\label{sec:AnnChannels}

\begin{table}
\small
\begin{center}
\begin{tabular}{lll} \hline 
  Initial state & Final state & Diagrams \\ \hline
   & $H_1 H_1$, $H_1 H_2$, $H_2 H_2$, $H_3 H_3$ &
  $t(\chi_k^0)$, $u(\chi_k^0)$, $s(H_{1,2})$ \\
   & $H_1 H_3$, $H_2 H_3$ &
  $t(\chi_k^0)$, $u(\chi_k^0)$, $s(H_{3})$, $s(Z^0)$ \\
   & $H^- H^+$ &
  $t(\chi_c^+)$, $u(\chi_c^+)$, $s(H_{1,2})$, $s(Z^0)$ \\
   & $Z^0 H_1$, $Z^0 H_2$ &
  $t(\chi_k^0)$, $u(\chi_k^0)$, $s(H_{3})$, $s(Z^0)$ \\
  $\chi_i^0 \chi_j^0$ & $Z^0 H_3$ &
  $t(\chi_k^0)$, $u(\chi_k^0)$, $s(H_{1,2})$ \\
   & $W^- H^+$, $W^+ H^-$ &
  $t(\chi_c^+)$, $u(\chi_c^+)$, $s(H_{1,2,3})$ \\
   & $Z^0 Z^0$ &
  $t(\chi_k^0)$, $u(\chi_k^0)$, $s(H_{1,2})$ \\
   & $W^- W^+$ &
  $t(\chi_c^+)$, $u(\chi_c^+)$, $s(H_{1,2})$, $s(Z^0)$ \\
   & $f \bar{f}$ &
  $t(\tilde{f}_{L,R})$, $u(\tilde{f}_{L,R})$, $s(H_{1,2,3})$,
  $s(Z^0)$ \\ \hline  
   & $H^+ H_1$, $H^+ H_2$ &
  $t(\chi_j^0)$, $u(\chi_d^+)$, $s(H^+)$, $s(W^+)$ \\
   & $H^+ H_3$ &
  $t(\chi_j^0)$, $u(\chi_d^+)$, $s(W^+)$ \\
   & $W^+ H_1$, $W^+ H_2$ &
  $t(\chi_j^0)$, $u(\chi_d^+)$, $s(H^+)$, $s(W^+)$ \\
   & $W^+ H_3$ &
  $t(\chi_j^0)$, $u(\chi_d^+)$, $s(H^+)$ \\
  $\chi_c^+ \chi_i^0$ & $H^+ Z^0$ &
  $t(\chi_j^0)$, $u(\chi_d^+)$, $s(H^+)$ \\
   & $\gamma H^+$ &
  $t(\chi_c^+)$, $s(H^+)$ \\
   & $W^+ Z^0$ &
  $t(\chi_j^0)$, $u(\chi_d^+)$, $s(W^+)$ \\
   & $\gamma W^+$ &
  $t(\chi_c^+)$, $s(W^+)$ \\
   & $u \bar{d}$ &
  $t(\tilde{d}_{L,R})$, $u(\tilde{u}_{L,R})$, $s(H^+)$, $s(W^+)$ \\
   & $\nu \bar{\ell}$ &
  $t(\tilde{\ell}_{L,R})$, $u(\tilde{\nu}_{L})$, $s(H^+)$, $s(W^+)$
  \\ \hline
   & $H_1 H_1$, $H_1 H_2$, $H_2 H_2$, $H_3 H_3$ &
  $t(\chi_e^+)$, $u(\chi_e^+)$, $s(H_{1,2})$ \\
   & $H_1 H_3$, $H_2 H_3$ &
  $t(\chi_e^+)$, $u(\chi_e^+)$, $s(H_{3})$, $s(Z^0)$ \\
   & $H^+ H^-$ &
  $t(\chi_i^0)$, $s(H_{1,2})$, $s(Z^0,\gamma)$ \\
   & $Z^0 H_1$, $Z^0 H_2$ &
  $t(\chi_e^+)$, $u(\chi_e^+)$, $s(H_{3})$, $s(Z^0)$ \\
   & $Z^0 H_3$ &
  $t(\chi_e^+)$, $u(\chi_e^+)$, $s(H_{1,2})$ \\
   & $H^+ W^-$, $W^+ H^-$ &
  $t(\chi_e^+)$, $s(H_{1,2,3})$ \\
  $\chi_c^+ \chi_d^-$ & $Z^0 Z^0$ &
  $t(\chi_e^+)$, $u(\chi_e^+)$, $s(H_{1,2})$ \\
   & $W^+ W^-$ &
  $t(\chi_i^0)$, $s(H_{1,2})$, $s(Z^0, \gamma)$ \\
   & $\gamma \gamma$ (only for $c=d$)&
  $t(\chi_c^+)$, $u(\chi_c^+)$ \\
   & $Z^0 \gamma$ &
  $t(\chi_d^+)$, $u(\chi_c^+)$ \\
   & $u \bar{u}$ &
  $t(\tilde{d}_{L,R})$, $s(H_{1,2,3})$, $s(Z^0, \gamma)$ \\
   & $\nu \bar{\nu}$ &
  $t(\tilde{\ell}_{L,R})$, $s(Z^0)$ \\
   & $\bar{d} d$ &
  $t(\tilde{u}_{L,R})$, $s(H_{1,2,3})$, $s(Z^0, \gamma)$ \\
   & $\bar{\ell} \ell$ &
  $t(\tilde{\nu}_{L})$, $s(H_{1,2,3})$, $s(Z^0, \gamma)$ \\ \hline
   & $H^+ H^+$ &
  $t(\chi_i^0)$, $u(\chi_i^0)$ \\
  $\chi_c^+ \chi_d^+$ & $H^+ W^+$ &
  $t(\chi_i^0)$, $u(\chi_i^0)$ \\
   & $W^+ W^+$ &
  $t(\chi_i^0)$, $u(\chi_i^0)$ \\ \hline
\end{tabular}
\caption{All two-body final states for which annihilation cross
sections are calculated. The indices $i,j,k=1,2,3,4$
and the indices $c,d,e=1,2$. Note that the coannihilation channels are
only important when the mass difference is not too big. $t$, $u$ and
$s$ refers to which channel the annihilation goes through and the
particle inside the brackets is the particle in the propagator. $u$, 
$\tilde{u}$, $d$, $\tilde{d}$, $\nu$, $\tilde{\nu}$, $\ell$, 
$\tilde{\ell}$, $f$ and $\tilde{f}$ are generic notations for any up-
and down-type (s)quark, (s)neutrino, (s)lepton and (s)fermion.
A sum of diagrams over (s)fermion generation indices and over the
neutralino and chargino indices $k$ and $e$ is understood.}
\label{tab:annchannels}
\end{center}
\end{table}

We have calculated all two-body final state cross sections at tree
level for neu\-tral\-ino-neutral\-ino, neutralino-chargino and
chargino-chargino annihilation. A complete list is given in
Table~\ref{tab:annchannels}. 

Since we have so many different diagrams contributing, we have to use 
some method where the diagrams can be calculated efficiently.  To 
achieve this, we classify diagrams according to their topology ($s$-, 
$t$- or $u$-channel) and to the spin of the particles involved.  We 
then compute the helicity amplitudes for each type of diagram 
analytically with {\sc Reduce}~\cite{reduce} using general expressions 
for the vertex couplings.  Further details will be found in 
Ref.~\cite{paolohel}.  

The strength of the helicity amplitude method is that the analytical
calculation of a given diagram only has to be performed once and the
summing of the contributing diagrams for each given set of initial and
final states can be done numerically afterwards.

\section{Numerical methods}
\label{sec:nummeth}

In this section we describe the numerical methods we use to evaluate
the effective invariant rate and its thermal average, and to integrate
the density evolution equation.

We obtain the effective invariant rate numerically as follows.  We
generate {\sc Fortran} routines for the helicity amplitudes of all
types of diagrams automatically with {\sc Reduce}, as explained in the
previous section. We sum the Feynman diagrams numerically for each
annihilation channel $ij\to kl$. We then sum the squares of the
helicity amplitudes and sum the contributions of all
annihilation channels. Explicitly, we compute
\begin{equation} \label{eq:helsum}
  {d W_{\rm eff} \over d \cos\theta } = 
\sum_{ijkl}
{p_{ij} p_{kl} \over 32 \pi S_{kl} \sqrt{s} }
\sum_{\rm helicities}
   \left| \sum_{\rm diagrams}  {\cal M}(ij \to kl) \right|^2 
\end{equation}
where $\theta$ is the angle between particles $k$ and $i$.  We finally
integrate numerically over $\cos\theta$ by means of adaptive gaussian
integration.

In rare cases, we find resonances in the $t$- or $u$-channels. For the
process $ij\to kl$, this can occur when $m_i < m_k$ and $m_j > m_l$:
at certain values of $\cos\theta$, the momentum transfer is time-like
and matches the mass of the exchanged particle.  We have regulated the
divergence by assigning a small width of a few GeV to the
neutralinos and charginos.  Our results are not sensitive to
the choice of this width, though.

The calculation of the effective invariant rate $W_{\rm eff}$ is the
most time-consuming part. Fortunately, thanks to the remarkable
feature of Eq.~(\ref{eq:sigmavefffin2}), $W_{\rm eff}(p_{\rm eff})$
does not depend on the temperature $T$, and it can be tabulated once
for each model.  We have to make sure that the maximum $p_{\rm eff}$
in the table is large enough to include all important resonances,
thresholds and coannihilation thresholds.  As an extreme case,
consider when the effective invariant rate at high $p_{\rm eff}$ is
$10^{10}$ times higher than at $p_{\rm eff}=0$. For a typical
freeze-out temperature of $T=m_\chi/20$, the Boltzmann suppression of
high $p_{\rm eff}$ contained in $K_1$ in Eq.~(\ref{eq:sigmavefffin2})
results in that contributions to the thermal average from values of
$p_{\rm eff}$ beyond $\sim 1.5 m_{\chi}$ are negligible. For
coannihilations, this value of $p_{\rm eff}$ corresponds to a mass of
the coannihilating particle of $\sim 1.8m_{\chi}$.  To be on the safe
side all over parameter space, we include coannihilations whenever the
mass of the coannihilating particle is less than $2.1m_\chi$, even if
typically coannihilations are important only for masses less than $1.4
m_\chi$. For extra safety, we tabulate $W_{\rm eff}$ from $p_{\rm
  eff}=0$ up to $p_{\rm eff}=20 m_\chi$, more densely in the important
low $p_{\rm eff}$ region than elsewhere.  We further add several
points around resonances and thresholds, both explicitly and in an
adaptive manner.

To perform the thermal average in Eq.~(\ref{eq:sigmavefffin2}), we
integrate over $p_{\rm eff}$ by means of adaptive gaussian
integration, using a spline routine to interpolate in the $(p_{\rm
  eff},W_{\rm eff})$ table. To avoid numerical problems in the
integration routine or in the spline routine, we split the integration
interval at each sharp threshold. We also explicitly check for each
MSSM model that the spline routine behaves well at thresholds and
resonances.

We finally integrate the density evolution
equation~(\ref{eq:Boltzmann4}) numerically from $x=2$, where the
density still tracks the equilibrium density, to $x_0=m_\chi/T_0$. We
use an implicit trapezoidal method with adaptive stepsize. The relic
density at present is then evaluated with Eq.~(\ref{eq:omegah2}).

A more detailed description of the numerical methods will be found in
a future publication \cite{gondoloedsjo}.

\section{Results}

\begin{figure}
  \centerline{\epsfig{file=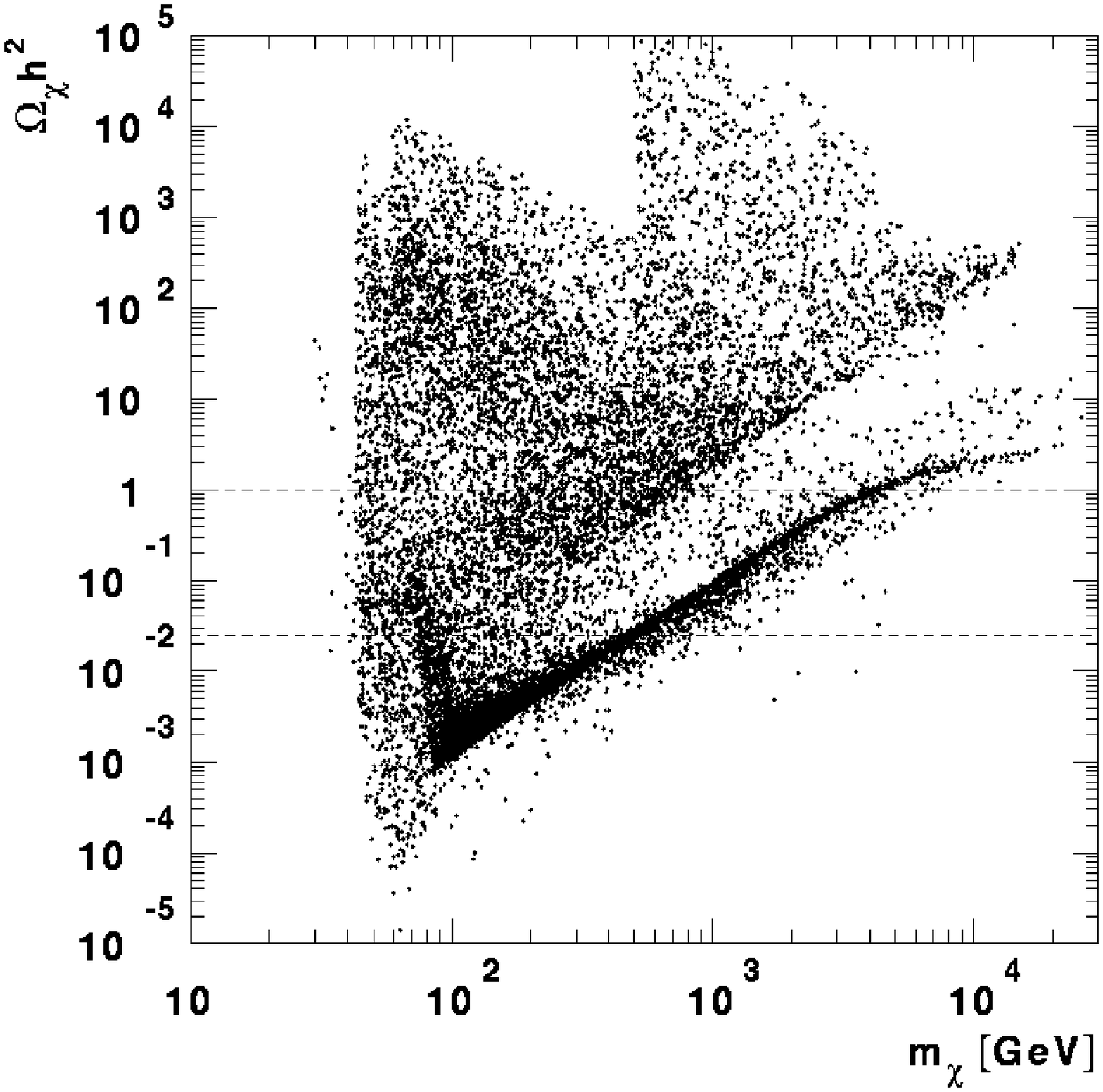,width=0.49\textwidth}
  \epsfig{file=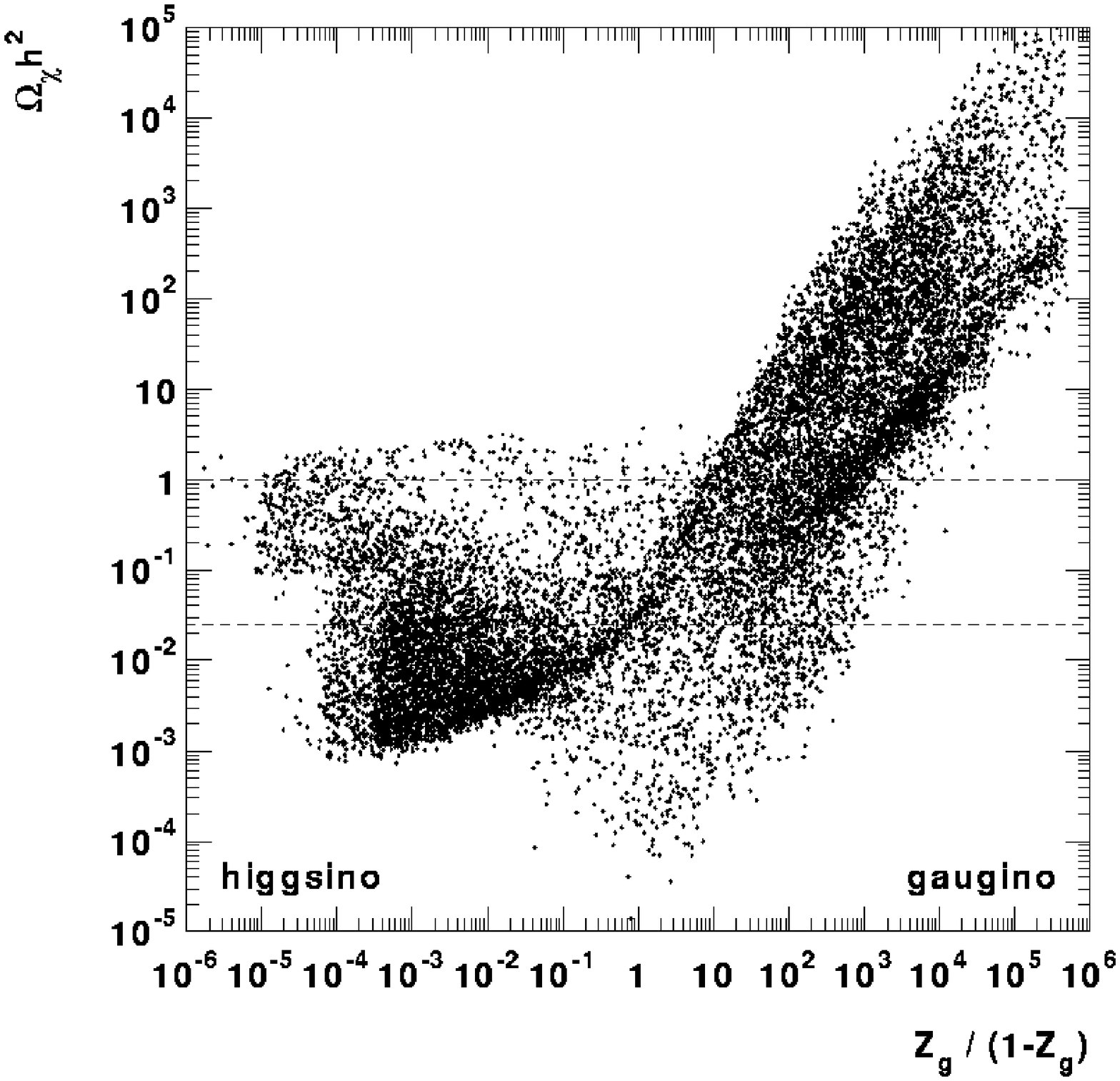,width=0.49\textwidth}}
  \caption{
    Neutralino relic density including neutralino and chargino
    coannihilations versus a) neutralino mass $m_\chi$ and b)
    neutralino composition $Z_g/(1-Z_g)$.  The horizontal lines
    indicate the cosmologically interesting region $0.025 < \Omega_\chi
    h^2 <1$.}
  \label{fig:oh2vsmx}
\end{figure}

\begin{figure}
  \centerline{\epsfig{file=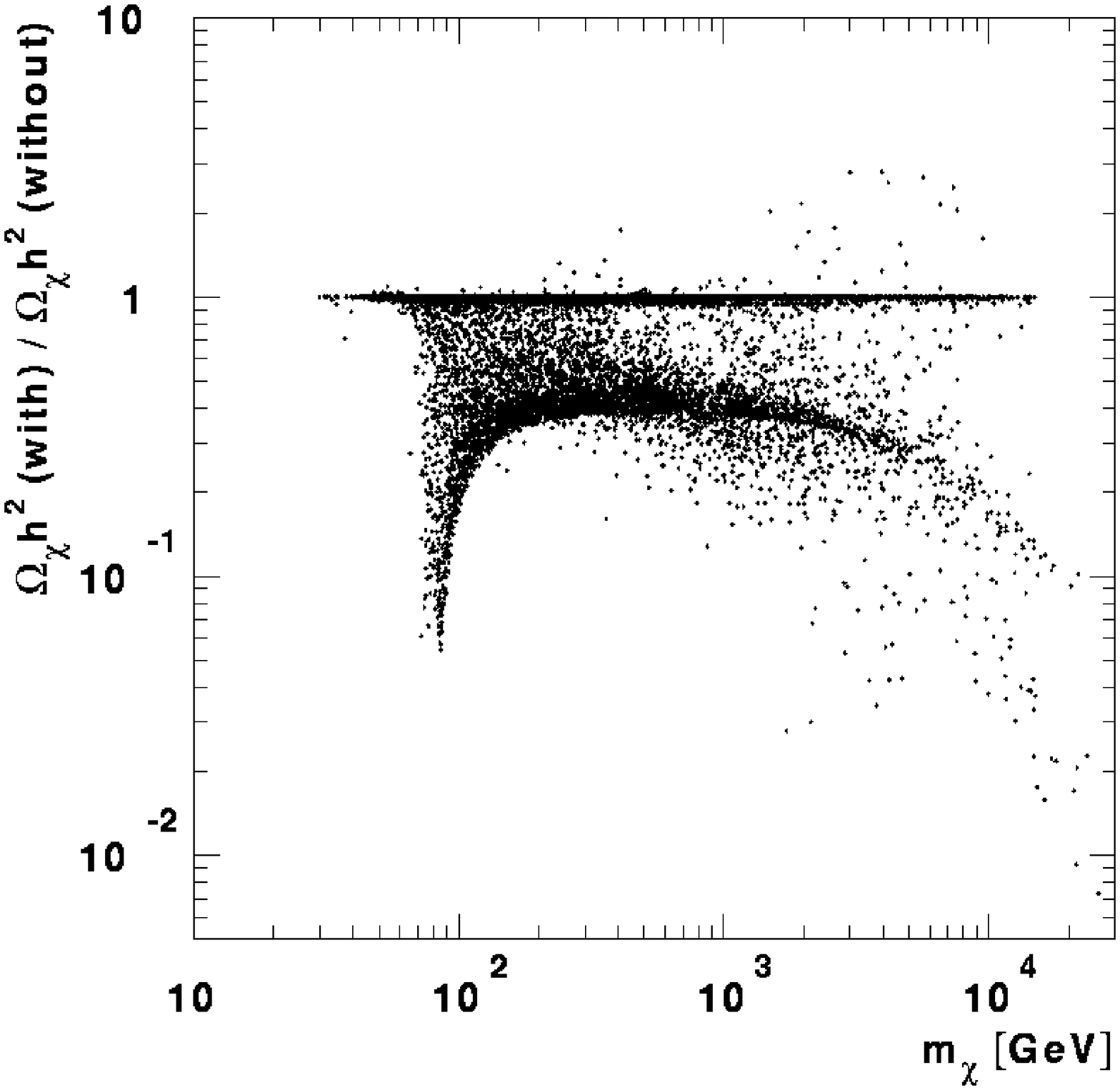,width=0.49\textwidth}
  \epsfig{file=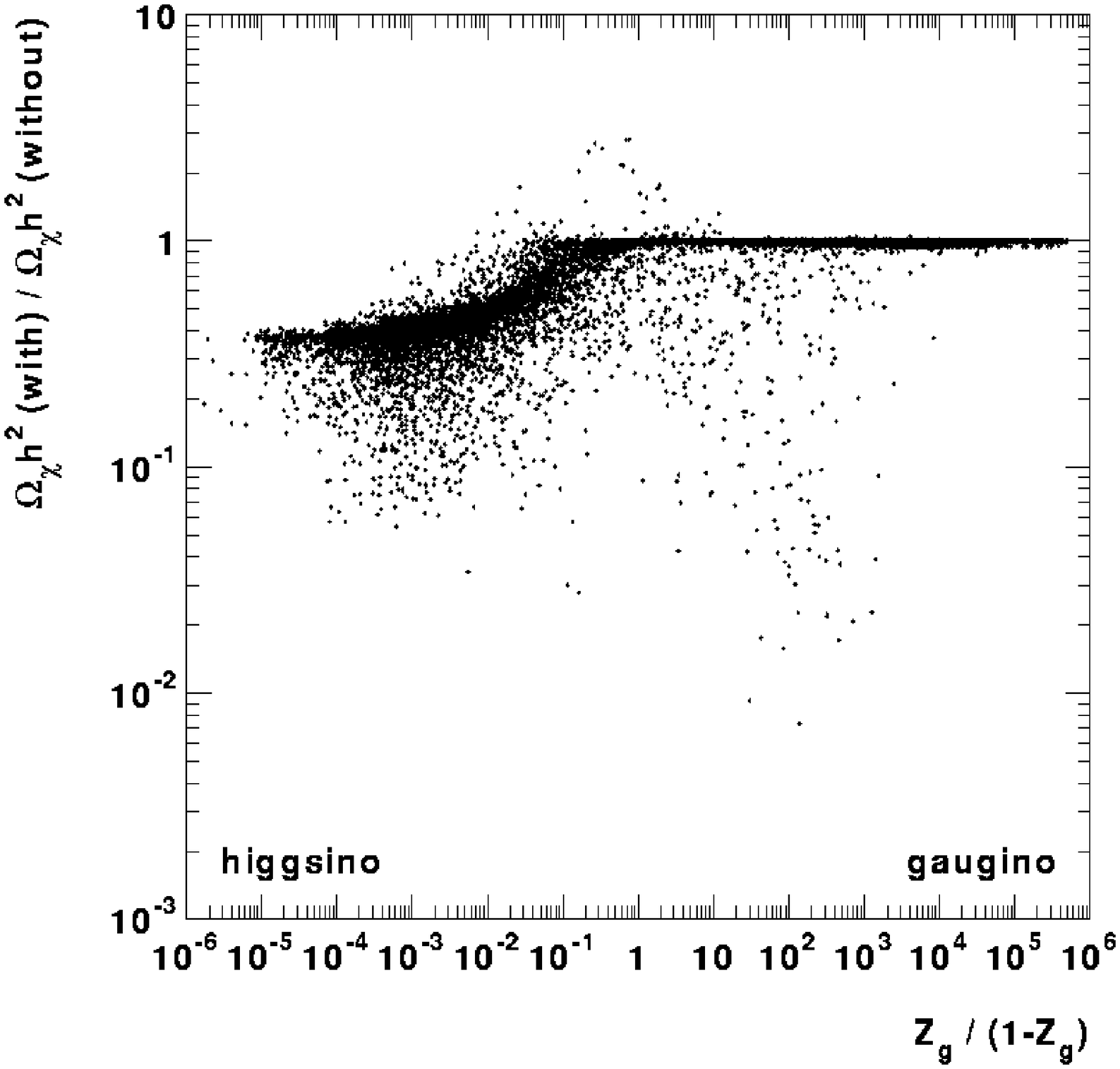,width=0.49\textwidth}}
  \caption{
    Ratio of the neutralino relic densities with and without
    neutralino and chargino coannihilations versus a) neutralino
    mass $m_\chi$ and b) neutralino composition $Z_g/(1-Z_g)$. }
  \label{fig:ratiovsmx}
\end{figure}

We now present the results of our relic density calculations for all
the models in Table~\ref{tab:scans}.  We will focus on the effect of
coannihilations, since this is the first time they are included for
general neutralino masses and compositions.

Fundamentally, we are interested in how the inclusion of
coannihilations modifies the cosmologically interesting region and the
cosmological bounds on the neutralino mass. We define the
cosmologically interesting region as $0.025 < \Omega_{\chi} h^2 < 1$.
In this range of $\Omega_\chi h^2$ the neutralino can constitute most
of the dark matter in galaxies and the age of the Universe is long
enough to be compatible with observations (see Chapter~\ref{Intro}).
The lower bound of 0.025 is somewhat arbitrary, and even if
$\Omega_{\chi}h^2$ would be less than $0.025$ the neutralinos would
still be relic particles, but only a minor fraction of the dark matter
in the Universe.

We start with a short general discussion and then present more details
in the following subsections.

Fig.~\ref{fig:oh2vsmx} shows the neutralino
relic density $\Omega_\chi h^2$ with coannihilations included versus
the neutralino mass $m_\chi$ and the neutralino composition $Z_g/(1-Z_g)$,
respectively.  The lower edge on neutralino masses comes essentially
from the LEP bound on the chargino mass, Eq.~(\ref{eq:LEPmcha}). 

The neutralino is a good dark matter candidate in the cosmologically
interesting region limited by the two horizontal lines. There are
clearly models with cosmologically interesting relic densities for a
wide range of neutralino masses and compositions.  The cosmologically
interesting region will be discussed more in
Section~\ref{sec:cosmregion}.

The effect of neutralino and chargino coannihilations on the value of
the relic density is summarized in Fig.~\ref{fig:ratiovsmx}, where we
plot the ratio of the neutralino relic densities with and without
coannihilations versus the neutralino mass $m_\chi$ and the neutralino
composition $Z_g/(1-Z_g)$. In many models, coannihilations reduce the
relic density by more than a factor of ten, and in some others they
increase it by a small factor. Coannihilations increase the relic
density if the effective annihilation cross section $\langle
\sigma_{\rm eff} v \rangle < \langle \sigma_{11} v_{11} \rangle$.
Recalling that $\langle \sigma_{\rm eff} v \rangle$ is the average of
the coannihilation cross sections (see Eq.~(\ref{eq:sigmaveffdef})), this
occurs when most of the coannihilation cross sections are smaller than
$\langle \sigma_{11} v_{11} \rangle$ and the mass differences of the
coannihilating particles are small.

Table~\ref{tab:reprmod} lists some representative models where
coannihilations are important plus one model where coannihilations
are negligible. Example 1 contains a light higgsino-like neutralino,
example 2 a heavy higgsino-like neutralino.  Examples 3 and 4 have
$|\mu| \sim |M_1|$, and example 5 has a very pure gaugino-like
neutralino. Example 6 is a model with a gaugino-like neutralino for
which coannihilations are not important.

\begin{table}
\footnotesize
\begin{center}
\begin{tabular}{lrrrrrr} \hline
        & \multicolumn{1}{c}{light}  & 
         \multicolumn{1}{c}{heavy} & 
        \multicolumn{2}{c}{$|\mu|\sim|M_1|$} &
       \multicolumn{1}{c}{$|\mu| \gg |M_1|$} &
        \multicolumn{1}{c}{gaugino}  \\ 
        & \multicolumn{1}{c}{higgsino}  & 
         \multicolumn{1}{c}{higgsino} & 
        & &
       \multicolumn{1}{c}{bino} &
        \\ \hline
Example No.          & \multicolumn{1}{c}{1}
        & \multicolumn{1}{c}{2}      & \multicolumn{1}{c}{3} & 
      \multicolumn{1}{c}{4} & \multicolumn{1}{c}{5} & 
        \multicolumn{1}{c}{6}  \\ \hline
$\mu$ [GeV]          & $77.7$   & 1024.3 & 358.7     & 
414.7 & $-7776.7$    & $-1711.1$ \\
$M_2$ [GeV]          & $-441.4$ & 3894.1 & $-691.1$  &
$-1154.6$ & $133.5$ & 396.6 \\
$\tan \beta$         & 1.31     & 40.0   & 2.00      &
7.30   & 37.0       & 22.8 \\
$m_A$ [GeV]          & 656.8    & 737.2  & 577.7     &
828.9 & 2039.5     & 435.1 \\
$m_0$ [GeV]          & 610.8    & 1348.3 & 1080.9    &
2237.9 & 4698.0   & 2771.6 \\
$A_b/m_0$            & $-1.77$  & $-1.53$& $-1.03$   &
$-1.26$ & $0.46$    & 1.97 \\
$A_t/m_0$            &  2.75    & $-2.01$& $-2.77$   &
$-0.80$  & $0.11$    & 0.52 \\ \hline
$m_{\chi_1^0}$ [GeV] &  76.3    & 1020.8 & 340.2     &
407.8  & 67.2      & 199.5 \\
$Z_g$                &  0.00160 & 0.00155 & 0.651    &
0.0262   & 0.999968  & 0.99933 \\
$m_{\chi_2^0}$ [GeV] &  96.3    & 1026.4 & 364.5     &
418.2  & 133.5     & 396.0 \\
$m_{\chi_1^+}$ [GeV] &  89.2    & 1023.7 & 362.2     &
414.1  & 133.5     & 396.0 \\
$\Omega_\chi h^2$ (no coann.) & 0.178  & 0.130  & 0.158  &
0.00522   & $1.33\times 10^4$       & 0.418 \\
$\Omega_\chi h^2$    & 0.0299 & 0.0388 & 0.0890 &
0.00905   & $1.15\times 10^4$      & 0.418 \\ \hline
\end{tabular}
\end{center}
\caption{
  Some representative models for which coannihilations are important
  (examples 1--5) and one model (example 6) for which they are not.  We
  give the seven model parameters, the masses of the lightest
  neutralinos and of the lightest chargino, the gaugino fraction of
  the lightest neutralino and the relic densities without
  coannihilations included and with.}
\label{tab:reprmod}
\end{table}

\begin{figure}
  \centerline{\epsfig{file=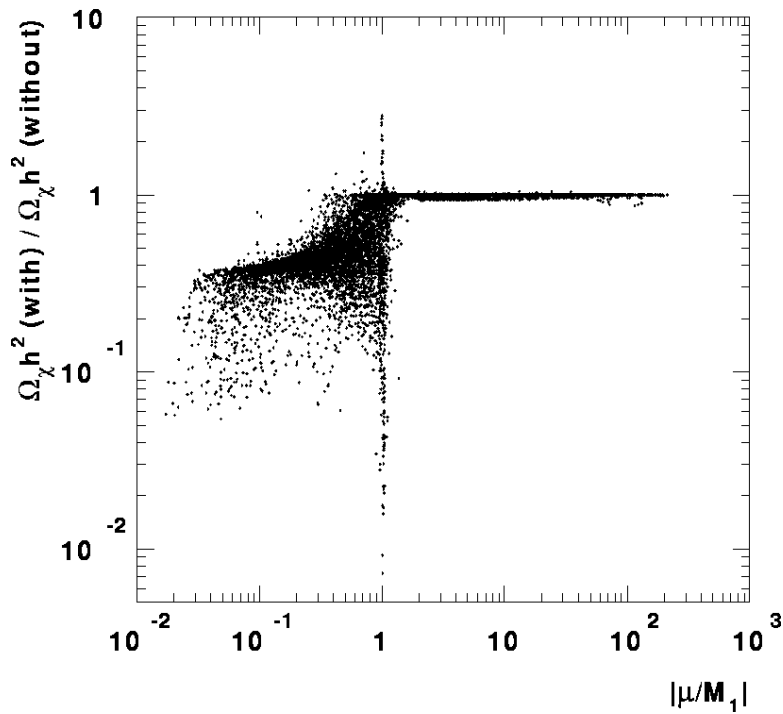,width=0.75\textwidth}} 
  \caption{ 
    Ratio of the relic densities with and without coannihilations
    versus $|\mu/M_1|$. Coannihilations are
    important when $|\mu/M_1|\lsim 2$.}
  \label{fig:ratiovsmum1}
\end{figure}

In Fig.~\ref{fig:ratiovsmum1} we show the reduction in relic density
due to the inclusion of coannihilations as a function of $|\mu/M_1|$.
A rule of thumb is that coannihilations are important when $|\mu/M_1|
\lsim 2$. But exceptions are found, as can be seen in
Fig.~\ref{fig:ratiovsmum1}. Notice that when $|\mu/M_1| \ll 1$, the
neutralino is higgsino-like; when $|\mu/M_1| \gg 1$, the neutralino is
gaugino-like; and when $|\mu/M_1| \sim 1$, the neutralino can be
higgsino-like, gaugino-like or mixed.

\begin{figure}
  \centerline{\epsfig{file=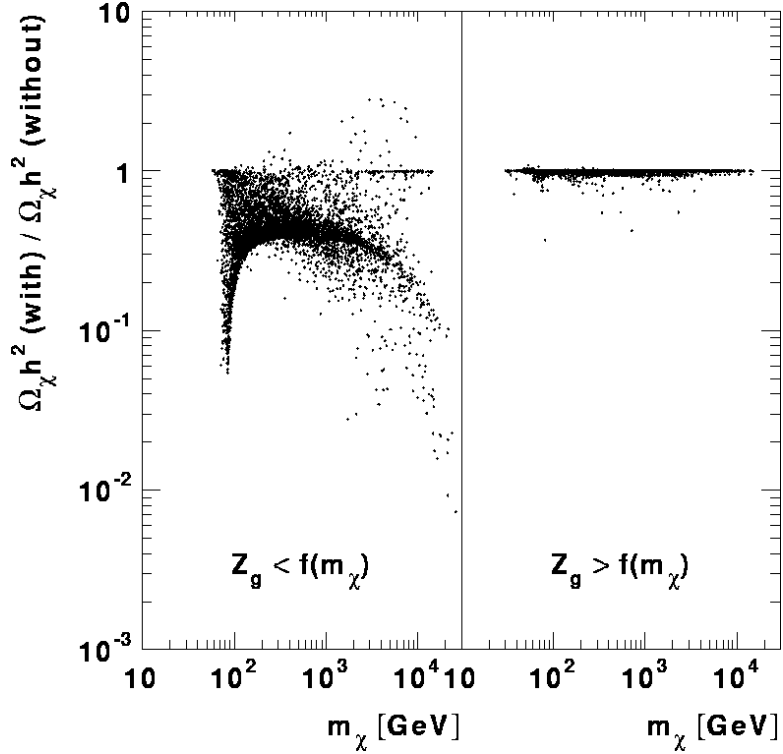,width=0.75\textwidth}} 
  \caption{ 
  Ratio of the relic densities with and without
  coannihilations versus neutralino mass $m_\chi$. 
  Coannihilations are generally not important when $Z_g >
  f(m_\chi)$, with $f$ given in the text.}
  \label{fig:ratiovsmx-zg}
\end{figure}

It can be convenient to have a criterium for when coannihilations are
important in terms of the composition as well. A rule of thumb is that
coannihilations are important when $Z_g < 0.23$ for $m_\chi < 200$ GeV
and when $Z_g/(1-Z_g) < (m_\chi/300\mbox{ GeV})^{3}$ for $m_\chi >
200$ GeV. There are exceptions to this rule as well, as can be seen in
Fig.~\ref{fig:ratiovsmx-zg} where the ratio of relic densities with
and without coannihilations is plotted versus the neutralino mass, the
left panel for points satisfying the present criterion, the right
panel for those not satisfying it.

In the following subsections, we present the cases where we found that
coannihilations are important and explain why. We first discuss
the already known case of light higgsino-like neutralinos, continue
with heavier higgsino-like neutralinos, the case $|\mu| \sim |M_1|$
and finally very pure gaugino-like neutralinos. We then end this
section by a discussion of the cosmologically interesting region.

\subsection{Light higgsino-like neutralinos}

We first discuss light higgsino-like neutralinos, $m_\chi<m_W$, $Z_g <
0.01$, since coannihilation processes for these have been investigated
earlier by other authors \cite{MizutaYamaguchi, DreesNojiri,
  NeuLoop1}.

Mizuta and Yamaguchi~\cite{MizutaYamaguchi} stressed the great
importance of including coannihilations for higgsinos lighter than the
$W$ boson.  For these light higgsinos, neutralino-neutralino
annihilation into fermions is strongly suppressed whereas
chargino-neutralino and chargino-chargino annihilations into fermions
are not.  Since the masses of the lightest neutralino and
the lightest chargino are of the same order, the relic density is
greatly reduced when coannihilations are included. Mizuta and
Yamaguchi claim that because of this reduction light higgsinos are
cosmologically of no interest.

Drees and Nojiri~\cite{DreesNojiri} included coannihilations between
the lightest and next-to-lightest neutralino, but did not include
those between the lightest neutralino and chargino, which are always
more important. In spite of this, they
concluded that the relic density of a higgsino-like neutralino will
always be uninterestingly small unless $m_\chi>500$ GeV or so.

Drees at al.~\cite{NeuLoop1} then re-investigated the relic density of
light higgsino-like neutralinos. They found that light higgsinos could
have relic densities as high as 0.2, and so be cosmologically
interesting, provided one-loop corrections to the neutralino masses
are included.

\begin{figure}
  \centerline{\epsfig{file=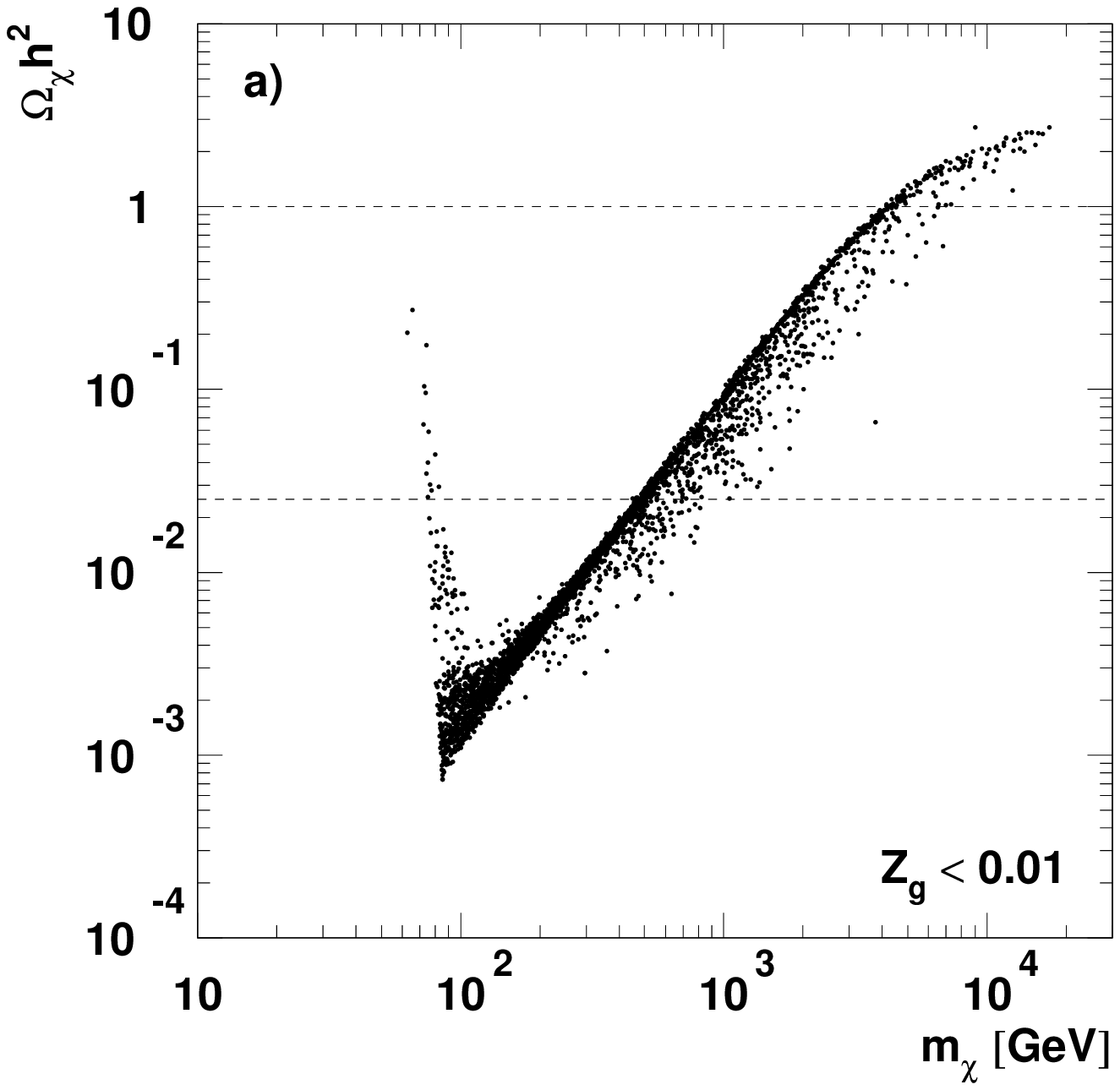,width=0.49\textwidth}
  \epsfig{file=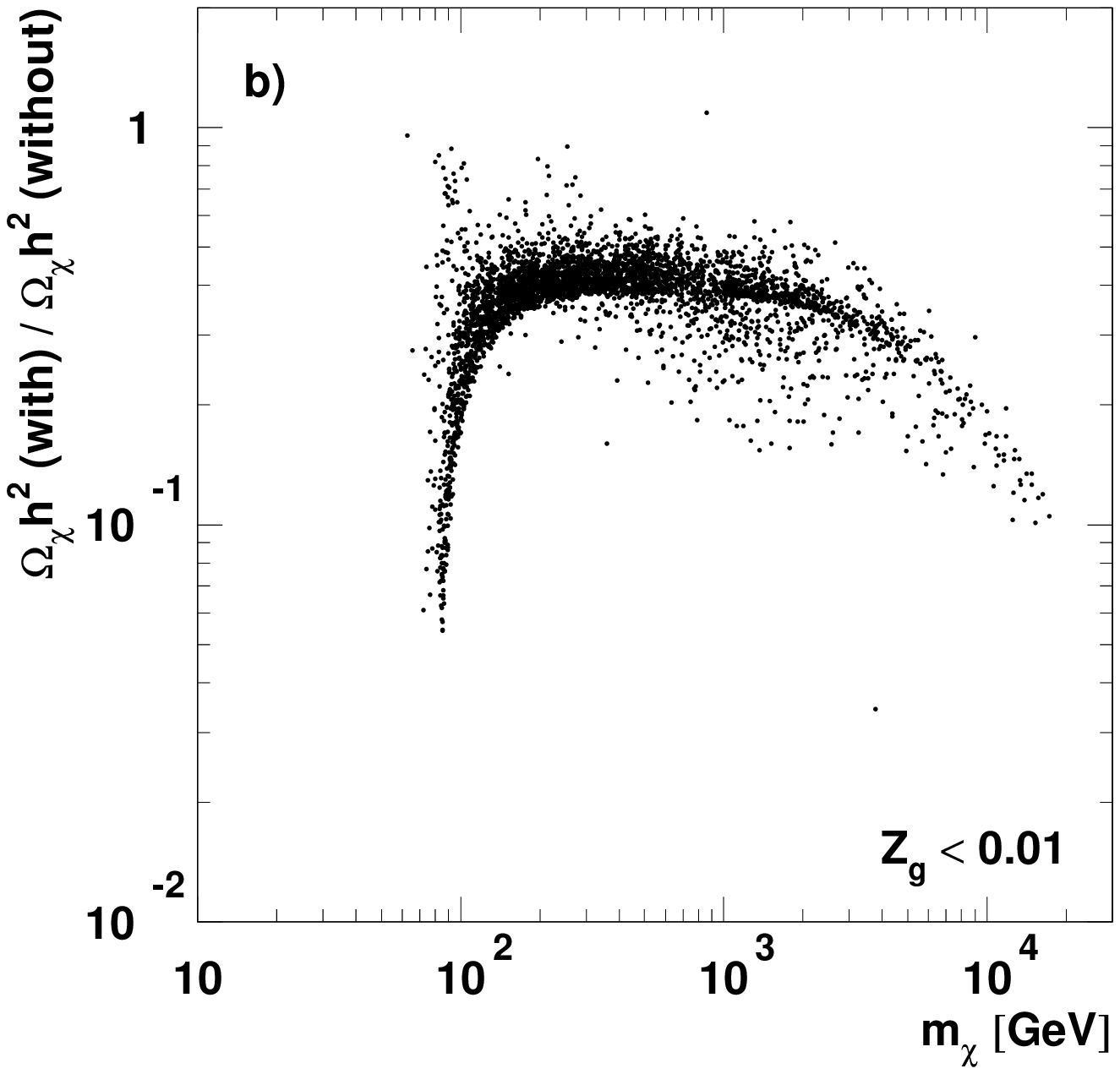,width=0.49\textwidth}}
  \caption{
    For higgsino-like neutralinos ($Z_g<0.01$) we show a) the relic
    density with coannihilations included and b) the ratio of the
    relic densities with and without coannihilations versus the
    neutralino mass.  The horizontal lines in a) limit the
    cosmologically interesting region $0.025 < \Omega_\chi h^2 <1$.}
  \label{fig:oh2vsmxh}
\end{figure}

We agree with these papers qualitatively, but we reach different
conclusions. We show our results in Fig.~\ref{fig:oh2vsmxh}, where we
plot the relic density of light higgsino-like neutralinos versus their
mass with coannihilations included, as well as the ratio between the
relic densities with and without coannihilations. The Mizuta and
Yamaguchi reduction can be seen in Fig.~\ref{fig:oh2vsmxh}b below 100
GeV, but due to the recent LEP2 bound on the chargino mass the effect
is not as dramatic as it was for them. If for the sake of comparison
we relax the LEP2 bound, the reduction continues down to $10^{-5}$ at
lower higgsino masses and we confirm qualitatively the Mizuta and
Yamaguchi conclusion --- coannihilations are very important for light
higgsinos --- but we differ from them quantitatively since we find
models in which light higgsinos have a cosmologically interesting
relic density.  For the specific light higgsino models in Drees et
al.~\cite{NeuLoop1} we agree on the relic density to within 20--30\%.
We find however other light higgsino-like models with higher
$\Omega_\chi h^2 \sim 0.3$, even without including the loop
corrections to the neutralino masses.

So there is a window of light higgsino models, $m_\chi \sim 75$ GeV,
that are cosmologically interesting. All these models have $\tan \beta
\lsim 1.6$ and those with the highest relic densities have $\tan \beta
\sim 1.2$.  These models escape the LEP2 bound on the chargino mass,
$m_{\chi^+} \sim 85$ GeV, because for $\tan \beta \lsim 2$ the mass of
the lightest neutralino can be lower than the mass of the lightest
chargino by tens of GeV\@.  By the same token, coannihilation
processes are not so important and the relic density in these models
remains cosmologically interesting.  Most of these models will be
probed in the near future when LEP2 runs at higher energies, but some
have too large a chargino mass ($m_\chi^+>95$ GeV) and too large
an $H_2^0$ boson mass ($m_{H_2^0}>90$ GeV) to be tested at LEP2\@. Thus
$\sim 75$ GeV higgsinos with $\tan\beta \lsim 2$ may remain good dark
matter candidates even after LEP2.

\subsection{Heavy higgsino-like neutralinos}

Coannihilations for higgsino-like neutralinos heavier than the $W$
boson have been mentioned by Drees and Nojiri~\cite{DreesNojiri}, who
argued that they should not change the relic density by much, and by
McDonald, Olive and Srednicki~\cite{McDonald}, who warn that they
might change it by an estimated factor of 2. We typically find a
decrease by factors of 2--5, and in some models even by a factor of 10
(see the right hand side Fig.~\ref{fig:oh2vsmxh}b).

For $m_{\chi}>m_{W}$, the lightest and next-to-lightest neutralinos
and the lightest chargino are close in mass, and they annihilate into
$W$ bosons besides fermion pairs. While the annihilation and
coannihilation cross sections into $W$ pairs are comparable, the
coannihilation of $\chi_1^0 \chi_2^0$, $\chi_1^0 \chi^+_1$ and
$\chi_2^0 \chi^+_1$ into fermion pairs is stronger than the $\chi_1^0
\chi_1^0\to f\bar{f}$ annihilation which is suppressed. This gives the
increase in the effective annihilation rate that we observe.

As a result, the smallest and highest masses for which higgsino-like
neutralinos heavier than the $W$ boson are good dark matter candidates
shift up from 300 to 450 GeV and from 3 to 7 TeV respectively.

Together with the result in the previous subsection, we conclude that
higgsino-like neutralinos ($Z_g<0.01$) can be good dark matter
candidates for masses in the ranges 60--85 GeV and 450--7000 GeV.

\subsection{Models with $|\mu| \sim |M_1|$}
\label{sec:mixedres}


Coannihilations for mixed or gaugino-like neutralinos have not been
included in earlier calculations. It has been believed that they are
not very important in these cases.  On the contrary, when $|\mu| \sim
|M_1|$ and $m_\chi \gsim m_W$ there is a very pronounced mass
degeneracy among the three lightest neutralinos and the lightest
chargino. The ensuing coannihilations can decrease the relic density
by up to two orders of magnitude or even increase it by up to a
factor of 3.  This is easily seen in Fig.~\ref{fig:ratiovsmum1} as the
vertical strip at $|\mu/M_1| \sim 1$.

If the lightest neutralino is mixed, $Z_g \sim 0.5$, coannihilations
can increase the relic density, whereas if it is more higgsino-like
or gaugino-like they will decrease it. This because the annihilation
cross section for mixed neutralinos is generally higher than those
for higgsino-like or gaugino-like neutralinos.

The largest decrease we see for this kind of models is when $|M_1|$
is slightly less than $|\mu|$ and both are in the TeV region. In
this case, the lightest neutralino is a very pure bino, and its
annihilation cross section is very suppressed since it couples
neither to the $Z$ nor to the $W$ boson. The chargino and other
neutralinos close in mass have much higher annihilation cross
sections, and thus coannihilations between them greatly reduce the
relic density. This big reduction suffices to lower $\Omega_\chi
h^2$ to cosmologically acceptable levels if $Z_g < 0.96$.  This
reduction does not occur for masses much lower than a TeV, because
the terms in the neutralino mass matrix proportional to the $W$ mass
prevent such pure bino states and the severe mass degeneracy.

To conclude, when $|\mu| \sim |M_1|$, coannihilations are very
important no matter if the neutralino is higgsino-like, mixed or
gaugino-like. The relic density can be cosmologically interesting for
these models as long as the gaugino fraction $Z_g<0.96$: these
neutralinos are good dark matter candidates.


\subsection{Gaugino-like neutralinos with $|\mu| \gg |M_1|$}


When $|\mu| \gg |M_1|$, the lightest neutralino is a very pure
gaugino.  According to the GUT relation Eq.~(\ref{eq:M1M2}), the
supersymmetric particles next in mass, the next-to-lightest neutralino
and the lightest chargino, are twice as heavy.  So we expect that
coannihilations between them are of no importance.\footnote{In
  models with non-universal gaugino masses, the lightest gaugino-like
  neutralino can be almost degenerate with the lightest chargino, and
  coannihilations can be important, as examined e.g. in
  Ref.~\cite{ChenDreesGunion}} In fact, as discussed in
section~\ref{sec:nummeth}, coannihilations would need to increase
the effective cross section by several orders of magnitude for these
large mass differences.

This actually happens in some cases.  They show up as the small spread
at high $|\mu/M_1|$ in Fig.~\ref{fig:ratiovsmum1}.  In these models,
the lightest neutralino is a very pure bino ($Z_g>0.999$) and the
squarks are heavy. Its annihilation to fermions is suppressed by the
heavy squark masses, and its annihilation to $Z$ and $W$ bosons is
either kinematically forbidden or extremely suppressed because a pure
bino does not couple to $Z$ and $W$ bosons.  On the other hand, the
lightest chargino annihilates to gauge bosons and fermions very
efficiently. The huge increase in the effective cross section,
compensated by the large mass difference, reduces the relic density by
10--20\%.  However, the relic density before introducing
coannihilations was of the order of $10^3$--$10^4$, and this small
reduction is not enough to make these special cases cosmologically
interesting.

\subsection{Cosmologically interesting region}
\label{sec:cosmregion}

\begin{figure}
  \epsfig{file=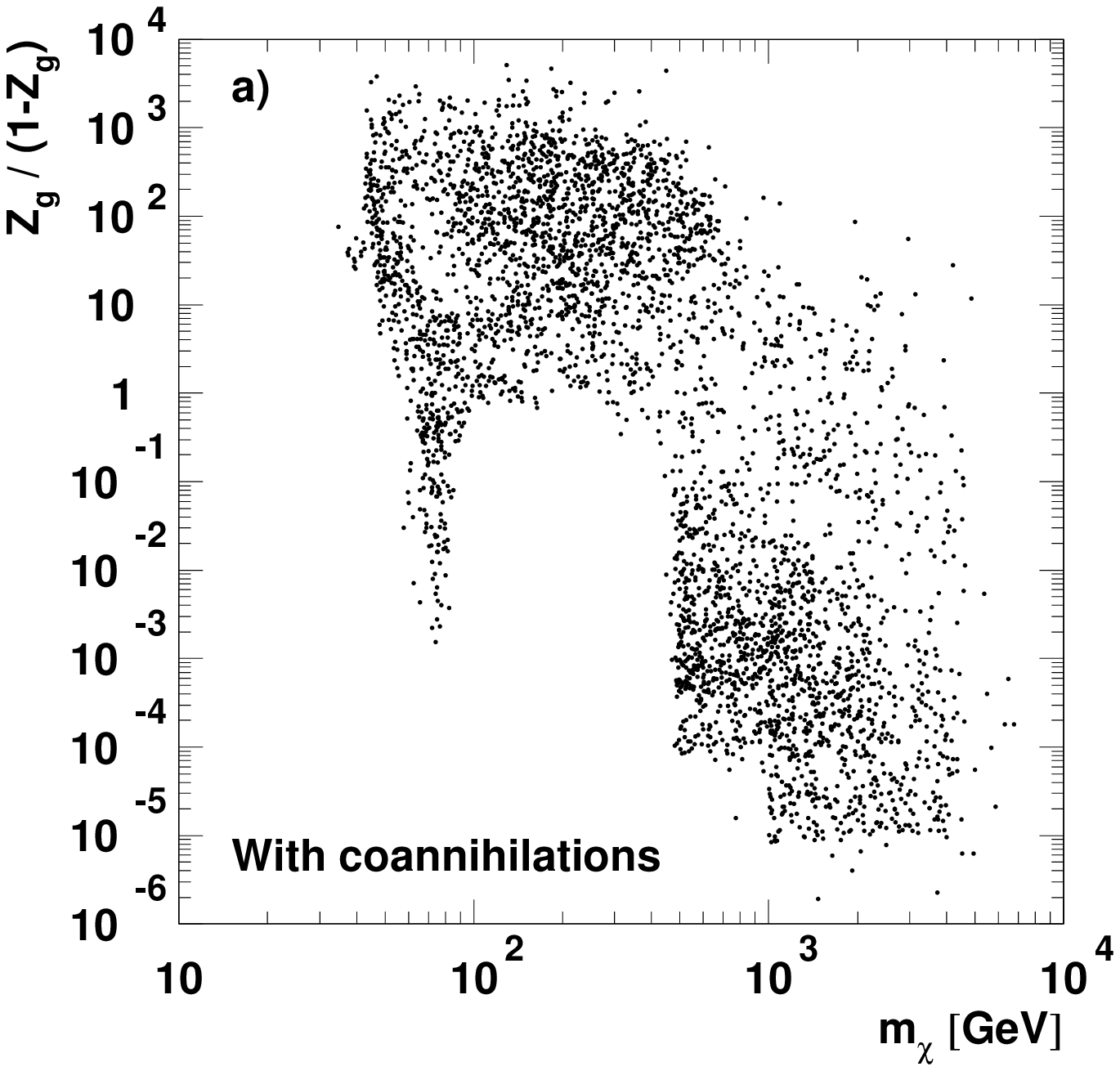,width=0.49\textwidth}
  \epsfig{file=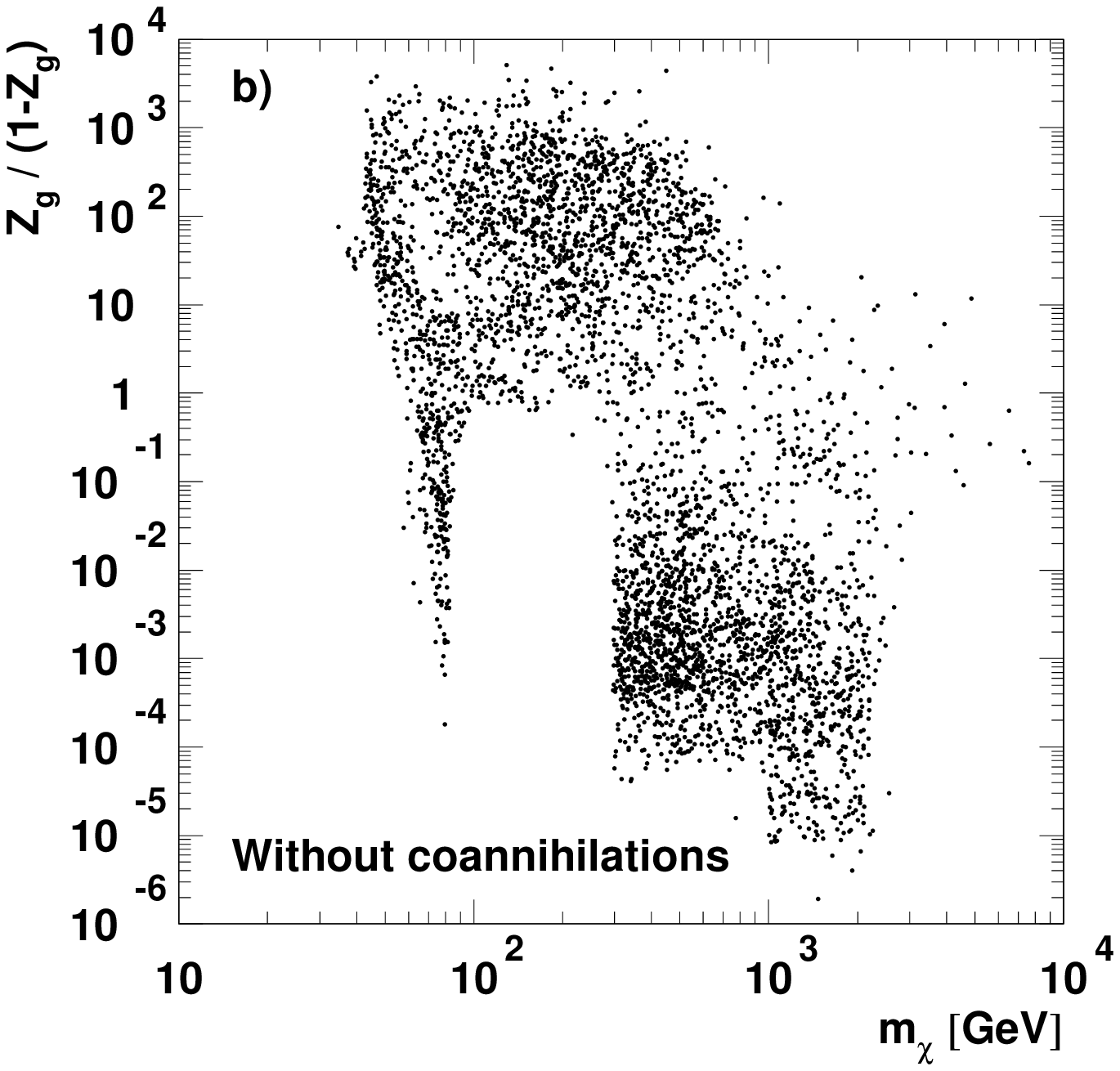,width=0.49\textwidth}
  \caption{Neutralino masses $m_\chi$ and compositions $Z_g/(1-Z_g)$
    for cosmologically interesting models a) with and b) without
    inclusion of coannihilations.}
  \label{fig:cosmregion}
\end{figure}

We now summarize when the neutralino is a good dark matter
candidate.  Fig.~\ref{fig:cosmregion} shows the cosmologically
interesting region $0.025 < \Omega_\chi h^2 <1$ in the neutralino
mass--composition plane $Z_g/(1-Z_g)$ versus $m_\chi$.

The light higgsino-like region does not extend to the left and down
due to the LEP2 bound on the chargino mass. The lower edge in 
gaugino fraction at $Z_{g} \sim 10^{-5}$ is the border of our survey
(how high $|M_{2}|$ is allowed to be). The upper limit on $Z_g$ and
the upper limit on the neutralino mass come from the requirement
$\Omega_\chi h^2<1$. The hole for higgsino-like neutralinos with
masses 85--450 GeV comes from the requirement $\Omega_\chi h^2>0.025$.

We see that coannihilations change the cosmologically interesting
region in the following aspects: the region of light higgsino-like
neutralinos is slightly reduced and the big region of heavier
higgsinos is shifted to higher masses, the lower boundary shifting
from 300 GeV to 450 GeV and the upper boundary from 3 TeV to 7 TeV.

The fuzzy edge at the highest masses is due to models in which the
squarks are close in mass to the lightest neutralino, in which case
$t$- and $u$-channel squark exchange enhances the annihilation cross
section. In these rather accidental cases, coannihilations with
squarks are expected to be important and enhance the effective cross
section even further. Thus, the upper bound on the neutralino mass
of 7 TeV is an underestimate.

\chapter{Neutralino Detection by Neutrino Telescopes}
\label{Indirect}

As mentioned in Chapter~\ref{Intro}, neutralinos can accumulate in the
Sun \cite{PressSpergel,SOS,earlysun} or the Earth \cite{earlyearth},
annihilate and produce detectable muon neutrinos.  In this chapter,
the method we have used to predict the muon fluxes resulting from
neutralino annihilation in the Sun or Earth will be described.  We
will also discuss, what signal flux levels that can be probed by
neutrino telescopes and what a detected signal will tell us about the
neutralino mass.

\begin{figure}
  \centerline{\epsfig{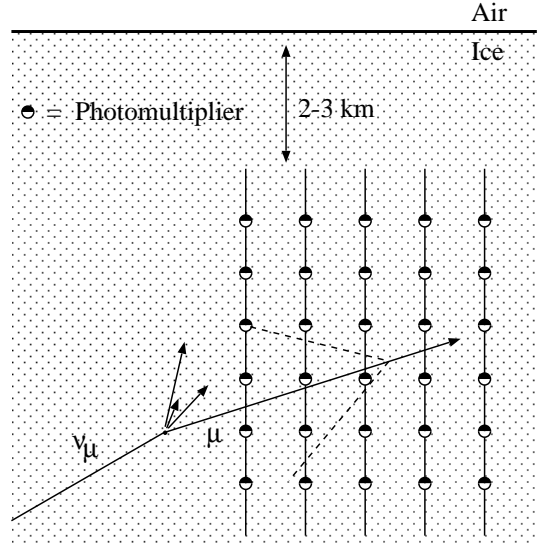}}
  \caption{Schematic view of a neutrino telescope like {\sc Amanda}. The
    dashed lines represent the \v{C}erenkov cone of light emitted by
    the muon when it traverses the ice. The \v{C}erenkov light is
    picked up by the photomultipliers from which the muon track can be
    reconstructed. Note that the muon scattering angle is exaggerated
    in the figure.}
  \label{fig:neutrinotelescope}
\end{figure}

There are many neutrino telescopes in use, e.g.\ Baksan \cite{Baksan},
{\sc Macro} \cite{Macro}, Kamio\-kande \cite{Kamiokande},
\cite{SuperKamiokande}, and others being built and/or proposed, {\sc
  Amanda} \cite{Amanda}, {\sc Nestor} \cite{Nestor}, and maybe others.
A neutrino telescope consists of water or ice situated well below the
ground (to minimize the background coming from atmospheric muons).
When a neutrino passes by it may interact with the ice or rock
surrounding the detector and produce a lepton.  If the neutrino is a
muon neutrino and hence the lepton a muon, it will neither decay too
fast or get stopped too fast and may travel several kilometers
(depending on its energy) before getting stopped.  As the muon moves
through the water (ice) it will emit \v{C}erenkov radiation, which can
be detected by photomultipliers.  From this signal, the direction of
the muon and thus the muon neutrino can be reconstructed.  In
Fig.~\ref{fig:neutrinotelescope} we show the principle of a neutrino
telescope as described above.

In the following sections, the steps performed to obtain a prediction
of the muon flux for a given set of MSSM parameters are described. 
Note however that except for Section \ref{sec:MuonPred}, 
the results obtained in this chapter are valid for any WIMP and not 
only neutralinos. 

\section{Neutralino capture and annihilation rates}

\subsection{Capture and annihilation rates}

When the Sun and Earth moves through our galactic halo, neutralinos 
may scatter off nuclei in them and lose enough energy to get 
gravitationally trapped \cite{PressSpergel}.  They will then oscillate 
back and forth, occasionally scatter and after a while accumulate in 
the center of the Sun 
\cite{SOS,earlysun} and the Earth 
\cite{earlyearth} where they can annihilate.  The evolution equation for 
the number of neutralinos, $N$, in the Sun or the Earth is given by
\begin{equation} \label{eq:Nevol}
  \frac{dN}{dt} = C - C_{A} N^2 - C_{E} N
\end{equation}
where the first term is the neutralino capture, the second term 
is twice the annihilation rate $\Gamma_{A} = 
\frac{1}{2} C_{A} N^2$ and the last term is neutralino evaporation. 
The evaporation term can be neglected for neutralinos heavier than 
about 5 GeV \cite{GriestSeckel87,Gould87} and 
since we are not interested in these low-mass 
neutralinos we can safely drop the last term in Eq.~(\ref{eq:Nevol}).
If we solve Eq.~(\ref{eq:Nevol}) for the annihilation rate 
$\Gamma_{A}$ we get
\begin{equation} \label{eq:GammaA}
  \Gamma_{A} = \frac{1}{2} C \tanh^2 \frac{t}{\tau} , \qquad \tau = 
  \sqrt{C C_{A}}
\end{equation}
where $\tau$ is the time scale for capture and annihilation 
equilibrium to occur.  In most cases where the muon fluxes are within 
reach of present and near-future telescopes, equilibrium will have 
occurred and the annihilation rate is at `full strength', $\Gamma_{A} 
\simeq \frac{1}{2} C$.  Note that in this case, the annihilation rate 
is determined by the elastic scattering cross sections, on which $C$ 
depends, and not by the annihilation cross section. In our calculation 
we have of course used the full expressions without assuming that 
the annihilation occurs at `full strength'.

The capture rate $C$ depends on, among other things, the local halo
mass density, $\rho_{\chi}$, the velocity dispersion of dark matter
particles in the halo, $\bar{v} = \sqrt{\langle v^2 \rangle}$, the
elastic scattering cross sections and the composition of the Earth and
Sun and we have used the convenient expressions given in
Ref.~\cite{jkg} based on the formulas in Ref.~\cite{Gouldcapt}.  The
main uncertainties in the capture rate come from the local halo mass
density, $\rho_{\chi}$, which is uncertain of about a factor of two
or so, and the velocity dispersion $\bar{v}$.  We have chosen
$\rho_{\chi} = 0.3$ GeV/cm$^3$ and $\bar{v}=270$ km s$^{-1}$.
Estimates of $\rho_{\chi}$ can be found in e.g.\ Ref.~\cite{rhochi}.

Note that if one-loop corrections to the neutralino coupling to Higgs
bosons \cite{NeuLoop1} are included, which we have not, this coupling
can for higgsino-like neutralinos either increase by more than two
orders of magnitude or in accidental cases be reduced to exactly 0.
This means that the spin-independent scattering cross sections for
higgsinos can get greatly increased or reduced which mainly effects
the capture rate in the Earth (and direct detection experiments which
we don't discuss here). For mixed or gaugino-like neutralinos these
one-loop corrections are expected to be small.

\subsection{Annihilation profiles}

\begin{figure}
  \centerline{\epsfig{file=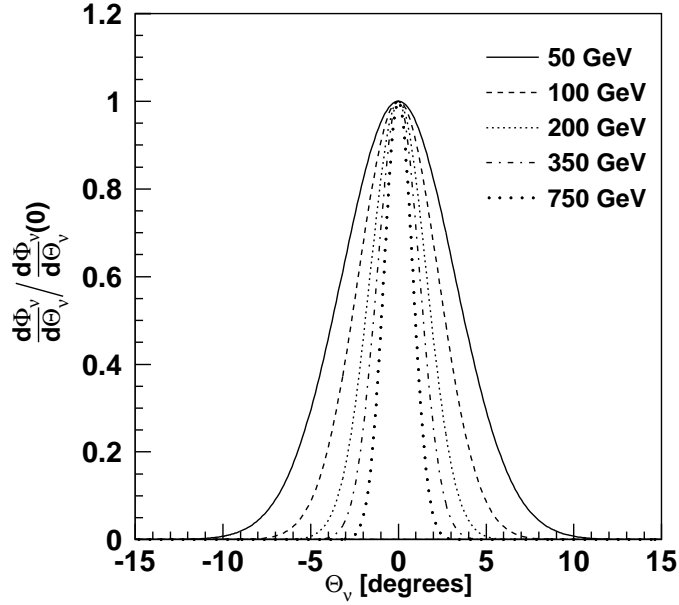,width=.75\textwidth}}
  \caption{Projected angular distributions of neutralino generated 
  neutrinos from the Earth for neutralino masses of 50 
  GeV (solid), 100 GeV (dashed), 200 GeV (dotted), 
  350 GeV (dash-dotted) and 750 GeV (wide dotted). The analogous 
  distributions from the Sun are simply narrow peaks at 
  $\theta_{\nu}=0^\circ$}
  \label{fig:nudistearth}
\end{figure}

The annihilation rate per volume element is given by
\begin{equation}
  \Gamma = n^2 \langle \sigma_A v \rangle 
\end{equation}
where $n_{\chi}$ is the number density of neutralinos and 
$\langle \sigma_A v \rangle$ is the thermally averaged annihilation 
cross section. The number density of neutralinos is given by 
\cite{JEdiploma,JEpaper2}
\begin{equation} \label{eq:aprof1}
  n(r) = n(0) e^{-r^2/2r_{\chi}^2}
\end{equation}
with
\begin{equation} \label{eq:aprof2}
  r_{\chi} = \left[ \frac{3kT}{4 \pi G \rho_\chi m_\chi} \right]^{1/2} \simeq
  \frac{0.56 R_{\otimes}}{\sqrt{m_{\chi}/{\rm GeV}}}
\end{equation}
where we in the last step have assumed that the annihilation takes
place in the Earth. We have used the Earth radius $R_{\otimes}$ = 6378
km, the core temperature $T \simeq 6000$ K and the core density $\rho
\simeq 13.1$ g cm$^{-3}$. In Fig.~\ref{fig:nudistearth} the resulting
projected neutrino angle is shown for neutralino annihilation in the
Earth.

For the Sun, the annihilation region is very concentrated to the core 
and subtends a negligible solid angle as seen from a neutrino 
telescope at the Earth.

\section{Muon fluxes - Monte Carlo simulations}

A detailed Monte Carlo simulation of the muon fluxes at neutrino
telescopes for different annihilation channels and neutralino masses
was first done by Ritz and Seckel \cite{RS}. We have followed their
approach, except for neutrino propagation through the Sun where we
have used detailed Monte Carlo simulations \cite{JEdiploma,JEpaper1}
instead of the approximate analytical formulae in Ref.~\cite{RS}.
With respect to calculations using Ref.~\cite{RS} (e.g.\ 
Ref.~\cite{neuprod}), this Monte Carlo treatment of the neutrino
propagation through the Sun increase the muon fluxes by 5--20\%\@.

\subsection{Annihilation channels and branching ratios}

From the previous section we know how to calculate the neutralino
annihilation rates and it is straightforward (by e.g.\ the same
methods as described in Section \ref{sec:AnnChannels}) to calculate
the annihilation branching ratios into different annihilation
channels.  Since the temperature in both the center of the Earth and
Sun is so small, the neutralinos are highly non-relativistic and we
can to a good approximation use the zero relative velocity
annihilation cross sections to calculate the branching ratios.  The
annihilation channels of any significance for the muon fluxes are $c
\bar{c}$, $b \bar{b}$, $t \bar{t}$, $\tau^+ \tau^-$, $W^+ W^-$, $Z^0
Z^0$, $Z^0 H_{1}^0$, $Z H_{2}^0$, $H_{1}^0 H_{3}^0$, $H_{2}^0 H_{3}^0$
and $H^\pm W^\pm$.  Note that the branching ratio into neutrinos
directly is zero in the non-relativistic limit and hence the only
neutrinos we get are those coming from decay of other annihilation
products.  We can also not have annihilation into $Z^0 H_{3}^0$ or
$H_{1,2}^0 H_{1,2}^0$ in this limit since the initial state is CP-odd
and so must the final state be.  Lighter quarks will not contribute
since the annihilation cross section into fermions is approximately
proportional to the mass of the fermion squared. Of the charged
leptons, only muons and tauons are interesting as potential muon
neutrino producers but as we will see in the next subsection, muons
will be stopped well before they decay whereas tau leptons do decay
before getting stopped so the only lepton channel we have to consider
is $\tau^+ \tau^-$.

\subsection{Charged lepton interactions in the Earth}
\label{sec:chaintearth}

For relativistic charged particles (other than electrons) the mean
energy loss is given by the Bethe-Bloch equation \cite{PDG}
\begin{equation} \label{eq:BetheBloch}
  -\frac{dE}{dx} = K z^2 \frac{Z}{A} \frac{1}{\beta^2} 
  \left[ \frac{1}{2} \ln \frac{2 m_e c^2 \beta^2 \gamma^2 T_{\rm max}}
 {I^2} - \beta^2 -\ln \frac{\hbar \omega_p}{I} - \ln \beta \gamma
  +\frac{1}{2} \right]
\end{equation}
where
\begin{eqnarray}
  K & = & 4 \pi N_A r_e^2 m_e c^2 \simeq 0.307 \mbox{ MeV cm$^2$ 
  mol$^{-1}$}
  \\
  T_{\rm max} & = & \frac{2 m_e c^2 \beta^2 \gamma^2}{1+2 \gamma m_e/M +
  (m_e/M)^2} \\
  \hbar \omega_p & = & 28.816 \sqrt{\rho \langle Z/A \rangle} \mbox{
  eV} \label{eq:omegap}
\end{eqnarray}
with $M$ being the mass of the particle, $\beta c$ its velocity, $Z$
the atomic number of the stopping material, $A$ its atomic weight,
$\rho$ its density and $I$ the mean excitation energy which is given
approximately by
\begin{equation}
  I \simeq 16 Z^{0.9} \mbox{ eV.}
\end{equation}
At high energies ($\gsim$ 500 GeV for muons in ice), 
Eq.~(\ref{eq:BetheBloch}) is only a lower limit for the energy loss 
due to that radiative effects not included in 
Eq.~(\ref{eq:BetheBloch}) start dominating at high energies. This is 
however not important in our present analysis, but will be included in 
Section~\ref{sec:MuonInt} where more accurate expressions are needed.

We want to see if muons and tau leptons have time to decay before 
they get stopped.
For muons in the core of the Earth (which consists of mainly liquid 
iron of density 13.1 g/cm$^3$) the minimum of 
Eq.~(\ref{eq:BetheBloch}) is
\begin{equation}
  \left( -\frac{dE}{dt}\right)^{\rm Earth}_{\rm min} \simeq 5.7
  \times 10^8 \mbox{ GeV/s} 
\end{equation}
from which the mean stopping time is given by $\tau_{\rm stop} =
E/(dE/dt)$. Since the decay time is given by $\gamma \tau_{\rm dec} =
E \tau_{\rm dec} / M$ we find that when the ratio
\begin{equation}
  \frac{\tau_{\rm stop}}{\gamma \tau_{\rm dec}} = \frac{M}{\frac{dE}{dt}
  \tau_{\rm dec}}
\end{equation}
is much smaller than one, the particles will get stopped well before
they have time to decay. For muons we get
\begin{equation}
 \left( \frac{\tau_{\rm stop}}{\gamma \tau_{\rm dec}} \right)^{\rm
 Earth} \lsim 8.4\times 10^{-5}
\end{equation}
and hence any muon produced will be stopped before having time to
decay producing any muon neutrinos. For tau leptons, on the other
hand, we can in the same way find that the maximal energy loss per
second (up to $E_{\tau}$ = 5000 GeV) is given by
\begin{equation}
  \left( -\frac{dE}{dt}\right)^{\rm Earth}_{\rm max} \simeq 9.3
  \times 10^8 \mbox{ GeV/s} 
\end{equation}
from which it is seen that the upper limit on the energy loss of a
tau lepton, $\gamma \tau_{\rm dec} dE/dt$ is less than a GeV even for TeV
energy tauons. Hence energy loss of tau leptons can safely be neglected.

Note that the simple estimates in this subsection could change by a
factor of 2--3 or so at high energies, $\gsim 500$ GeV, if radiative
effects are included. The conclusion that muons get stopped and that
tau lepton interactions can be neglected would be the same though.

\subsection{Charged lepton interactions in the Sun}
\label{sec:chaintsun}

For charged lepton interactions in the core of the Sun, which is a 
plasma, Eq.~(\ref{eq:BetheBloch}) has to be replaced by  \cite{Jackson}
\begin{equation}
  - \frac{dE}{dx} = - \frac{e^2}{4 \pi \epsilon_{0}} 
    \frac{\omega_{p}^2}{(\beta c)^2} \ln
    \left[ \frac{\Lambda m_{e}c^2 \gamma \beta^2}{\hbar \omega_{p}} 
    \right]
\end{equation}
where $\omega_{p}$ is the plasma frequency as given by 
Eq.~(\ref{eq:omegap}) and $\Lambda$ is a number of order unity which 
we put equal to 1\@. By using that the composition of the core of the 
Sun is 24\% $^1$H and 64\% $^4$He with a density of 148 
g/cm$^3$ \cite{Bahcall} we then get the minimal energy 
loss for muons in the core of the Sun to be
\begin{equation}
  \left( -\frac{dE}{dt}\right)^{\rm Sun}_{\rm min} \simeq 7.5
  \times 10^9 \mbox{ GeV/s} 
\end{equation}
and hence
\begin{equation}
 \left( \frac{\tau_{\rm stop}}{\gamma \tau_{\rm dec}} \right)^{\rm
 Sun} \lsim 6.4\times 10^{-6}.
\end{equation}
Muons are thus stopped well before they decay and can hence be 
considered as absorbed. For tau leptons we get the maximal energy 
loss (up to $E_{\tau}$ = 5000 GeV) to be
\begin{equation}
  \left( -\frac{dE}{dt}\right)^{\rm Sun}_{\rm max} \simeq 1.4
  \times 10^{10} \mbox{ GeV/s} 
\end{equation}
which implies that the energy loss is only a few GeV even for TeV tau 
leptons. Hence tau lepton energy loss can be neglected.

To conclude the previous subsection and this one, muons get stopped 
well before they decay in both the Earth and Sun and the energy loss 
of tau leptons can be neglected in both the Earth and Sun.

\subsection{Heavy hadron interactions in the Sun and Earth}

The heavy quarks produced in neutralino annihilation will form mesons
and baryons which may interact before they decay. The top quarks will
decay before they even have time to form any hadrons and their
interactions with the surrounding medium can be neglected.  For $c$
and $b$ quarks however, interactions with the surrounding medium has
to be taken care of.  This is done in an approximate fashion where the
decay/hadronization is simulated as if in vacuum as described below 
and the interactions that may have occurred are introduced afterwards 
as a general energy decrease. This approximation is reasonable when the 
number of interactions is not more than a few which is the case for 
moderately heavy neutralinos, $m_{\chi} \lsim 500$ GeV. For heavier 
neutralinos neither the $c \bar{c}$ nor the $b \bar{b}$ channel will 
dominate and hence the approximation is justified.

The cross sections for $c$ and $b$ hadron scattering off a nucleon can 
be estimated by noting that the scattering cross section for any 
hadron off a proton is approximately given by \cite{Povh}
\begin{equation} \label{eq:crosshadron}
  \sigma_{{\rm hadron}-p} \simeq 6 \langle r_{\rm st}^2 \rangle_{\rm 
  hadron}
\end{equation}
where $\langle r_{\rm st}^2 \rangle_{\rm hadron}$ is the mean squared 
strong interaction radius of the hadron. Povh et al.\ found 
that 
$\langle r_{\rm st}^2 \rangle_{\Lambda} \simeq 0.58\pm0.02$ fm$^2$, 
$\langle r_{\rm st}^2 \rangle_{\pi} \simeq 0.41\pm0.02$ fm$^2$ and
$\langle r_{\rm st}^2 \rangle_{J/\Psi} \simeq 0.04\pm 0.02$ fm$^2$.
If we assume that the decrease in mean squared radius is constant for 
each light quark we change to a $c$ or $b$ quark (justified by 
experiments) we find that the 
mean squared strong interaction radii for $c$ or $b$ mesons and 
baryons are given by
\begin{eqnarray}
  \langle r_{\rm st}^2 \rangle_{c/b-{\rm meson}} \simeq 0.23 \mbox{ 
  fm$^2$} \\
  \langle r_{\rm st}^2 \rangle_{c/b-{\rm baryon}} \simeq 0.40 \mbox{ 
  fm$^2$} 
\end{eqnarray}
which together with Eq.~(\ref{eq:crosshadron}) yields
\begin{eqnarray}
  \sigma_{c/b-{\rm meson}} & \simeq & 14 \mbox{ mb} \\
  \sigma_{c/b-{\rm baryon}} & \simeq & 24 \mbox{ mb} 
\end{eqnarray}

From the simulations it is known how far the leading hadron has moved
before decaying and the probability that it should have undergone one or
more interactions is then easily calculated given the cross sections
above.  If an interaction is found to have occurred the leading hadron
of the new jet takes the energy fraction $z$ of the initial hadron.
The average energy transfers, $\langle z \rangle $, used are those
calculated by Ritz and Seckel \cite{RS},
\begin{eqnarray}
  \langle z \rangle & \simeq & 0.6 \frac{m_{c}}{m_{i}} \quad \mbox{for 
  $c$ hadrons} \\
  \langle z \rangle & \simeq & 0.7 \quad \mbox{for $b$ hadrons}
\end{eqnarray}
where $m_{i}$ is the mass of the initial $c$ hadron and $m_{c}$ is the 
mass of the $c$ quark.  The same amount of energy decrease is assumed 
to apply to the produced neutrino.  For the $b\bar{b}$ channel, 
interactions are only significant in the Sun and only for heavier 
neutralinos, $m_{\chi}\gsim 50$ GeV. The effect is very dramatic for 
even more massive neutralinos but for those other annihilation 
channels will dominate.

\subsection{Monte Carlo simulations}
\label{sec:MonteCarloSim}

Of the annihilation channels mentioned above, gauge bosons and tau 
leptons can decay directly to neutrinos but the quarks will hadronize 
and eventually give rise to muon neutrinos.  The Higgs bosons will 
decay mainly to quarks which will hadronize as well.  All decays and 
hadronizations are simulated with the Lund Monte Carlo {\sc Jetset 
7.4} and {\sc Pythia 5.7} \cite{Jetset} for each of the annihilation 
channels $c \bar{c}$, $b \bar{b}$, $t \bar{t}$, $\tau^+ \tau^-$, $W^+ 
W^-$ and $Z^0 Z^0$ for the different neutralino masses $m_{\chi}$ = 
10, 25, 50, 80.2, 91.3, 100, 150, 175, 200, 250, 350, 500, 750, 1000, 
3000 and 5000 GeV. Note that the annihilation channels containing 
Higgs bosons do not need to be simulated separately since the Higgs 
bosons decay to particles contained in these six `fundamental' 
channels mentioned above and their contribution to the muon neutrino 
flux can thus be calculated as soon as the Higgs masses and their 
decay channels are known.  For each mass and annihilation 
channel, $2.5\times 10^5$ events have been simulated and all muon 
neutrinos produced have been kept.  Hence, a neutrino flux is obtained 
for any of the given annihilation channels and neutralino masses.

\subsection{Neutrino interactions and cross sections}

The neutrino-nucleon charged and neutral current cross sections are
approximately given by
\begin{eqnarray} 
  \sigma_{\rm CC} & \simeq & a E_\nu \label{eq:sigmacc} \\
  \sigma_{\rm NC} & \simeq & b E_\nu \label{eq:sigmanc}
\end{eqnarray}
where the coefficients $a$ and $b$ are given by 
\begin{equation} \label{eq:a-params}
  \left\{ \begin{array}{lcl}
  a_{\nu n} & = & 8.81 \times 10^{-39} \mbox{ cm$^2$ GeV$^{-1}$} \\
  a_{\nu p} & = & 4.51 \times 10^{-39} \mbox{ cm$^2$ GeV$^{-1}$} \\
  a_{\bar{\nu} n} & = & 2.50 \times 10^{-39} \mbox{ cm$^2$ GeV$^{-1}$} \\
  a_{\bar{\nu} p} & = & 3.99 \times 10^{-39} \mbox{ cm$^2$ GeV$^{-1}$}
  \\
  \end{array} \right.
\end{equation}
\begin{equation} \label{eq:b-params}
  \left\{ \begin{array}{lcl}
  b_{\nu n} & = & 2.20 \times 10^{-39} \mbox{ cm$^2$ GeV$^{-1}$} \\
  b_{\nu p} & = & 1.97 \times 10^{-39} \mbox{ cm$^2$ GeV$^{-1}$} \\
  b_{\bar{\nu} n} & = & 1.15 \times 10^{-39} \mbox{ cm$^2$ GeV$^{-1}$} \\
  b_{\bar{\nu} p} & = & 1.14 \times 10^{-39} \mbox{ cm$^2$ GeV$^{-1}$}
  \\
  \end{array} \right.
\end{equation}
where GRV structure functions \cite{GRV} have been used down to 
$Q^2=0.3 \mbox{ GeV}^2$.  These cross sections agree well with 
neutrino experiments on isoscalar targets as well as with those 
obtained using other structure functions, like CTEQ2D \cite{CTEQ}.  

\subsection{Neutrino interactions in the Sun}

\begin{figure}
  \centerline{\epsfig{file=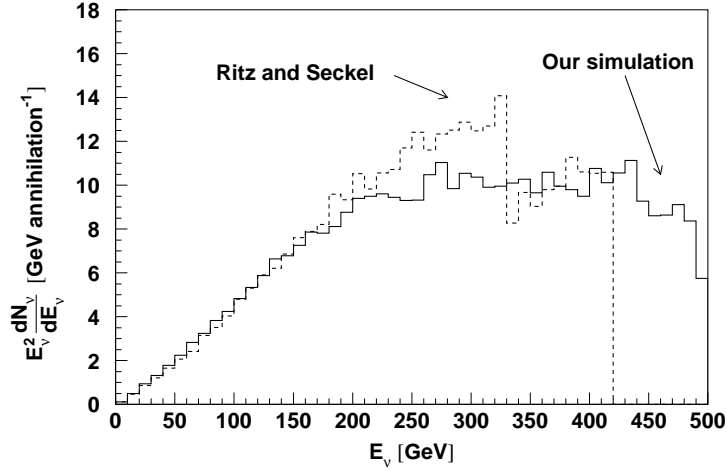,width=0.75\textwidth}}
  \caption{The neutrino spectrum ($\nu_{\mu} + \bar{\nu}_{\mu}$) weighted by
    $E_{\nu}^2$ at the surface of the Sun for the $W^+ W^-$ channel
    with a neutralino mass of 500 GeV\@. The two peaks in the Ritz and
    Seckel spectrum correspond to $\nu_{\mu}$ and $\bar{\nu}_{\mu}$
    respectively.}
  \label{fig:JEvsRS}
\end{figure}

Ritz and Seckel \cite{RS} considered neutrino interactions on the way
out of the Sun in an approximate way where neutral current
neutrino-nucleon interactions were assumed to be much weaker than
charged current interactions and the energy loss was assumed to be
continuous. Neither of these approximations are very good and hence we
have instead simulated the neutrino interactions on the way out of the
Sun with {\sc Pythia} 5.7 \cite{Jetset}. For the Earth, the effective
thickness is not big enough to be of any importance and neutrino
energy loss or absorption on the way to the detector can thus be
neglected.

The effective thickness for the Sun is calculated by using the solar
model in Ref.~\cite{Bahcall}. The effective thickness of the Sun is
given by $R_{\odot} I_0$ of which the part in hydrogen is $R_{\odot}
I_H$ where $R_{\odot}$ is the radius of the Sun.  The $I$'s are given
by
\begin{eqnarray}
  I_0 & = & \int_0^1 \rho(x) dx \simeq 21.4 \mbox{ g cm$^{-3}$} \\
  I_H & = & \int_0^1 \rho(x) X_H (x) dx \simeq 11.0 \mbox{ g cm$^{-3}$}
\end{eqnarray}
where $x=r/R_{\odot}$, $\rho(x)$ is the solar density and $X_H(x)$ is
the hydrogen mass fraction. In terms of protons and neutrons the
corresponding integrals would be
\begin{eqnarray}
  I_p & = & \frac{1}{2} (I_0 + I_H) = 16.2 \mbox{ g cm$^{-3}$} \\
  I_n & = & \frac{1}{2} (I_0 - I_H) = 5.2 \mbox{ g cm$^{-3}$}.
\end{eqnarray}

With these effective thicknesses and the neutrino-nucleon cross
sections, Eqs.\ (\ref{eq:sigmacc})--(\ref{eq:b-params}), at hand it is
then straightforward to calculate the probability that a given
(anti)neu\-trino has participated in an interaction (charged or neutral
current), and if it has (and the interaction is a neutral current
interaction) simulate it with {\sc Pythia} and proceed with the same
procedure until the neutrino has reached the surface of the Sun.

In principle one should also take into account the fact that all 
annihilations don't occur exactly at the center of the Sun which 
will introduce a smearing of the effective thickness. This effect can 
be estimated \cite{JEdiploma} and is found to be very small. Hence it 
is a very good approximation to assume that all neutrinos from the 
Sun originate from the center.

The difference between this approach and the Ritz and Seckel approach 
is shown in Fig.~\ref{fig:JEvsRS} where the neutrino flux weighted by 
the neutrino energy squared is shown.  Note that $E_{\nu}^2 
dN_{\nu}/dE_{\nu}$ is approximately proportional to the muon flux 
since both the neutrino-nucleon cross sections and the muon range are 
approximately proportional to the energy of the neutrino/muon.  The 
mean energy of the neutrinos at the surface of the Sun is about the 
same with the Ritz and Seckel approach and this more detailed analysis 
but the \emph{distribution} is different and since the muon flux is 
proportional to the second moment of the distribution (as explained 
above) one should 
expect a difference in the predicted muon fluxes at a detector.  In 
fact, the total muon flux with this method is about 5--20\% higher 
than with the Ritz and Seckel
approach (with the higher
difference at higher masses). Except for this difference in neutrino
interactions our results agree well with the Ritz and Seckel results
\cite{RS} as well as with the analytical results in 
Ref.~\cite{JungmanKam}.

\subsection{Neutrino interactions at the detector}

When a neutrino comes close to the detector it may interact and
produce a muon via a charged current interaction. This process is also
simulated with {\sc Pythia} 5.7 \cite{Jetset} where information of not
only the muon energy but also the muon angle with respect to the
neutrino is kept. For neutralino annihilations in the Earth, the size
of the annihilation region has also been included according to the
distributions in Eqs.~(\ref{eq:aprof1})--(\ref{eq:aprof2}).

\subsection{Muon interactions}
\label{sec:MuonInt}

When a muon is produced it can travel several kilometers (depending on 
energy) before reaching a detector. When the muons travel through the 
ice or rock surrounding the detector they may interact and lose 
energy, where the energy loss is approximately given by
\begin{equation} \label{eq:demudx}
  \frac{dE_{\mu}}{dx} \simeq - \alpha - \beta E_{\mu}
\end{equation}
where the coefficients $\alpha$ and $\beta$ are fitted to the energy
losses calculated in  
Ref.~\cite{Lohmann} and are given by 
\begin{equation} \label{eq:alphabetaice}
  \left\{ \begin{array}{lcl}
   (\alpha/\rho)_{\rm ice} & \simeq & 0.00260 \mbox{ GeV g$^{-1}$ cm$^2$} \\
   (\beta/\rho)_{\rm ice} & \simeq & 3.49 \times 10^{-6} \mbox{ g$^{-1}$ cm$^2$}
   \end{array} \right.
\end{equation}
for muons propagating in water or ice and 
\begin{equation} \label{eq:alphabetarock}
  \left\{ \begin{array}{lcl}
   (\alpha/\rho)_{\rm rock} & \simeq & 0.00221 \mbox{ GeV g$^{-1}$ cm$^2$} \\
   (\beta/\rho)_{\rm rock} & \simeq & 4.40 \times 10^{-6} \mbox{ g$^{-1}$ cm$^2$}
   \end{array} \right.
\end{equation}
for muons propagating in rock.  The errors of these parameterizations
are less than 2\% in the region 30--10000 GeV and less than 6\% in the
region 10--30 GeV. Hence they are sufficiently accurate for our needs.
By integrating Eq.~(\ref{eq:demudx}) we get the mean energy of the
muons after having traversed a distance $x$ of the detector
surroundings to be
\begin{equation} \label{eq:emux}
  E_\mu (x) = \left(E_\mu^0+\frac{\alpha}{\beta}\right)e^{-\beta x} -
  \frac{\alpha}{\beta} 
\end{equation}
where $E_\mu^0$ is the initial muon energy.
From this relation we find that the range of a muon of
energy $E_\mu^0$ is
\begin{equation}
  L_\mu = \frac{1}{\beta} \ln \left[
  \frac{E_\mu^0+\alpha/\beta}{\alpha/\beta} \right] 
\end{equation}
which for low energies, $E_\mu^0 \lsim \alpha/\beta \simeq 500$ GeV, is
approximated by $L_\mu = E_\mu^0/\alpha$. Note that 
the radiative effects contained in the $\beta$-term in
Eq.~(\ref{eq:demudx}) are actually stochastic and will lead to energy
straggling, i.e.\ some muons will lose more energy and some less. This
is however only important when $E_\mu^0 \gsim \alpha/\beta \simeq 500$
GeV and for our needs the use of the mean energy as given by
Eq.~(\ref{eq:emux}) is good enough.

The muons will also undergo multiple Coulomb scattering on their way to 
the detector in which process they don't lose energy but the angular 
distribution of the muons gets smeared. For the multiple Coulomb 
scattering we have used the formulas in Ref.~\cite{PDG}.

Since the neutrino flux to a very good approximation is constant in
the region surrounding the detector where neutrino-interactions
producing detectable muons occur, we for each produced muon choose a
distance between 0 and $L_\mu$ away from the detector where the muon
was produced and degrade its energy on its way to the detector
according to Eq.~(\ref{eq:emux}). We also take care of the Multiple
Coulomb scattering occurring during this passage of matter as
described above.

\subsection{Resulting muon fluxes}

With the methods described above, we have what we need to calculate 
the muon flux at a detector for a given neutralino (or any WIMP) mass 
and annihilation channel.  As described above, $2.5\times10^5$ 
annihilations are simulated for each neutralino mass and annihilation 
channel.  The produced neutrinos are let to interact on their way to 
the detector and the charged current interactions close to the 
detector where the muons are produced are simulated.  The muons can 
then interact and scatter on the way to the detector.  The 
(differential) muon flux at the detector is then given by summing up 
all these muons and weighting each muon by the probability that such a 
muon would have been created and detected,
\begin{equation}
  P_{\rm det} = \frac{\sigma_{CC} (E_\nu) N L_\mu(E_\mu^0)}{4 \pi D^2}
\end{equation}
where $\sigma_{CC}$ is the charged current cross section,
Eqs.~(\ref{eq:sigmacc}) and (\ref{eq:a-params}), $N$ is the number of
nucleons per cm$^3$ in the material surrounding the detector, $L_\mu$
is the range of a muon produced with energy $E_\mu^0$ and $D$ is the
distance from the source (the Sun or the center of the Earth) to the
detector.

This way the muon fluxes in units of m$^{-2}$ annihilation$^{-1}$ are 
obtained for the set of masses and annihilation channels given in 
Section~\ref{sec:MonteCarloSim}.  When a muon flux is needed for 
another mass, an interpolation is performed and when muon fluxes from 
other than these 'fundamental' annihilation channels are needed, e.g.\ 
$Z H_2^0$, the flux is easily calculated based on the 'fundamental' 
annihilation channels.  A Higgs boson will decay in 
flight to any of the particles for which the muon fluxes are 
calculated, e.g.\ $b \bar{b}$.  The Higgs bosons are let to decay in 
flight and the fluxes are obtained by integrating over the production 
angle of the decay products with respect to the Higgs momentum.  The 
total flux from the Higgs boson is then obtained by summing over all the 
Higgs decay channels.

In the next section, it is described how the muon fluxes for specific 
MSSM parameters are obtained.

\section{Muon fluxes -  predictions}
\label{sec:MuonPred}

\begin{figure}
\centerline{\epsfig{file=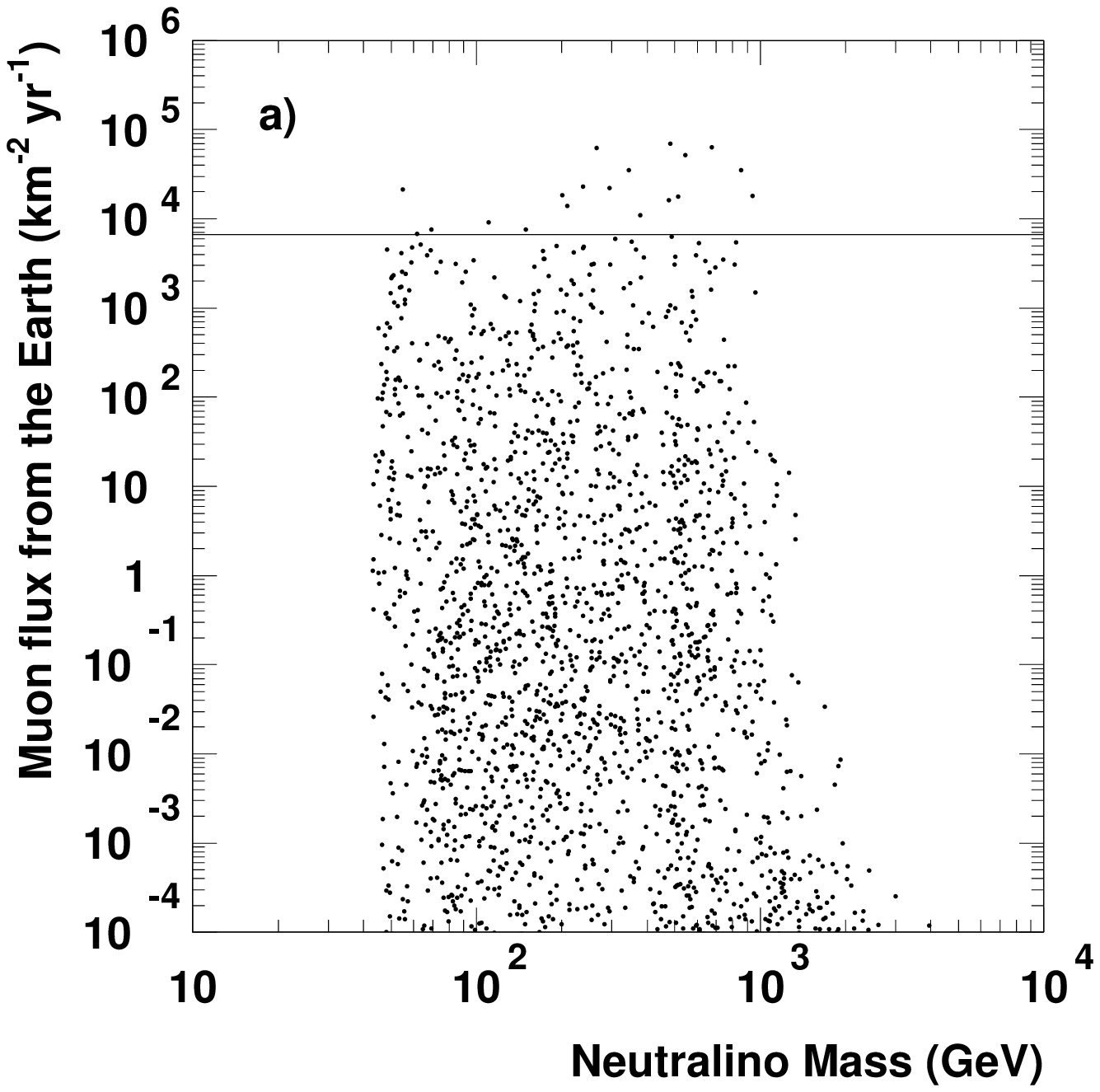,width=0.49\textwidth}
            \epsfig{file=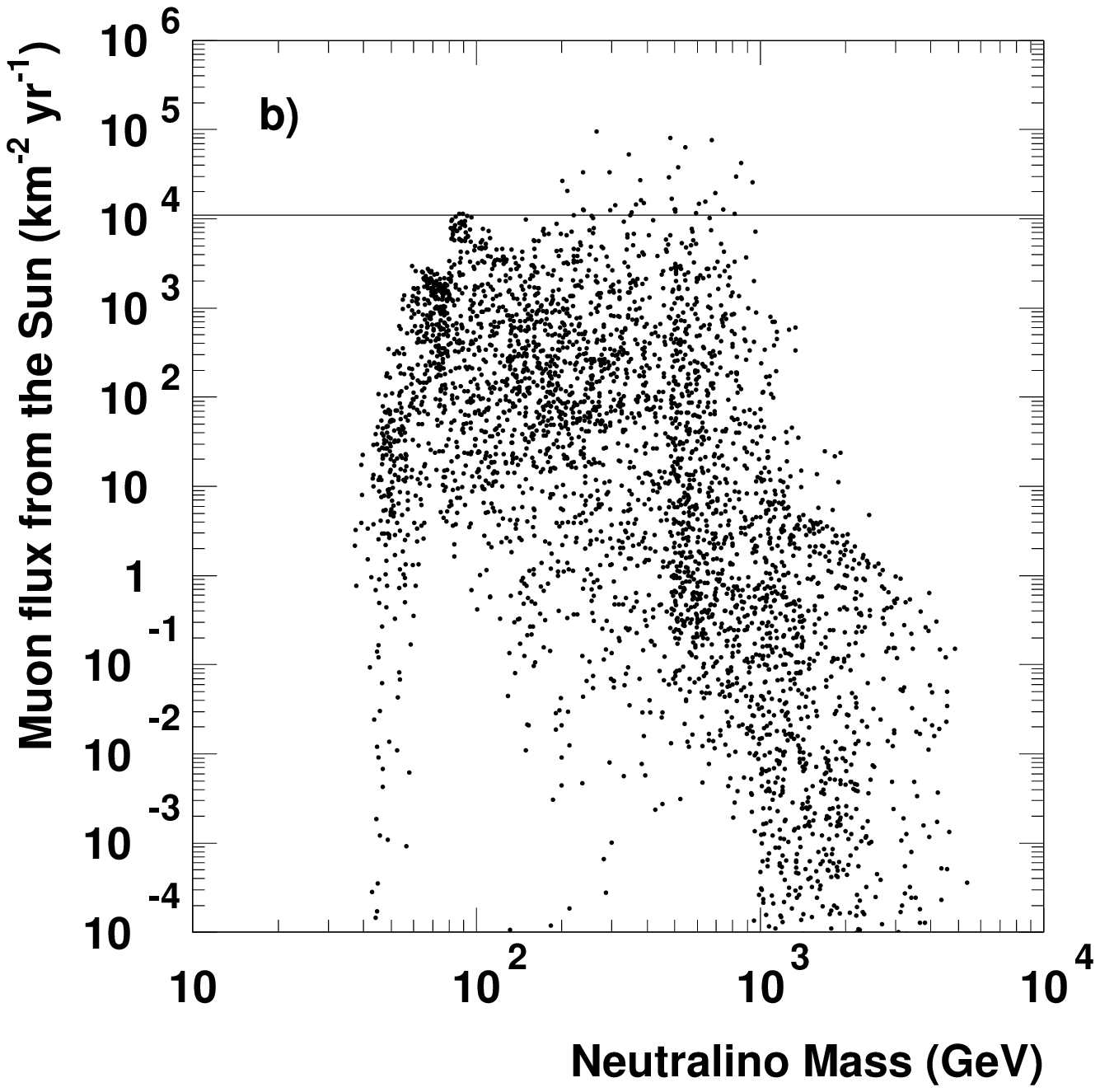,width=0.49\textwidth}}
\caption{The predicted muon flux versus the neutralino mass coming 
from neutralino annihilation in a) the Earth and b) the Sun. The
horizontal line is the Baksan limit \protect\cite{Baksan}. Only 
models with $0.025 < \Omega_{\chi} h^2 <1$ are shown and the muon 
energy threshold has been assumed to be 1 GeV\@. }
\label{fig:muonvsmx}
\end{figure}

\begin{figure}
\centerline{\epsfig{file=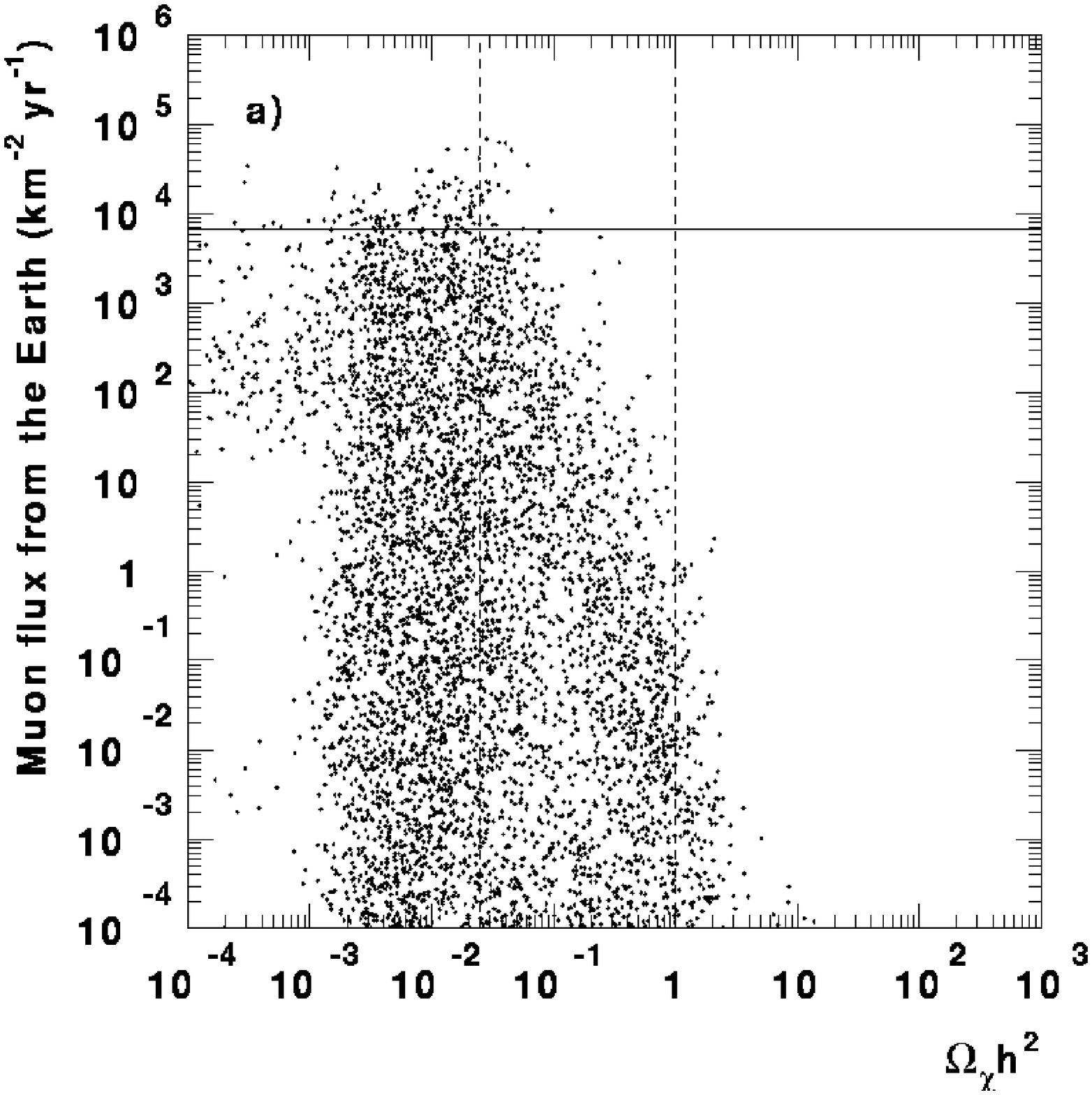,width=0.49\textwidth}
            \epsfig{file=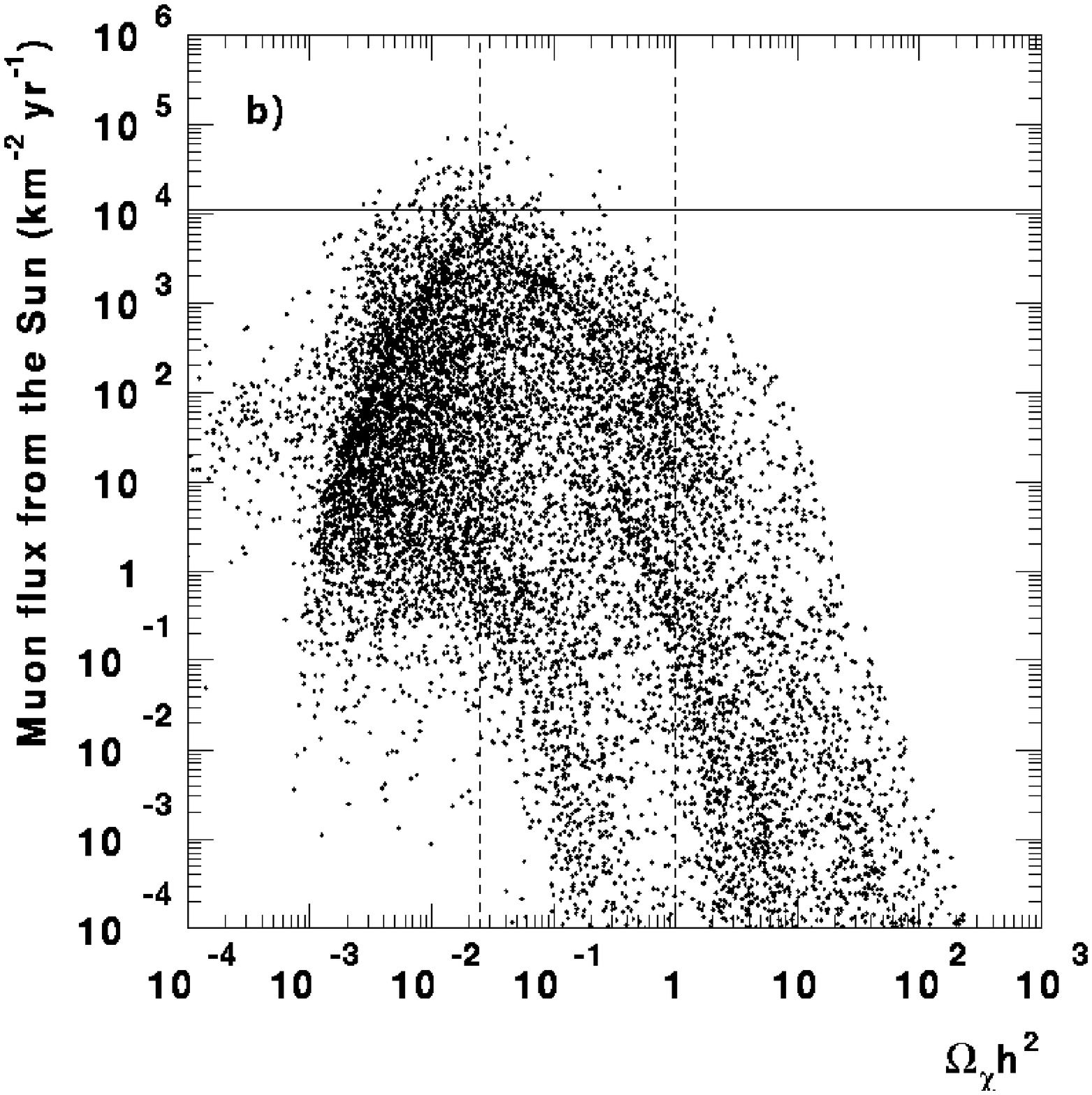,width=0.49\textwidth}}
\caption{The predicted muon flux versus the relic density,
  $\Omega_\chi h^2$, for neutralino annihilation in a) the Earth and
  b) the Sun. The horizontal line is the Baksan limit
  \protect\cite{Baksan} and the vertical dashed lines indicate the
  cosmologically interesting region, $0.025 < \Omega_\chi h^2 < 1$. 
  The muon energy threshold has been assumed to be 1 GeV\@.}
  \label{fig:muonvsoh2}
\end{figure}

\begin{figure}
\centerline{\epsfig{file=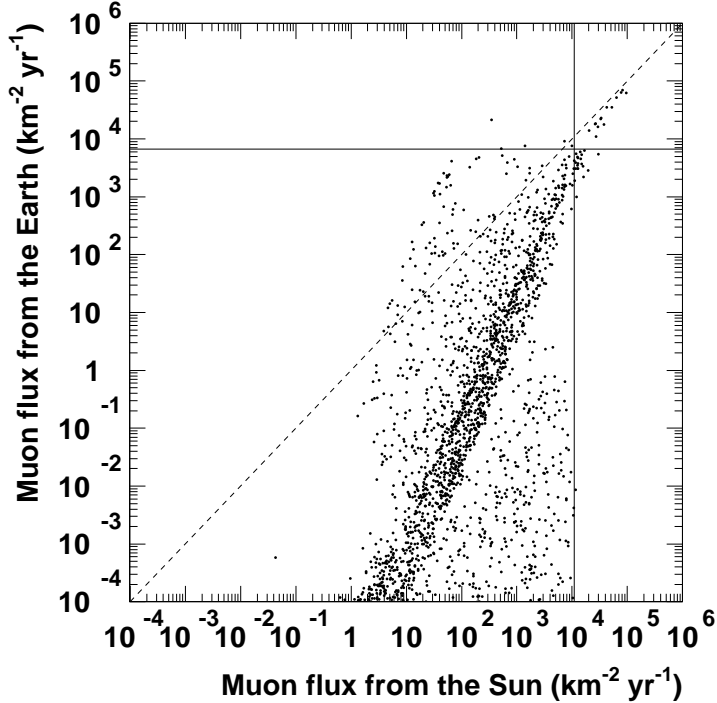,width=.75\textwidth}}
\caption{The predicted muon flux from neutralino annihilation in the
  Earth versus the flux from the Sun. The horizontal and vertical
  lines are the Baksan limits \protect\cite{Baksan}. The dashed line,
  indicating equal rates, is shown just for convenience. Only 
models with $0.025 < \Omega_{\chi} h^2 <1$ are shown and the muon 
energy threshold has been assumed to be 1 GeV\@.}
\label{fig:mufluxeavssu}
\end{figure}

\begin{figure}
\centerline{\epsfig{file=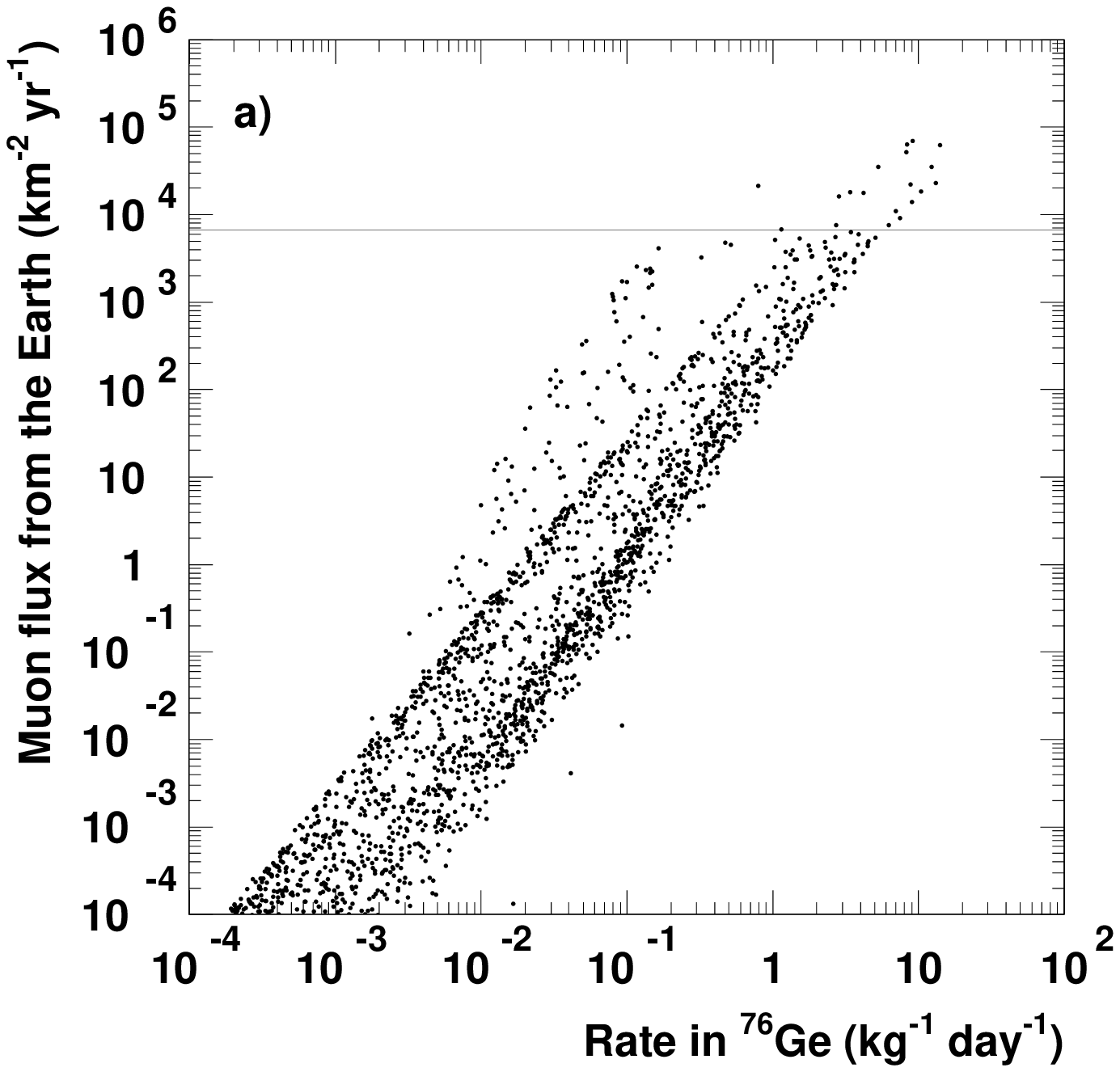,width=0.49\textwidth}
            \epsfig{file=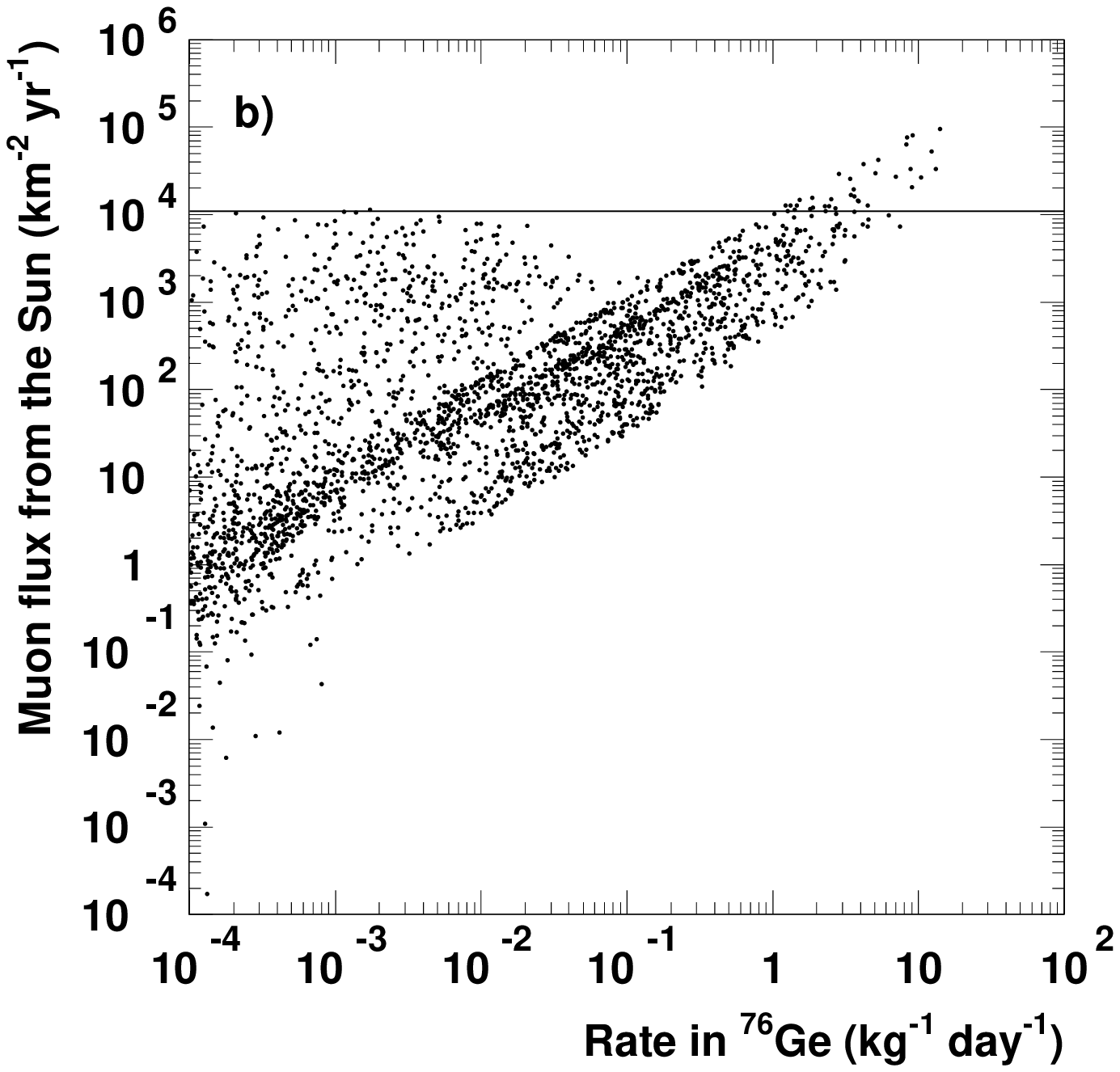,width=0.49\textwidth}}
\caption{The indirect detection rates from neutralino annihilations 
in a) the Earth and b) the Sun versus the direct detection rates in 
$^{76}$Ge \protect\cite{bsgLarsPaolo,JEpaper3}. 
The horizontal line is the Baksan limit \protect\cite{Baksan}. Only 
models with $0.025 < \Omega_{\chi} h^2 <1$ are shown and the muon 
energy threshold has been assumed to be 1 GeV\@.}
\label{fig:muonvsrge}
\end{figure}

\begin{figure}
\centerline{\epsfig{file=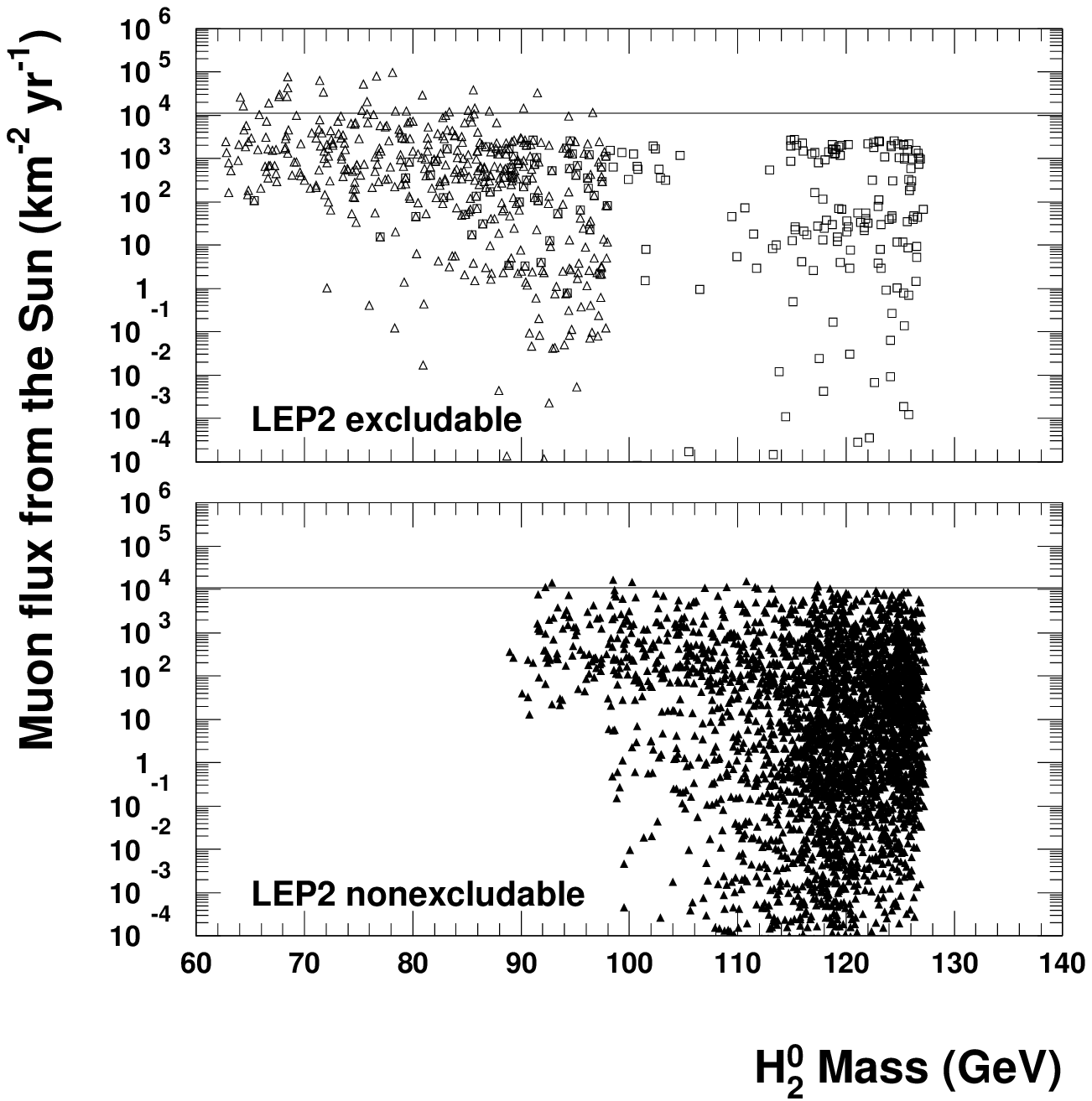,width=0.75\textwidth}}
\caption{The predicted muon flux versus the mass of the lightest Higgs 
  boson, $m_{H_2^0}$ for neutralino annihilation in the Sun.  The 
  horizontal line is the Baksan limit \protect\cite{Baksan}.  The upper 
  part are models which can be excluded at LEP2 after 150 pb$^{-1}$ of 
  running at 192 GeV and the lower part are models which cannot be 
  excluded by LEP2.  The open triangles are models that can be excluded 
  due to no Higgs discovery and the open squares are models that can be 
  excluded due to no chargino discovery.  Only models with $0.025 < 
  \Omega_{\chi} h^2 <1$ are shown and the muon energy threshold has been 
  assumed to be 1 GeV\@.}
\label{fig:muonvsmh2}
\end{figure}

We are now ready to apply the results obtained earlier in this chapter
and calculate the expected muon fluxes at neutrino telescopes
\cite{JEpaper3}. For other references on predicted rates in neutrino
telescopes, see e.g.\ Ref.~\cite{neuprod}.

The differential muon flux at a neutrino telescope is given by
\begin{equation} \label{eq:muflux}
  \frac{\partial^2 \phi_{\mu}}{\partial \theta \partial E_{\mu}} = 
  \Gamma_{A} A_{\rm eff} \sum_{i} B_{i}
  \frac{\partial^2 \phi^i_{\mu}}{\partial \theta \partial E_{\mu}} 
\end{equation}
where $\Gamma_{A}$ is the annihilation rate, Eq.~(\ref{eq:GammaA}), 
$A_{\rm eff}$ is the effective area of the detector in the direction 
of the source, $B_{i}$ is the branching ratio 
for annihilation into annihilation channel $i$ and $\partial^2 
\phi_{\mu}^i/\partial \theta \partial E_\mu$ is the differential muon 
flux from annihilation channel $i$ as obtained in the previous 
section.

In this chapter, we have up till now only assumed that the dark matter 
candidate we consider is a WIMP, but when we want to make detailed 
predictions of the muon fluxes at neutrino telescopes we have to 
specify what WIMP candidate we have, in our case the neutralino.  We 
have performed several different scans of the supersymmetric parameter 
space as given in Table~\ref{tab:scans} in Section~\ref{sec:scans}.  
For each given model we have checked against all experimental bounds 
given in Section~\ref{sec:AccCon} and only kept models not excluded by 
any experiment.

As soon as the MSSM parameters are chosen, the mass of the neutralino, 
its composition, the annihilation rate and annihilation branching 
ratios can be calculated.  The annihilation branching ratios $B_{i}$ 
in Eq.~(\ref{eq:muflux}) are evaluated with the same methods as those 
in Section~\ref{sec:AnnChannels} but for the relative velocity $v=0$ since the 
neutralinos are highly non-relativistic when they annihilate in the 
Earth and Sun.

We are mainly interested in models where the relic density of
neutralinos is cosmologically interesting, i.e.\ $0.025 < \Omega_\chi
h^2 < 1$, and in the figures shown here (except for
Fig.~\ref{fig:muonvsoh2}) we will only show models with a relic
density in this desired range.  The relic densities are calculated
with coannihilations between neutralinos and charginos included as
described in Chapter~\ref{RelDens}.

In Fig.~\ref{fig:muonvsmx} we show the predicted muon 
fluxes coming from neutralino annihilation in the Earth and Sun versus 
neutralino mass. As seen the expected rates in neutrino telescopes
vary over several orders of magnitude. The spread is bigger for
annihilation in the Earth since the capture rate in the Earth only
depends on the spin-independent scattering cross sections whereas the
capture rate in the Sun also gets a contribution from the
spin-dependent scattering cross sections (due mainly to hydrogen). 
We also show the limit on the muon fluxes coming from the Baksan 
experiment \cite{Baksan} and
we see that present neutrino telescopes already have started to
explore the MSSM parameter space. As will be seen in
Section~\ref{sec:ExpMin} an $\mathcal{O}(1 \mbox{ km$^2$})$ neutrino
telescope can explore muon fluxes down to about 50-100 km$^{-2}$
yr$^{-1}$ which at least for the Sun is a substantial fraction of the
models in Fig.~\ref{fig:muonvsmx}. Note however that the density of
points in these figures does not have any physical meaning, they are just
artifacts of how the scanning is performed. 

Most of our models with very high muon fluxes (i.e.\ those probed by 
Baksan) come from the 'light Higgs'
scan in Table~\ref{tab:scans} where the mass of the $A$ boson is low
and hence the spin-independent scattering cross sections are
high (which means that the capture rates are high).

In Fig.~\ref{fig:muonvsoh2} we show the expected muon fluxes versus
the neutralino relic density. In this figure only we also show models
with non-interesting relic densities. For $\Omega_\chi h^2 <0.025$ the
local halo mass density $\rho_\chi$ is rescaled as $\rho_\chi
\Omega_\chi h^2 /0.025$ since for these low relic densities the
neutralinos cannot make up all of the galactic halo. The general trend
of getting lower muon fluxes for higher $\Omega_\chi h^2$ is clearly
seen.  This is due to that the relic density is approximately
inversely proportional to the annihilation cross section (see e.g.\ 
the approximate expression Eq~(\ref{eq:omegasimp})) and due to the
crossing symmetry also to the scattering cross section and hence to
the annihilation rate. The bend over for $\Omega_\chi h^2<0.025$ is
due to the scaling of $\rho_\chi$ explained above.

In Fig.~\ref{fig:mufluxeavssu} we show the muon flux from neutralino
annihilations in the Earth versus the flux from neutralino
annihilations in the Sun. As seen, in most cases the flux from the Sun
is higher due to that the capture rate in the Sun also gets
contributions from spin-dependent scatterings since there is so much
hydrogen in the Sun whereas the capture rate in the Earth only depends
on the spin-independent scattering cross sections. 

In Fig.~\ref{fig:muonvsrge} we compare the muon fluxes in neutrino 
telescopes with the direct detection rates in $^{76}$Ge 
\cite{bsgLarsPaolo,JEpaper3}. Future direct detection experiments may 
probe event rates down to about 0.01 events kg$^{-1}$ day$^{-1}$.  We see 
that for a given factor of improvement in sensitivity, indirect 
detection from the Sun generally gains more than direct detection 
which in turn usually gains more than indirect detection from the 
Earth.  Be aware of the huge spread of the points though.  Especially in 
Fig.~\ref{fig:muonvsrge}b we see a nice complementarity between the two 
search methods. The reason for this is 
that spin-dependent scatterings contribute to the flux from the Sun 
but not to either the flux from the Earth or to the direct detection 
rates. There are however some direct detection materials which have 
spin, but the spin-dependent scatterings are even in these cases not 
contributing much to the direct rates.

We now want to compare these rates in neutrino telescopes with the
search potentials of LEP2\@. LEP2 will mainly put new constraints on
the mass of the lightest Higgs boson, $m_{H_2^0}$, and on the chargino
masses. In Fig.~\ref{fig:muonvsmh2} we show the muon fluxes from
neutralino annihilation in the Sun versus the $H_2^0$ mass. In the
upper panel we show models that can be probed by LEP2 after 150
pb$^{-1}$ of running at 192 GeV \cite{HiggsLEP2} and in the lower
panel, models not excludable by LEP2 are shown. For the Earth, the
corresponding figure would be about the same but with a wider spread
of the muon fluxes.  Clearly, there is a very nice complementarity
between neutrino telescope search capabilities and LEP2 search
capabilities.  If we are unlucky, however, the Nature has chosen the
parameters such that SUSY will escape both neutrino telescopes and
LEP2. Note that LEP2 will be able to exclude all our models where
$m_{H_2^0} \lsim 90$ GeV\@.

\section{Backgrounds}

There are a few backgrounds for neutrino telescopes that need to be 
handled.  The largest background is down-going muons produced by 
cosmic ray particle interactions in the Earth's atmosphere.  These 
muons don't survive too far in the Earth though and can thus be 
reduced by putting the neutrino telescope deep underground.  At 
several kilometers depth there are still a lot of down-going muons 
present, but since these are down-going and by looking for 
up-going muons (i.e.\ when the source is below the horizon) this 
background does not present a big problem.  In the cosmic ray particle 
interactions in the atmosphere there are however also neutrinos 
produced \cite{AtmMu,honda} and these will constitute the main 
background since they can not be removed by going deeper underground.  
By using the atmospheric neutrino fluxes in Ref.~\cite{honda} and the 
neutrino to muon conversion formulas in 
Ref.~\cite{GaisserStanev} we find the background muon fluxes 
underground given in Table~\ref{tab:atmbgd}.  To obtain these 
backgrounds we have used the values of $\alpha$ and $\beta$ valid for 
ice, Eq.~(\ref{eq:alphabetaice}), and rock, 
Eq.~(\ref{eq:alphabetarock}), respectively. We have integrated 
over a cone of half-aperture angle $5^\circ$ which in many cases is a 
reasonable angular width of the signal flux.

\begin{table}
  \begin{center}
  \begin{tabular}{lrrrr} \hline
  $E_\mu^{\rm th}$ [GeV]   &    1 &   10 &   25 &   50 \\ \hline
  Vertical, ice    &  903 &  506 &  359 &  257 \\
  Horizontal, ice  & 2098 & 1333 &  993 &  742 \\ \hline
  Vertical, rock   & 1030 &  564 &  393 &  275 \\
  Horizontal, rock & 2372 & 1474 & 1077 &  788 \\ \hline
  \end{tabular}
  \caption{The background fluxes of upward-going muons coming from
    atmospheric muon neutrinos. The units are km$^{-2}$ yr$^{-1}$ and
    the fluxes are within a cone of half-aperture angle 5$^\circ$,
    i.e.\ a solid angle of 0.0239 sr.} \label{tab:atmbgd}
  \end{center}
\end{table}

To discriminate the signal from the background one has to use the fact
that the signal has an angular (and energy) dependence different from
the background as will be discussed in the next section.  From the Sun
there is also a small background coming from cosmic ray particle
interactions in the Sun's corona \cite{SunBgd}.  This background will
have a similar angular dependence as the signal but a different energy
dependence.  This background is of the order of 10 events/km$^2$ which
is usually below the signals accessible with neutrino telescopes of
order 1 km$^2$ being currently planned (see next section).  However,
in the most sensitive cases where the neutralino mass is high and we
have a specific model to test for, this background should be taken
into account.  Since it will mostly effect the analysis when even
bigger neutrino telescopes than $\cal O$(1 km$^2$) are built, we will
for the moment neglect this background.

\section{What signal levels can be probed?}
\label{sec:ExpMin}

If we now imagine having a specific neutrino telescope, how high must 
the signal flux be to be detectable?  First of all it 
must be high enough so that we get any events and secondly it must be 
high enough to be seen above the background.  Since the atmospheric 
background has both a completely different angular and energy 
distribution than the signal one would expect that it should not be 
too difficult to observe the signal above this background.  In the 
simplest data analysis one would just look for an excess of events in 
an angular cone centered around the Earth or Sun with a given 
half-aperture angle $\theta_{\rm max}$.  The angle $\theta_{\rm max}$ 
would be chosen large enough to include as much signal as possible and 
small enough to exclude as much background as possible.  This is what 
is usually done with neutrino telescopes today.  In 
Ref.~\cite{BottinoMoscoso} was investigated what the optimal 
$\theta_{\rm max}$ would be for different neutralino masses and 
compositions.  Here we go one step further \cite{JEpaper4} and 
investigate what could be done if the full angular and/or energy 
resolution is actually used to discriminate against the atmospheric 
background.  Note, though, that currently designed neutrino 
telescopes have a very poor energy resolution in the energy region 
where the signal from neutralino annihilation is.

In the following subsections, we describe how we calculate the expected
sensitivities for neutrino detectors with angular and/or energy
resolution and present numerical results for some representative
models.


\subsection{'Single-bin' analysis}

Consider a theory which predicts an upward-muon flux $\phi_s$ (where
the $s$ stands for `signal') with an angular distribution
$d\phi_s/d\theta=\phi_s^0 f_s(\theta) \sin \theta$, where $\theta$ is
the angle the muon makes with the direction of the source of interest,
and $\int f_s(\theta)\,\sin\theta\,d\theta=1$ (i.e.\ $f(\theta)$ is
constant for an isotropic distribution).  We would like to disentangle
this signal {}from a background of atmospheric neutrino-induced muons
which has a flux $\phi_b$ with an angular distribution
$d\phi_b/d\theta = \phi_b^0 f_b(\theta) \sin \theta$, which is nearly
isotropic (at least on small angular scales).

Consider first an experiment that can only tell that a muon has been 
detected with an angle $\theta\leq\theta_{\rm max}$ and an energy 
$E_{\mu}>E_{\mu}^{\rm th}$, but no further information on the muon 
direction or energy is available.  Then, the angular acceptance cone 
around the source must be large enough to include all (or most) of the 
muons produced by neutrinos from the source.  One would therefore have 
some number of muons detected with an angle $\theta\leq\theta_{\rm 
max}$.  For example, in their searches for energetic neutrinos from 
the Earth and Sun, the Baksan collaboration \cite{Baksan} reports the 
flux of muons within an angle $\theta_{\rm max}=30^\circ$ of the Sun 
or the center of the Earth.  The Kamiokande collaboration 
\cite{Kamiokande} reports the flux of muons within an angle varying 
between $\theta_{\rm max}=5^\circ$--$30^\circ$. We will refer to this 
way of analyzing data as the `single-bin' (or `0D') approach.

With such an experiment, the number of background events after an exposure
$\cE$ (for example, in units of km$^2$ yr) is $N_b=\cE \phi_b^0
\int_0^{\theta_{\rm max}}\,f_b(\theta)\,\sin\theta\,d\theta$.  
The number of expected events from the
source of interest is $N_s=\cE\phi_s^0 \int_0^{\theta_{\rm max}}\,
f_s(\theta)\, \sin\theta\, d\theta$.  A
$3\sigma$ detection would require an excess of $3\sqrt{N_b+N_s}$ events
over the number expected.  Then, a $3\sigma$ excess will be observable
only if $\phi_s > 3\sigma$ where $\sigma = \sqrt{N_b+N_s}/\cE$.  

{}From such a simple experiment described above where no energy or
angular distributions are used, we can conclude that the minimal
exposure required for a $3\sigma$ discovery is
\begin{equation} \label{eq:expminsimple}
  \cE_{\rm min} = \frac{9 \left(\phi_b+\phi_s \right)}{\phi_s^2}
\end{equation}
where $\phi_b$ and $\phi_s$ are the background and signal fluxes above
threshold and within the angular cone of acceptance $\theta_{\rm
  max}$.  Note that Eq.~(\ref{eq:expminsimple}) is only valid when the
fluxes are high.  This minimal exposure is relevant to the way, e.g.,
Baksan and Kamiokande have analyzed their data (with different values
of $\theta_{\rm max}$). 

In the following more detailed examples, 
Eq.~(\ref{eq:expminsimple}) with either $\theta_{\rm max}=5^\circ$ or 
the optimal $\theta_{\rm max}$ (which maximizes $\phi_{s}/\sigma$) 
will be used for comparison.  Note that for low masses ($\lsim 100$ 
GeV) the optimal $\theta_{\rm max}$ will be higher than $5^\circ$ and 
for high masses it will be lower.  However, one cannot know in advance 
what the optimal cut will be, unless we have a specific model we want 
to test.  If we don't have a specific model 5$^\circ$ is a reasonable 
choice of $\theta_{\rm max}$ giving decent results both for low and 
high masses.  In principle it is however possible to extract the 
optimal angular cut from data by varying $\theta_{\rm max}$ and for 
the optimal value the signal should be most clearly visible.  We have 
not investigated further how well this method of finding $\theta_{\rm 
max}$ would do, but it will never do better than the single-bin 
approach where the optimal angular cut is known in advance.


\subsection{Covariance-matrix analysis}
\label{sec:covmatrix}

Now consider a slightly more sophisticated experiment which has
angular and/or energy resolution.  It is possible that we will
actually be fitting for both a background and a signal flux of muons
where the background flux is given by
\begin{equation}\label{eq:eaatmbg}
     \frac{d^2 \phi_b}{dE d\theta}(E,\theta) = \phi_b^0 f_b(E,\theta),
\end{equation}
which we assume to be isotropic (at least over small angular patches), 
$f_b(E,\theta)=f_b(E)$.  We will only consider the atmospheric 
neutrino background resulting from cosmic-ray interactions in the 
Earth's atmosphere.  In addition, we will want to fit data for an 
annihilation signal which generally depends on the neutralino (or any 
WIMP) mass $m_\chi$ and its composition.  We may parameterize this as
\begin{equation} \label{eq:param}
     \frac{d^2 \phi_s}{dE d\theta} (E,\theta) = \phi_s^0
     \left[ a f_{\rm hard}(m_\chi,E,\theta)+(1-a) 
     f_{\rm soft}(m_\chi,E,\theta) \right] ,
\end{equation}
where $a$ parameterizes the relative contributions of a `hard' and 
`soft' annihilation spectrum.  As a `hard' annihilation spectrum 
we have used the $\tau^+ \tau^-$ channel below the $W$ mass and $W^+ 
W^-$ above and as a `soft' spectrum we have used $b \bar{b}$.  These 
channels represent the extreme hardnesses of the spectrum for any 
given neutralino mass.  For the evaluation of the neutrino and muon 
flux for these channels we have used the method given earlier in this 
chapter.

Therefore, we are assuming that the muon angular and/or energy
distribution from both background and signal will be described by the
set of parameters ${\bf s}= \{\phi_b^0,\phi_s^0,m_\chi,a\}$ (one
could also envision more parameters).  We now want to ask, with what
precision can we measure these parameters with a given experiment,
assuming the true distribution is given by some set of parameters,
${\bf s}_0$?

To answer this, we assume the data is binned into a number of angle/energy
bins, and each bin $i$ is centered on angle $\theta_i$ and energy
$E_i$ with widths $\Delta E_i$ and $\Delta\theta_i$.  Therefore, for a
given set ${\bf s}$ of parameters, the flux will be
\begin{equation} \label{eq:totflux}
  \frac{d^2 \phi}{dE d\theta} (E,\theta;{\bf s}) = \frac{d^2\phi_b}{dE
    d\theta} (E,\theta;{\bf s})+\frac{d^2 \phi_s}{dE d\theta}
    (E,\theta;{\bf s}).
\end{equation}
The probability distribution for the number of events expected
in each bin is a Poisson distribution with mean $N_i = {\cal E}
\frac{d^2 \phi}{dE d\theta} (E_i,\theta_i) \Delta E_i\Delta \theta_i
$, so it has a width $\sigma_i =\sqrt{N_i}$.

So, suppose the true parameters are ${\bf s}_0$.  Then the probability
distribution for observing an angle/energy distribution which is best
fit by the parameters ${\bf s}$ is
\begin{equation}
     P({\bf s}) \propto \exp \left[ -\frac{1}{2} ({\bf s} - {\bf
     s}_0) \cdot [ \alpha ] \cdot ({\bf s} - {\bf s}_0) \right],
\end{equation}
where the curvature matrix $[\alpha]$ is given approximately by
\begin{eqnarray}
     \alpha_{ab} & = & {\cal E}\sum_i\, \frac{1}{\sigma_i^2} \, 
     \frac{\partial N_i}{\partial s_a}\, \frac{\partial N_i}
     {\partial s_b}   \nonumber \\
     & = &  4{\cal E}\sum_i\, \frac{\partial \sqrt{N_i}}
     {\partial s_a}({\bf s}_0) \, \frac{\partial \sqrt{N_i}}{\partial
     s_b}({\bf s}_0),
\end{eqnarray}
where the partial derivatives are evaluated at ${\bf s}= {\bf s}_0$,
and we used $\sigma_i = \sqrt{N_i}$ in the second line.  In a
realistic experiment, the width of the bins would be comparable to the
angular and/or energy resolution of the experiment.  In the limit of
perfect angular and energy resolution, the sum becomes an integral,
\begin{equation}
     \alpha_{ab}  = 4{\cal E} \int\!\int\, dE\, d\theta \, \frac{\partial
     \sqrt{d^2\phi(E,\theta;{\bf s})/dE d\theta}}{\partial s_a}\,
     \frac{\partial \sqrt{d^2 \phi(E,\theta;{\bf s})/dE d\theta}}
     {\partial s_b}.
\label{integralalpha}
\end{equation}
The covariance matrix, $[{\cal C}] = [\alpha]^{-1}$ gives an estimate
of the standard errors that would be obtained from a
maximum-likelihood fit to data: The standard error in measuring the
parameter $s_a$ (after marginalizing over all the other undetermined
parameters) is approximately $\sigma_a \simeq {\cal C}_{aa}^{1/2}$.
If three times the standard error in the parameter $\phi_s^0$ is less than
$\phi_s^0$, for a given underlying model ${\bf s}_0$ and for a given
experiment, then this model will be distinguishable from background at
the $3\sigma$ level.

\begin{figure}
  \centerline{\psfig{file=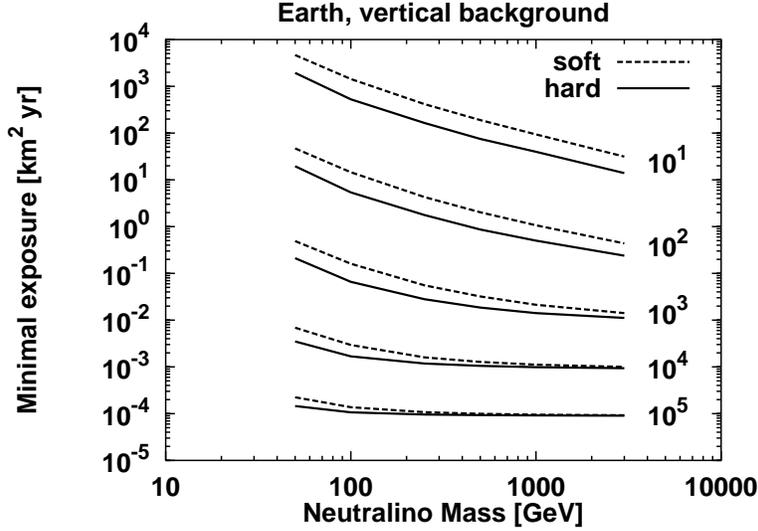,width=0.75\textwidth}}
  \caption{ The exposures needed for a 3$\sigma$ discovery for 
  different signal fluxes (indicated to the right in the figure in 
  units of km$^{-2}$ yr$^{-1}$) as a function of neutralino (or any 
  WIMP) mass assuming perfect angular resolution but no energy 
  resolution (and with a muon energy threshold of 1 GeV).  The muon 
  fluxes are evaluated for annihilation in the Earth with vertical 
  background.  The solid (dashed) lines correspond to hard (soft) muon 
  spectra.  The three signal parameters $\{\phi_s^0,m_\chi,a\}$ in 
  Eq.~(\protect\ref{eq:param}) are assumed to be unknown while the 
  background flux is assumed to be known.  Note that only exposures 
  less than, say, 25 km$^{2}$ yr are realistic in the near future.  }
\label{fig:minexpearth1}
\end{figure}

If all of the parameters except for $\phi_s^0$ are fixed, then
$[\alpha]$ is a $1\times1$ matrix, i.e.\ $1/\sigma^2$.  In this
case, Eq. (\ref{integralalpha}) reduces to
\begin{equation}
    \frac{1}{\sigma^2} = \cE \int\!\int\, 
    \frac{[f_s(\theta,E)]^2 }{\phi_b^0 f_b(E) + \phi_s^0
    f_s (\theta,E)}
    \, \sin\theta\, d\theta\, dE.
\label{energysensitivitytwo}
\end{equation}
To illustrate, if there were no background,
Eq.~(\ref{energysensitivitytwo}) says that the statistical uncertainty
in the number of events is the square root of the number of events,
and this makes sense.  However, if the total number of events is
nonzero, then a signal has been discovered.

In fact, if there is no background, and an event is seen, it
constitutes discovery.  On the other hand, if nothing is seen, the
95\% CL upper limit to the number is 3.

\subsection{Results}

We are now ready to perform some actual calculations using the
techniques described in the previous subsection for the specific example
of neutralino annihilation in the Sun and Earth. We
assume that the neutrino energy spectra are either the hard or soft
spectra described above; energy spectra from specific neutralino 
models
should fall somewhere between these two extremes. Since the muon flux
is proportional to the neutrino energy squared, the hard annihilation
channels will generally be more important and hence in general the
muon spectra will be more hard than soft. Because of the steep fall
with energy of the atmospheric background, hard spectra generally
require less exposure.  In all integrations with angular (and energy)
distribution the integration in Eq.~(\ref{integralalpha}) is performed
up to $\theta=30^\circ$.
For the atmospheric background we have used the results given in Ref.\ 
\cite{honda}. 

\begin{figure}
  \centerline{\epsfig{file=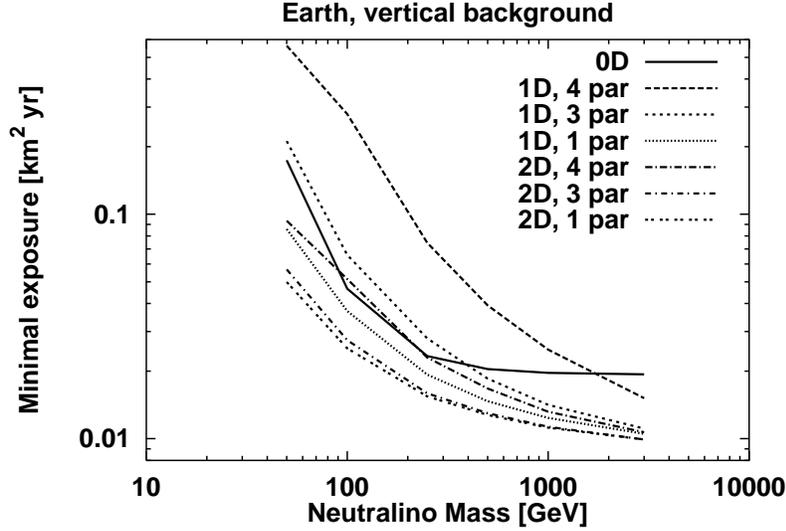,width=0.75\textwidth}}
  \caption{The exposures needed for a 3$\sigma$ discovery for the signal flux
  $\phi_s^0 = 10^3$ km$^{-2}$ yr$^{-1}$ coming from neutralino annihilation
  in the Earth.  The minimal exposures needed for a detector with
  neither angular nor energy resolution (0D), only angular but no
  energy resolution (1D) and both angular and energy resolution (2D)
  is shown. For the 1D and 2D cases, results are given for all four
  parameters in Eqs.~(\protect\ref{eq:eaatmbg}) and
  (\protect\ref{eq:param}) being free (4 par), only the three signal
  flux parameters being free (3 par) and only the normalization of the
  signal flux, $\phi_s^0$ being free (1 par). An energy threshold of 1
  GeV is used in all cases and for the 0D case an integration of the
  fluxes up to $\theta_{\rm max}=5^\circ$ is performed. All curves are
  for hard annihilation spectra.}
  \label{fig:minexpearth2}
\end{figure}

In Fig.~\ref{fig:minexpearth1} the minimal exposures needed to make a
$3\sigma$ discovery are shown for a detector with perfect angular
resolution but no energy resolution. The two extreme cases of soft and
hard annihilation spectra are shown. In producing
Fig.~\ref{fig:minexpearth1} we have assumed that only the signal flux
parameters $\{\phi_s^0,m_\chi,a\}$ in Eq.~(\protect\ref{eq:param}) are
unknown. This is not unreasonable since the background can be expected
to be measured well by an off-source measurement.  For annihilation in
the Sun the curves are similar but less steep. We find that a
neutrino telescope with exposures of about 1--25 km$^2$ yr would be
able to detect signal fluxes down to about 50--100 km$^{-2}$ yr$^{-1}$.

A comparison with what one could detect if one also has energy
resolution is given in Fig.~\ref{fig:minexpearth2} where a comparison
between how many parameters in Eqs.~(\protect\ref{eq:eaatmbg}) and
(\protect\ref{eq:param}) that are known is also shown. We find that by having
energy resolution we can gain as much as another factor of two in what
signal fluxes we can detect.

In case of a detector with angular resolution but no energy resolution
we have also investigated what could be gained by varying the energy
threshold and the gain is very small and only for neutralino masses above
100 GeV. Increasing the threshold above 10 GeV gives no further
improvements. On the other hand, large detectors (like {\sc Amanda}) which
have a threshold of tens of GeV will not lose much sensitivity either,
for neutralino masses above 100 GeV.

\subsection{Example of a neutrino telescope}

\begin{figure}
\centerline{\epsfig{file=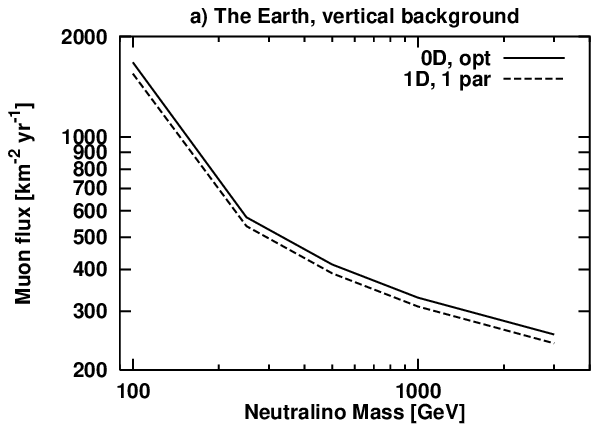,width=0.49\textwidth}
            \epsfig{file=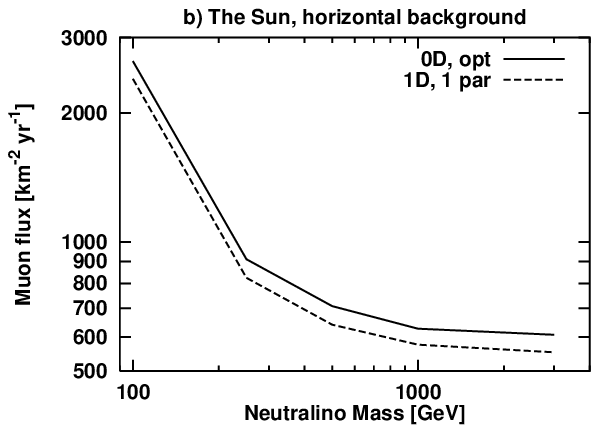,width=0.49\textwidth}}
\caption{The muon fluxes coming from neutralino annihilation in a) the
  Earth and b) the Sun that can be discovered (at the $3\sigma$ level)
  with a neutrino telescope of present {\sc Amanda} size. The
  exposures assumed are a) ${\cal E}=0.05$ km$^2$ yr and b) ${\cal
    E}=0.03$ km$^2$ yr. The neutralino annihilation spectrum is
  assumed to be hard.  The angular cut for the 0D case has been set to
  the optimal one and only the signal parameter $\phi_s^0$ in
  Eq.~(\protect\ref{eq:param}) is assumed to be unknown in the 1D
  case. The background flux is assumed to be a) vertical and
  b) horizontal as is the case for {\sc Amanda}. The muon energy 
  threshold has been assumed to be $E_{\mu}^{\rm th}=25$ GeV\@.}
  \label{fig:flxamanda}
\end{figure}

Let us now consider a more specific example of a neutrino telescope.
We will consider a neutrino telescope with a size of {\sc Amanda} (at
present $\sim10000$ m$^2$ in the direction of the Earth and $\sim6000$
m$^2$ in the direction of the Sun) run for 5 years, i.e.\ with an
exposure of ${\cal E} = 0.05$ km$^2$ yr and ${\cal E} = 0.03$ km$^2$
yr for the Earth and Sun respectively. In Fig.~\ref{fig:flxamanda} the
muon fluxes that can be probed by such a detector is shown. We have
here assumed that the muon energy threshold is 25 GeV and that we have
an a priori model we want to test, i.e.\ we have used the optimal
angular cut for the 'single-bin' analysis (0D) and we have only
assumed that the signal flux normalization $\phi_s^0$ in
Eq.~(\ref{eq:param}) is unknown for the `1D' case. We show the muon
fluxes that can be probed when the neutralino annihilation spectrum is
hard.  We see that at higher masses it is for {\sc Amanda} possible to
probe fluxes down to a few hundred km$^{-2}$ yr$^{-1}$ at high masses.

\begin{figure}
\centerline{\epsfig{file=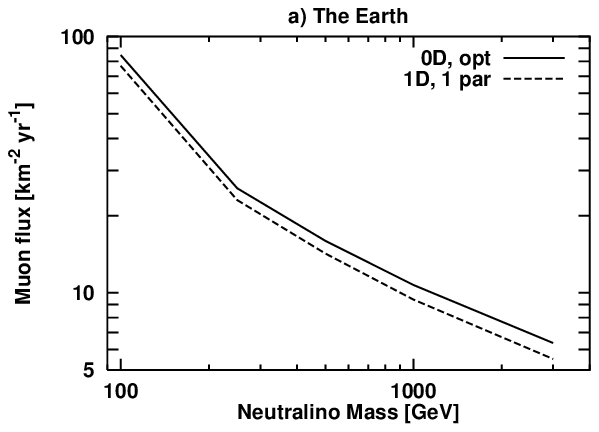,width=0.49\textwidth}
  \epsfig{file=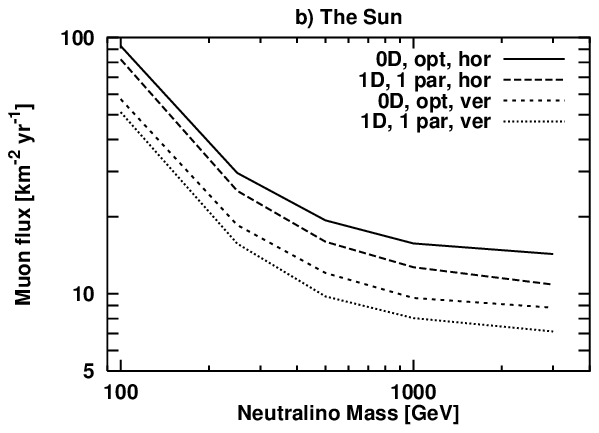,width=0.49\textwidth}} \caption{The 
  muon fluxes coming from neutralino annihilation in a) the Earth and 
  b) the Sun that can be discovered (at the $3\sigma$ level) with a 
  neutrino telescope with an exposure of ${\cal E}=10$ km$^2$ yr.  The 
  neutralino annihilation spectrum is assumed to be hard.  The angular 
  cut for the 0D case has been set to the optimal one and only the 
  signal parameter $\phi_s^0$ in Eq.~(\protect\ref{eq:param}) is 
  assumed to be unknown in the 1D case.  In a) the atmospheric 
  background is vertical and in b) curves for both vertical and 
  horizontal backgrounds are shown. The muon energy 
  threshold has been assumed to be $E_{\mu}^{\rm th}=25$ GeV\@.}
  \label{fig:flxkm2}
\end{figure}

In Fig.~\ref{fig:flxkm2} we show the muon fluxes that can be probed
with a neutrino telescope with an area of 1 km$^2$ run for 10 years,
i.e.\ with an exposure of ${\cal E}=10$ km$^2$ yr. As above, the
optimal angular cut has been used and only $\phi_s^0$ is assumed to be
unknown. The muon energy threshold has again been assumed to be 
$E_\mu^{\rm th} = 25$ GeV\@. We see that such a detector can probe
muon fluxes down to about 10 km$^{-2}$ yr$^{-1}$ at high masses. In
Fig.~\ref{fig:flxkm2}b we see that one would gain about a factor of 1.5
by having vertical atmospheric background instead of horizontal. When
looking towards the Sun, a
neutrino telescope at the poles will have almost horizontal background
and a neutrino telescope close to the equator will have horizontal
background in the beginning/end of the night and nearly vertical in the
middle of the night (how vertical depends on the latitude and season).

In case all three signal flux parameters in
Eq.~(\ref{eq:param}) are unknown, the curves in
Figs.~\ref{fig:flxamanda}--\ref{fig:flxkm2} would be about a factor of
1.5--2 higher. If the energy threshold is higher than 25 GeV, the
curves will be higher at low masses.

Note that for especially the $\cal O$(1 km$^2$) telescope, as shown in 
Fig.~\ref{fig:flxkm2}, we are in the background dominated regime which 
means that if we increase the exposure by a given factor, the muon 
fluxes we can probe decrease by the square root of that factor.

\subsection{Discussion}

We can conclude that if want to test a specific model we only gain
about 10--25\% by using the full angular resolution compared to the
`single-bin' approach. If we don't have a specific model to test for,
which is usually the case, we have to choose an angle $\theta_{\rm
  max}$ in the `single-bin' approach compared to which we can gain up
to a factor of 2 by using the angular resolution as proposed here.  By
using the parameterization of the signal flux, Eq.~(\ref{eq:param}) we
can also gain some information on the neutralino mass and the hardness
of the spectrum though.

By varying the energy threshold not much more is gained, but by having
energy resolution about a factor of 1.5--2 can be gained, slightly
more at low signal fluxes and slightly less at high signal fluxes.

We also note that for neutrino telescopes with a size of about 1
km$^2$ and only angular reslution,
the signal fluxes we can expect to probe is in the region of
50--100 km$^{-2}$ yr$^{-1}$ when all three signal parameters in
Eq.~(\ref{eq:param}) are unkonwn and about a factor of two lower if
only the normalization is unkown. If we in addition have energy
resolution we gain about a factor of 1.5--2 more.
The signal fluxes within reach are almost an order
of magnitude larger than the expected background coming from cosmic
ray interactions in the Sun (about 10 events km$^{-2}$ yr$^{-1}$
\cite{SunBgd}). Therefore it is quite safe to neglect this background
at the present stage. When detectors are getting even bigger it will
however be a severe limitation when looking for the neutrino flux from
the Sun since this background is also highly directional as the
signal. The energy dependence is quite different though, so having
energy resolution may be beneficial in this case.

\section{What does a detected signal tell?}

\begin{figure}
  \centerline{\epsfig{file=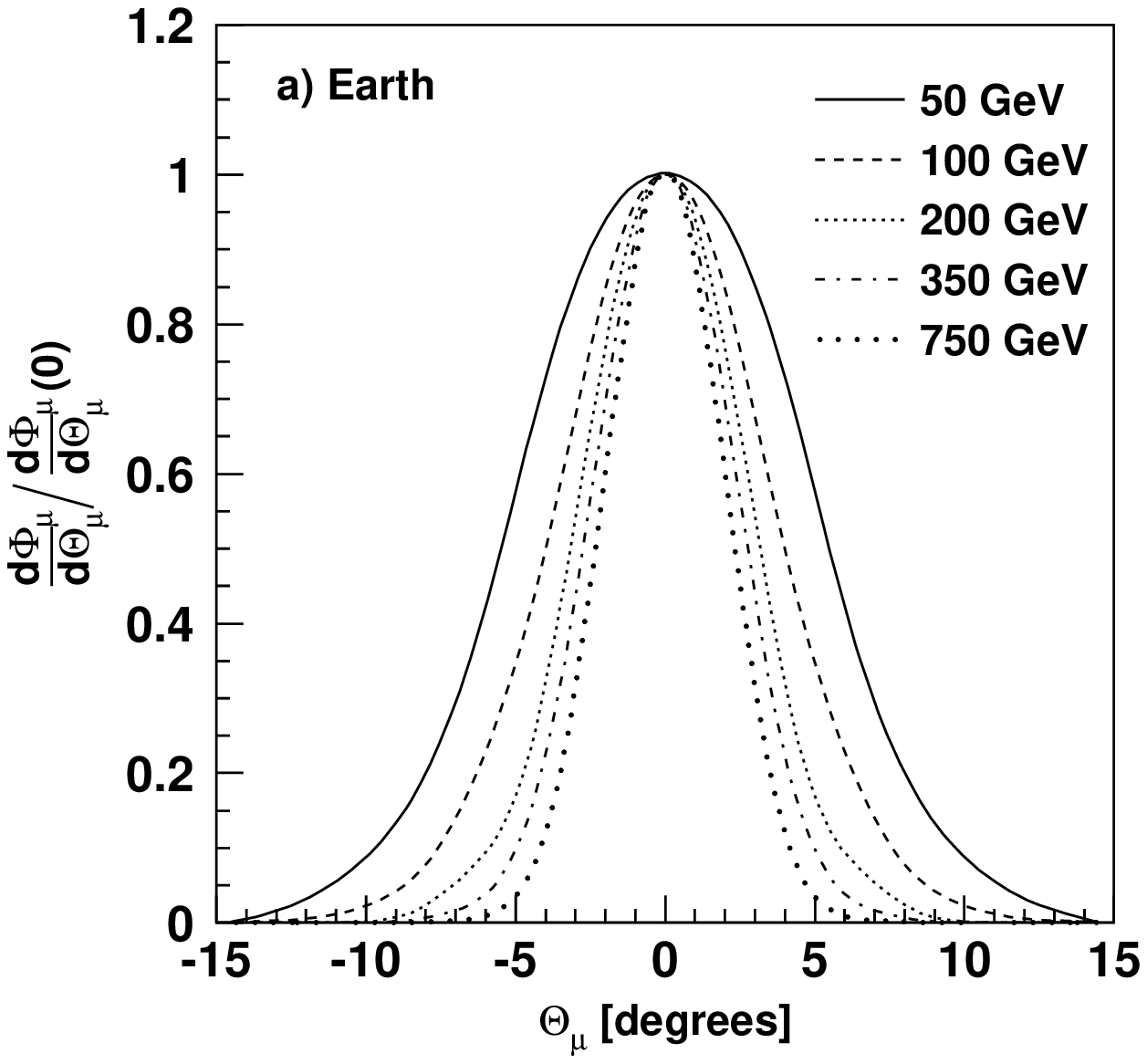,width=.49\textwidth}
    \epsfig{file=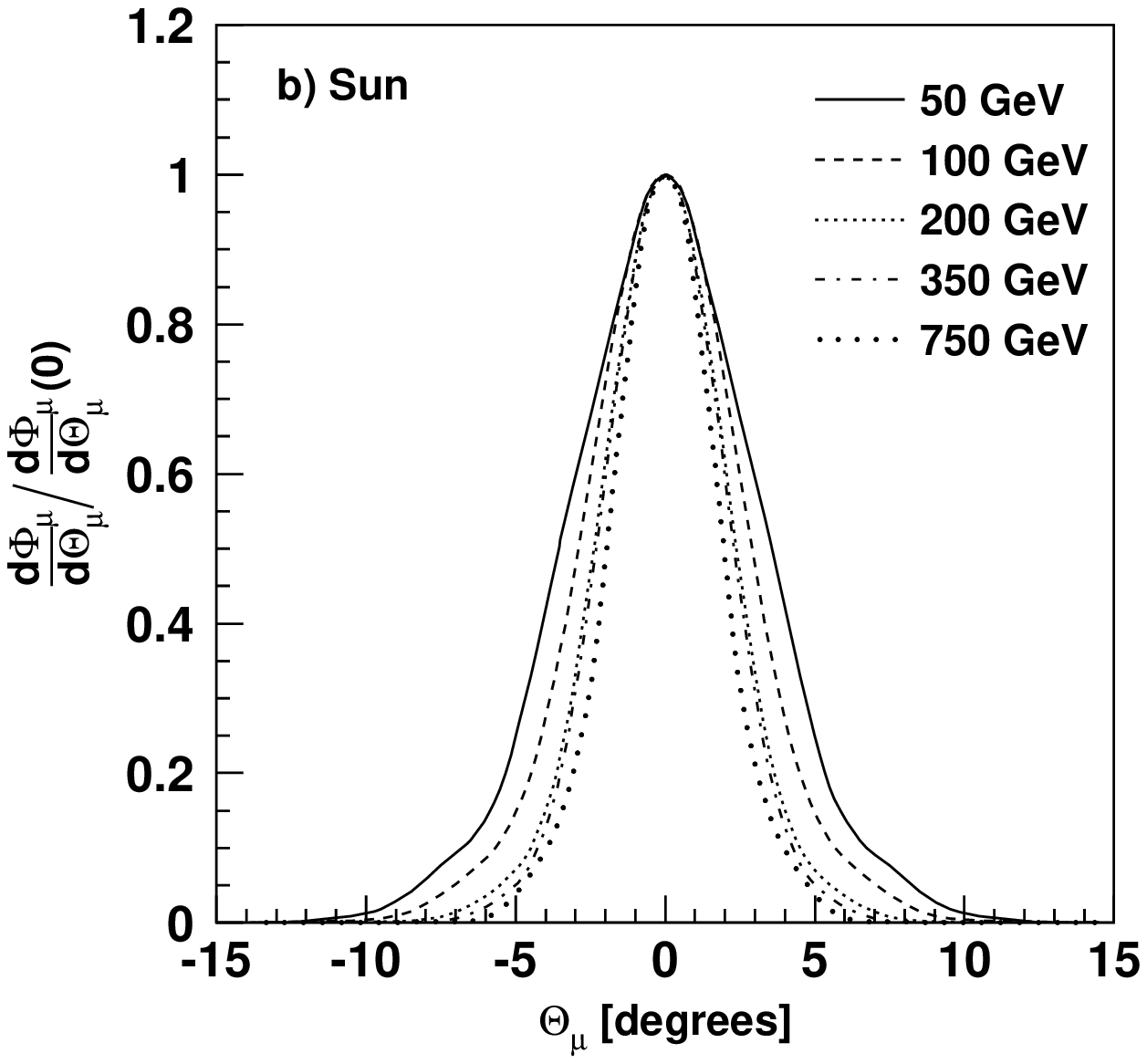,width=.49\textwidth}}
  \caption{The projected angular distributions of neutrino-induced 
  muons from neutralino (or any WIMP) annihilations in a) the Earth 
  and b) the Sun for neutralino masses of 50 GeV (solid), 100 GeV 
  (dashed), 200 GeV (dotted), 350 GeV (dash-dotted) and 750 GeV (wide 
  dotted). The distributions shown are for hard annihilation 
  channels, $W^+W^-$ for 100--750 GeV and $\tau^+\tau^-$ for 50 GeV, 
  and with a detector muon threshold of $E_{\mu}^{\rm th}=10$ GeV\@ 
  and a detector (projected) angular resolution of 1.4$^\circ$.}
    \label{fig:mudisteasu}
\end{figure}

\begin{figure}
  \vspace{7.5cm}
  \includegraphics{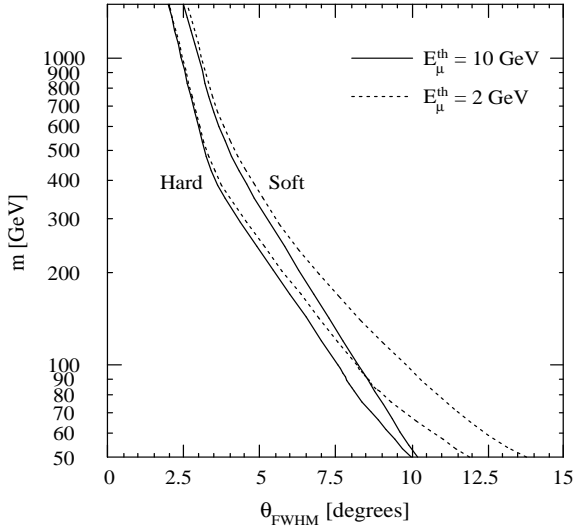}
  \caption{The neutralino mass versus full width half maximum
    $\theta_{\rm FWHM}$ of the neutrino-induced muon distribution for
    soft and hard channels. The solid line corresponds to $E_\mu^{\rm
    th}$ = 10 GeV and the dashed line corresponds to $E_\mu^{\rm th}$ =
    2 GeV\@. The curves given are for neutralinos annihilating in the
    Earth.}
    \label{fig:mxvsfwhmearth}
\end{figure}

We have in Eqs.~(\ref{eq:aprof1})--(\ref{eq:aprof2}) and 
Fig.~\ref{fig:nudistearth} seen that the width of the neutrino 
distribution depends on the neutralino mass. From 
Eqs.~(\ref{eq:aprof1})--(\ref{eq:aprof2}) we can derive that the root 
mean square value of the projected angular distribution given in 
Fig.~\ref{fig:nudistearth} is 
\begin{equation}
  \theta_{\nu}^{\rm rms} \simeq \frac{1}{\sqrt{2}} 
  \frac{r_{\chi}}{R_{\otimes}} \mbox{ rad} \simeq 
  \frac{23^\circ}{\sqrt{m_{\chi}/\mbox{GeV}}} \quad , \quad m_{\chi} 
  \gsim 10 \mbox{ GeV.}
\end{equation}
If we could measure this width we would, since the width only depends 
on the neutralino (or any WIMP) mass, be able to determine the 
neutralino mass in a model-independent fashion.  It is however not 
possible to measure this width directly since we measure the muons and 
not the neutrinos in neutrino telescopes.  When converting the 
neutrino flux to a muon flux, we will introduce a model dependence, 
since we need to know the neutrino energy distribution which depends 
on the branching ratios to different annihilation channels.  The angle 
the muon makes with respect to the neutrino in the charged current 
scattering decreases as the square root of the neutrino energy, i.e.\ 
we will get a smearing of the neutrino distribution which is smaller 
the harder the annihilation spectrum is and the heavier the 
neutralino is. The muon will also undergo multiple Coulomb scattering 
on it way to the detector which further smears the angular 
distribution. Note that both the intrinsic neutrino angular 
distribution and the charged current scattering angle tends to widen 
the angular distributions the lower the neutralino mass is.

As we did in the previous section, we will consider the extreme cases 
of `hard' and `soft' annihilation spectra where we for the hard 
spectra use $W^+W^-$ when $m_{\chi}>m_{W}$ and $\tau^+ \tau^-$ when 
$m_{\chi}<m_{W}$ and for the soft spectra we use the $b\bar{b}$ 
annihilation channel. These cases should represent the extreme cases 
for any neutralino (or any WIMP).

We have calculated the resulting muon angular distributions for hard
and soft spectra and for different masses using the methods described
earlier in this chapter. In Fig.~\ref{fig:mudisteasu} we show the
resulting muon distributions for hard annihilation spectra, a muon
energy threshold of $E_{\mu}^{\rm th}=10$ GeV and a detector
(projected) angular resolution of 1.4$^\circ$. If we compare
Fig.~\ref{fig:mudisteasu}a with Fig.~\ref{fig:nudistearth} we clearly
see the widening of the distributions that the charged current and the
multiple Coulomb scatterings give rise to. For neutralino annihilation
in the Sun, the neutrino distributions are very narrow peaks at
$\theta_{\nu}=0^\circ$ and hence the distributions in
Fig.~\ref{fig:mudisteasu} reflects the smearing due to the charged
current and multiple Coulomb scatterings only.

Using these kind of muon distributions we can now investigate if it 
is possible to extract the neutralino mass from the width of the muon 
angular distributions. 
In Fig.~\ref{fig:mxvsfwhmearth} we show the neutralino mass versus the
full width half maximum of the angular distribution for neutralino
annihilation in the Earth. These curves are evaluated for a neutrino
telescope with perfect angular resolution. When the resolution is more
typical, 1--2$^\circ$, the curves bend upwards at the lower-angle end
thus reducing the upper limit on the mass for which the mass can be
inferred. We find that for neutralino masses $\lsim 400$ GeV, the
Earth angular distribution can be used to infer the neutralino
mass. If the threshold is low, $\lsim 5$ GeV, the Sun angular
distribution can also be used.

If one would observe a signal from both the Earth and Sun, one could 
imagine subtracting the Sun angular distribution from the Earth one 
and thus achieve the desired model-independent neutrino angular 
distribution.  This is however not possible to do in a completely 
model-independent way since the neutrino interactions in the Sun depend 
on the neutrino spectrum and hence on model parameters.  It is probably 
possible, however, to reduce the model dependence with this technique.

\chapter{Conclusions}
\label{Concl}

We have performed a detailed evaluation of the relic density of
neutralinos (in the framework of the MSSM) including all two-body
final states at tree level and coannihilations between all neutralinos
and charginos lighter than $2.1m_{\chi}$.  We have found that the
neutralino density is cosmologically interesting ($0.025 < \Omega_\chi
h^2 < 1$) for a wide range of neutralino masses and compositions.

The coannihilation processes we have included are
important not only for light higgsinos, but whenever $|\mu/M_1|\lsim
2$, which includes all higgsino-like neutralinos as well as some mixed
and gaugino-like neutralinos. In these cases coannihilations can
reduce the relic density by typically a factor of 2--5, but sometimes
even with up to a factor of 100. We have also found that
coannihilations can increase the relic density up to a factor of 3.

There has been claims in the literature
\cite{MizutaYamaguchi,DreesNojiri} that light higgsino-like
neutralinos are never cosmologically interesting, but we found, in
agreement with Ref.~\cite{NeuLoop1}, that neutralinos with masses
$m_{\chi} \sim 75$ GeV and $\tan \beta \lsim 2$ can be cosmologically
interesting.

We have also shown that higgsino-like neutralinos with masses $m_\chi
\gsim 450$ GeV can have $\Omega_{\chi} h^2 >0.025$ and that the upper
limit on the neutralino mass for the neutralinos not to overclose the
Universe increases from 3 to about 7 TeV when coannihilations are
included.  It may be noted, though, that ${\cal O} (\mbox{TeV})$
massive neutralinos may appear unnatural in the sense that they
require fine-tuning of the parameters.

To evaluate the relic density correctly will be even more important in
the near future when the cosmological parameters can be expected to be
measured quite accurately \cite{CMBdettheory,CMBdetexp} and we might want
to draw some conclusion on the MSSM parameters from such measurements.

We have also evaluated the indirect detection rates in neutrino
telescopes coming from neutralino annihilations in the Sun and Earth.
The detection rates in neutrino telescopes are found to, for many
models, be explorable by the next generation neutrino telescopes, with
sizes of $\mathcal{O}(\mbox{1 km$^2$})$ especially for neutralino
annihilation in the Sun. We have also shown that there is a nice
complementarity between which models neutrino telescopes can probe and
which can be probed at e.g.\ LEP2\@.  There are however models giving
rise to very small detection rates in neutrino telescopes and that are
not explorable by LEP2\@.

We have also investigated what muon fluxes a given neutrino telescope
can probe and we have found that neutrino telescopes with an exposure
of about 1--25 km$^2$ yr can probe signal fluxes down to about 50--100
km$^{-2}$ yr$^{-1}$ if the full angular distribution of the signal is
used. If supersymmetry is found in some other experiment and we know
the relevant MSSM parameters, neutrino telescopes of this size will be
able to search for such a specific signal down to about 10--50
km$^{-2}$ yr$^{-1}$.

We have also shown that if a signal is seen, the width of the angular
distribution can be used to infer the neutralino mass.  If a signal is
seen from the Earth, neutralino masses $m_\chi \lsim 400$ GeV can be
determined. For a signal from the Sun, the mass can be inferred only
if the muon energy threshold is small ($\lsim$ 5 GeV).

We can thus at this stage conclude that very exciting times will soon
come when the new bigger neutrino telescopes start operating.

\appendix
\chapter{Feynman Rules for the MSSM}
\label{FeynApp}




\section{Introduction}

This is a collection of Feynman rules based on the rules given in
Ref.~\cite{HaberKane,GuHa86,Mandl}. They are slightly rewritten in a
format more suitable for numerical implementation and we have used the
conventions given in Chapter~\ref{MSSMdef}, except for the convention
on the matrix that diagonalize the neutralino mass matrix,
Eq.~(\ref{eq:neumass}).

In our actual calculations we have used the complex matrix $N_{ij}$
with positive mass eigenvalues as explained in Chapter~\ref{MSSMdef}.
Another convention is to instead have a real matrix $Z_{ij}$ in which
case the mass eigenvalues can be either positive or negative.  In this
convention the neutralinos are instead of Eq.~(\ref{eq:neulinear})
given by
\begin{equation}
  \tilde{\chi}_i^0 = Z_{i1} \tilde{B}^0 + Z_{i2} \tilde{W}^3
    + Z_{i3} \tilde{H}^0_1 + Z_{i4} \tilde{H}^0_2.
\end{equation}
The relation between the $Z$-matrix and the $N$-matrix is \cite{GuHa86}
\begin{equation}
  N_{ij} = \sqrt{\varepsilon_i} Z_{ij}
\end{equation}
where $\varepsilon_i$ is the sign of the $i$:th mass eigenvalue (i.e.\
the mass eigenvalues are $\varepsilon_i m_i$ with $m_i>0$).

The Feynman rules below are given in a form where it is very easy to
choose one convention or the other. If the convention with a real
matrix $Z_{ij}$ is desired, one only has to drop the complex
conjugation of $Z_{ij}$ whenever that appears and remember that
$\varepsilon_i$ denotes the sign of the $i$:th mass eigenvalue
with $m_i$ being the absolute value of the mass eigenvalue.  If one
instead wants to use the convention with a complex matrix $N_{ij}$ and
positive mass eigenvalues (which is the one we have used), one just has
to replace $Z_{ij}$ by $N_{ij}$ and put all $\varepsilon_i=1$.

Compared to Ref.~\cite{HaberKane,GuHa86,Mandl} all vertices are divided
by $i$ and the propagators are multiplied by $i$.

In these rules, the elementary charge $e>0$ and hence the electron
charge is $-e$. Quark charges are given as fractions of $e$ and are
denoted by $e_q$, i.e.\ $e_u = +2/3$ and $e_d=-1/3$.

Where relevant, the vertices are given as $g_{ijk}^L P_L + g_{ijk}^R
P_R$ where
\begin{eqnarray}
  P_L & = & \frac{1}{2} (1-\gamma_5) \\
  P_R & = & \frac{1}{2} (1+\gamma_5)
\end{eqnarray}
are the left- and righthanded projection operators.  The indices on
the $g$'s indicate the particles involved, for which we have used the
following shorthand notation,
\begin{itemize}
  \item[$\tilde{\chi}$] any neutralino or chargino.
  \item[$q$] any quark (we only give the rules for one generation, but 
  following Chapter~\ref{MSSMdef} it is straightforward to 
  extend this to three generations),
  \item[$\tilde{q}$] any squark,
  \item[$H$] any Higgs boson,
  \item[$V$] any gauge boson.
\end{itemize}
Note that the Feynman rules for leptons and sleptons are not given 
explicitly. They are however easily obtained from the quark and 
squark Feynman rules by the following substitutions
\begin{equation}
  \left\{ \begin{array}{lcl}
    u  & \rightarrow & \nu_l \\
    d  & \rightarrow & l^- \\
    \tilde{u}_L & \rightarrow & \tilde{\nu}_L \\
    \tilde{d}_{L/R} & \rightarrow & \tilde{l}_{L/R}
  \end{array} \right. 
\end{equation}
Note that $\tilde{\nu}_R$ does not exist.

We want to be able to evaluate different diagrams with general vertex
couplings and only in the numerical code insert the values of the
vertices for the actual particles involved. It is then convenient to
have a convention for the order of the indices on the vertices. We
have chosen to work with the following convention \cite{paoloprivate}:
\begin{itemize}
  \item The first fermion has a bar and is hence outgoing.  
  
  \item If two Higgses are present, the order of the indices is the 
  order in which they appear in the Lagrangian with the first one 
  complex conjugated.  This means for example that for the $V H_i 
  H_j$ vertex, $g_{VH_iH_j}$ is defined as
  \begin{equation}
    \mathcal{L} = g_{VH_i H_j} V_\mu H_i^\dagger i
    \lrpartialmu H_j
  \end{equation}
  which gives rise to the following Feynman rules
  \vertexmom{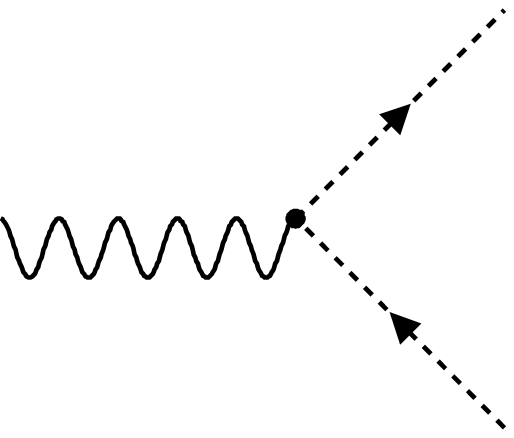}{$V$}{$H_j$}{$H_i$} {$g_{VH_iH_j} (p_i +
    p_j)^\mu$}{$p_j$}{$p_i$} 
  or
  \vertexmom{fig/gss.eps}{$V$}{$H_i$}{$H_j$} {$-g_{VH_iH_j} (p_i +
    p_j)^\mu$}{$p_i$}{$p_j$} 
  For the charged Higgses, the charge depends on the direction of the
  momentum of the Higgs.  If $H_i$ or $H_j$ is a charged Higgs it is
  an $H^+$ in the first rule above and an $H^-$ in the second one.
  For the rules given in the following sections the first Higgs is
  outgoing and the second one is ingoing and the charges are assigned
  accordingly.
  
  \item If one Higgs boson is present it is complex conjugated, i.e.~if it 
  is charged it is an $H^+$ moving out or an $H^-$ moving in.  
  
  \item If three Higgs bosons are present of which two are charged, 
  the first charged one is complex conjugated.
\end{itemize}

In the following sections, the Feynman rules for all three-particle
vertices are given and in Section~\ref{sec:Props} the propagators are
given.

\newlength{\temp}
\setlength{\temp}{\parindent}
\setlength{\parindent}{0cm}

\section{$H$-$\neu$-$\neu$ vertices}

\subsection{$H^0 \neu_j^0 \neu_i^0$}

First let us define
\begin{eqnarray}
  g \tilde{Q}_{ij}'' & = & \frac{1}{2} \left[ 
    Z_{i3} (g Z_{j2} - g' Z_{j1}) + 
    Z_{j3} (g Z_{i2} - g' Z_{i1}) \right] \\
  g \tilde{S}_{ij}'' & = & \frac{1}{2} \left[ 
    Z_{i4} (g Z_{j2} - g' Z_{j1}) + 
    Z_{j4}(g Z_{i2} - g' Z_{i1}) \right]
\end{eqnarray}

\subsubsection{$H_1^0 \neu_i^0 \neu_j^0$}

The vertex is \cite[Fig.\ 21]{GuHa86}
\vertex{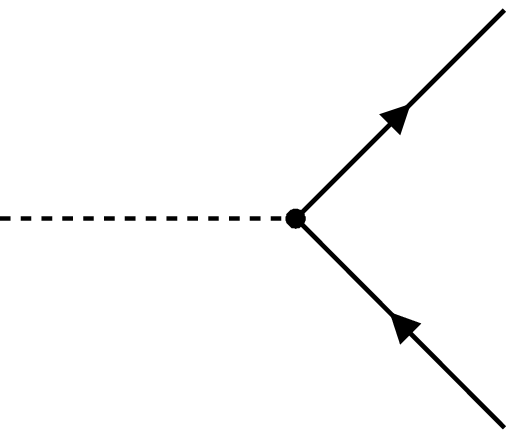}{$H_1^0$}{$\neu_j^0$}{$\neu_i^0$}
  {$g^L_{H_1ij} P_L + g^R_{H_1ij} P_R$}
where
\begin{eqnarray}
  g^L_{H_1ij} & = & g_{H_1ij}^* \varepsilon_j \\
  g^R_{H_1ij} & = & g_{H_1ij} \varepsilon_i \\
  g_{H_1ij} & = & g
    \left(- \tilde{Q}_{ij}'' \cos \alpha + 
    \tilde{S}_{ij}'' \sin \alpha \right)
\end{eqnarray}

\subsubsection{$H_2^0 \neu_i^0 \neu_j^0$}

The vertex is \cite[Fig.\ 21]{GuHa86}
\vertex{fig/sff.eps}{$H_2^0$}{$\neu_j^0$}{$\neu_i^0$}
  {$g^L_{H_2ij} P_L + g^R_{H_2ij} P_R$}
where
\begin{eqnarray}
  g^L_{H_2ij} & = & g_{H_2ij}^* \varepsilon_j \\
  g^R_{H_2ij} & = & g_{H_2ij} \varepsilon_i \\
  g_{H_2ij} & = & g
    \left(\tilde{Q}_{ij}'' \sin \alpha + 
    \tilde{S}_{ij}'' \cos \alpha \right)
\end{eqnarray}

\subsubsection{$H_3^0 \neu_i^0 \neu_j^0$}

The vertex is \cite[Fig.\ 21]{GuHa86}
\vertex{fig/sff.eps}{$H_3^0$}{$\neu_j^0$}{$\neu_i^0$}
  {$g^L_{H_3ij} P_L + g^R_{H_3ij} P_R$}
where
\begin{eqnarray}
  g^L_{H_3ij} & = & i g_{H_3ij}^* \varepsilon_j \\
  g^R_{H_3ij} & = & - i g_{H_3ij} \varepsilon_i \\
  g_{H_3ij} & = & g
    \left(\tilde{Q}_{ij}'' \sin \beta -
    \tilde{S}_{ij}'' \cos \beta \right)
\end{eqnarray}

\subsection{$H^0 \neu_i^+ \neu_j^+$}

First let us define
\begin{eqnarray}
  \tilde{Q}_{ij} & = & \sqrt{\frac{1}{2}} U_{i2} V_{j1} \\
  \tilde{S}_{ij} & = & \sqrt{\frac{1}{2}} U_{i1} V_{j2}
\end{eqnarray}

\subsubsection{$H_1^0 \neu_i^+ \neu_j^+$}

The vertex is \cite[Fig.\ 19]{GuHa86}
\vertex{fig/sff.eps}{$H_1^0$}{$\neu_j^+$}{$\neu_i^+$}
  {$g^L_{H_1ij} P_L + g^R_{H_1ij} P_R$}
where
\begin{eqnarray}
  g^L_{H_1ij} & = & - g
    \left( \tilde{Q}_{ij}^* \cos \alpha +
    \tilde{S}_{ij}^* \sin \alpha \right) \\
  g^R_{H_1ij} & = & - g
    \left( \tilde{Q}_{ji} \cos \alpha +
    \tilde{S}_{ji} \sin \alpha \right)
\end{eqnarray}

\subsubsection{$H_2^0 \neu_i^+ \neu_j^+$}

The vertex is \cite[Fig.\ 19]{GuHa86}
\vertex{fig/sff.eps}{$H_2^0$}{$\neu_j^+$}{$\neu_i^+$}
  {$g^L_{H_2ij} P_L + g^R_{H_2ij} P_R$}
where
\begin{eqnarray}
  g^L_{H_2ij} & = & g
    \left( \tilde{Q}_{ij}^* \sin \alpha -
    \tilde{S}_{ij}^* \cos \alpha \right) \\
  g^R_{H_2ij} & = & g
    \left( \tilde{Q}_{ji} \sin \alpha -
    \tilde{S}_{ji} \cos \alpha \right)
\end{eqnarray}

\subsubsection{$H_3^0 \neu_i^+ \neu_j^+$}

The vertex is \cite[Fig.\ 19]{GuHa86}
\vertex{fig/sff.eps}{$H_3^0$}{$\neu_j^+$}{$\neu_i^+$}
  {$g^L_{H_3ij} P_L + g^R_{H_3ij} P_R$}
where
\begin{eqnarray}
  g^L_{H_3ij} & = & i g
    \left( \tilde{Q}_{ij}^* \sin \beta +
    \tilde{S}_{ij}^* \cos \beta \right) \\
  g^R_{H_3ij} & = & - i g
    \left( \tilde{Q}_{ji} \sin \beta +
    \tilde{S}_{ji} \cos \beta \right)
\end{eqnarray}

\subsection{$H^- \neu_i^+ \neu_j^0$}
\label{sec:clashingarrows}

The vertex is \cite[Fig.\ 20]{GuHa86}
\vertex{fig/sff.eps}{$H^-$}{$\neu_j^+$}{$\neu_i^0$}
  {$g^L_{H^-ij} P_L + g^R_{H^-ij} P_R$}
where
\begin{eqnarray}
  g^L_{H^-ij} & = & - g \cos \beta \left[
    Z_{i4}^* V_{j1}^* + \sqrt{\frac{1}{2}} 
    (Z_{i2}^* + Z_{i1}^* \tan \theta_W) V_{j2}^*
    \right] \\
  g^R_{H^-ij} & = & - g \sin \beta \left[
    Z_{i3} U_{j1} - \sqrt{\frac{1}{2}} 
    (Z_{i2} + Z_{i1} \tan \theta_W) U_{j2}
    \right] \varepsilon_i
\end{eqnarray}

Given the vertex above one can derive \cite{archange} the following
Feynman rules with changed direction of the arrows
\vertexind{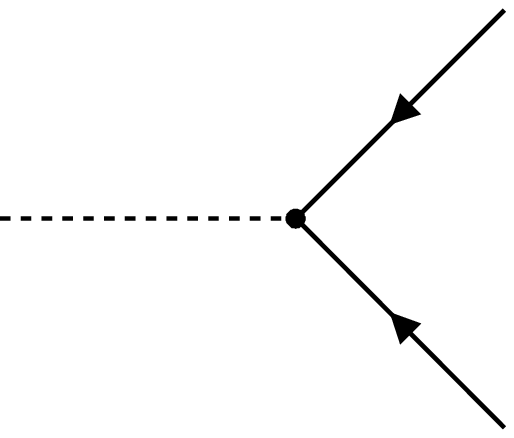}{$H^-$}{$\neu_j^+$}{$\neu_i^0$}
{$\left[-C^{-1}\left(g^L_{H^-ij} P_L + g^R_{H^-ij}
      P_R\right)\right]_{\beta \alpha}$} {$\alpha$}{$\beta$}
\vertex{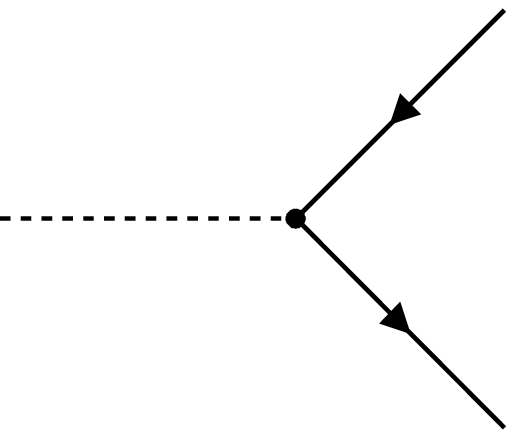}{$H^-$}{$\neu_j^+$}{$\neu_i^0$}
{$g^{R*}_{H^-ij} P_L + g^{L*}_{H^-ij} P_R$}
\vertexind{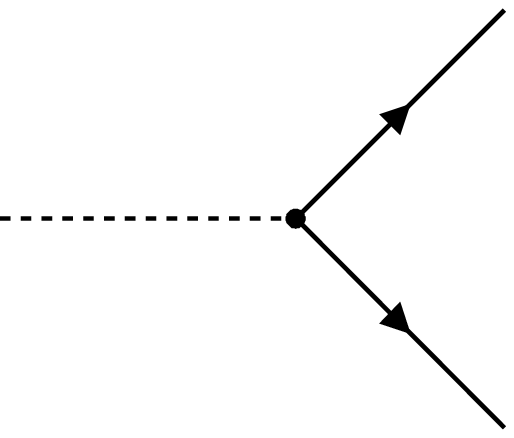}{$H^-$}{$\neu_j^+$}{$\neu_i^0$}
{$\left[\left(g^{R*}_{H^-ij} P_L + g^{L*}_{H^-ij}
      P_R\right)C\right]_{\alpha \beta}$} {$\alpha$}{$\beta$} 
where $C$ is the charge conjugation matrix and $\alpha$ and $\beta$
are spinor indices explicitly given for the cases of clashing arrows.
When working with clashing arrows one should either use the spinor
indices explicitly or note that if one writes down the Feynman
amplitude according to the direction of the Dirac fermion (the
chargino), the vertices with clashing arrows will appear as given
above and if the direction is instead chosen as the one given by the
neutralino the vertices will appear transposed.


\section{$V$-$\neu$-$\neu$ vertices}

\subsection{$W^+ \neu_i^0 \neu_j^+$}

First let us define
\begin{eqnarray}
  \tilde{O}_{ij}^L & = & - \frac{1}{\sqrt{2}}
    Z_{i4} V_{j2}^* + Z_{i2} V_{j1}^* \\
  \tilde{O}_{ij}^R & = & \frac{1}{\sqrt{2}}
    Z_{i3}^* U_{j2} + Z_{i2}^* U_{j1}
\end{eqnarray}

The vertex is \cite[Fig.\ 75]{HaberKane}
\vertex{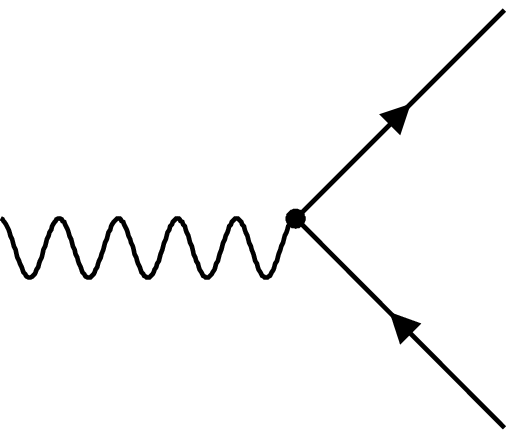}{$W^+$}{$\neu_j^+$}{$\neu_i^0$}
  {$\gamma^{\mu} \left( g^L_{Wij} P_L + g^R_{Wij} P_R \right)$}
where
\begin{eqnarray}
  g^L_{Wij} & = & g \tilde{O}_{ij}^L 
    \varepsilon_i \\
  g^R_{Wij} & = & g \tilde{O}_{ij}^R
\end{eqnarray}

Given the vertex above one can \cite{archange} derive the following
Feynman rules with changed direction of the arrows
\vertexind{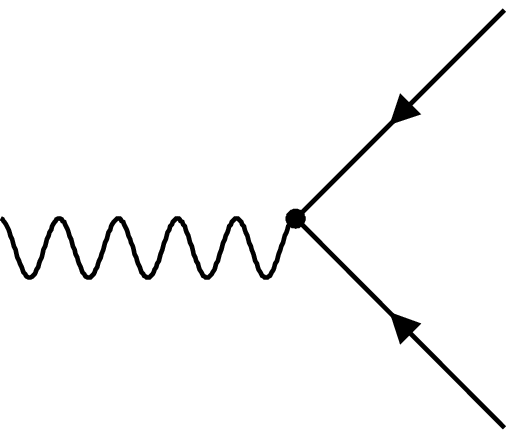}{$W^+$}{$\neu_j^+$}{$\neu_i^0$}
  {$\left[-C^{-1}\gamma^{\mu} \left( g^L_{Wij} P_L + g^R_{Wij} P_R 
   \right)\right]_{\beta \alpha}$}{$\alpha$}{$\beta$}
\vertex{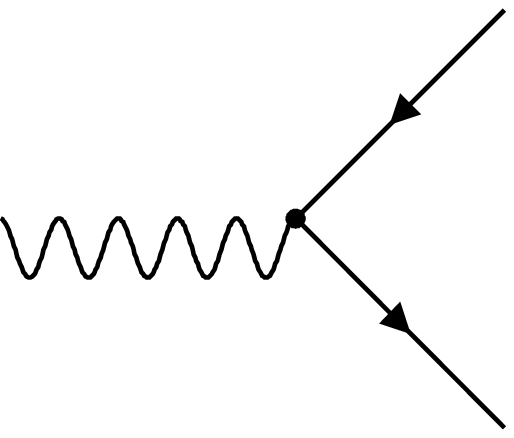}{$W^+$}{$\neu_j^+$}{$\neu_i^0$}
  {$\gamma^{\mu} \left( g^{L*}_{Wij} P_L + g^{R*}_{Wij} P_R 
  \right)$}
\vertexind{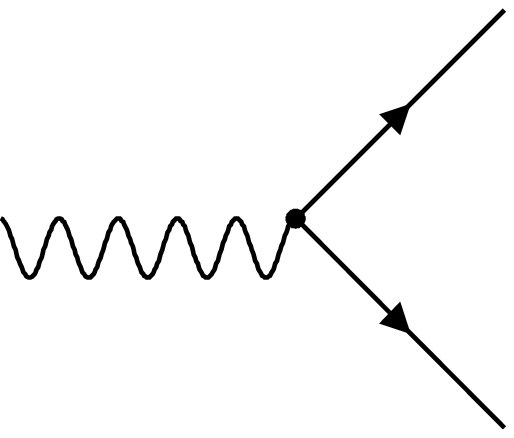}{$W^+$}{$\neu_j^+$}{$\neu_i^0$}
  {$\left[\gamma^{\mu} \left( g^{L*}_{Wij} P_L + g^{R*}_{Wij} P_R 
  \right)C\right]_{\alpha \beta}$}{$\alpha$}{$\beta$}
where the note at the end of Section~\ref{sec:clashingarrows} is
applicable here as well.

\subsection{$Z^0 \neu_-^+ \neu_j^+$}

First let us define
\begin{eqnarray}
  {\tilde{O'}_{ij}}^L & = & -V_{i1} V_{j1}^* - 
    \frac{1}{2} V_{i2} V_{j2}^* + 
    \delta_{ij} \sin^2 \theta_W \\
  {\tilde{O'}_{ij}}^R & = & - U_{i1}^* U_{j1} -
    \frac{1}{2} U_{i2}^* U_{j2} + 
    \delta_{ij} \sin^2 \theta_W
\end{eqnarray}

The vertex is \cite[Fig.\ 75]{HaberKane}
\vertex{fig/gff.eps}{$Z^0$}{$\neu_j^+$}{$\neu_i^+$}
  {$\gamma^{\mu} \left( g^L_{Zij} P_L + g^R_{Zij} P_R \right)$}
where
\begin{eqnarray}
  g^L_{Zij} & = & \frac{g}{\cos \theta_W}
    {\tilde{O'}_{ij}}^L \\
  g^R_{Zij} & = & \frac{g}{\cos \theta_W}
    {\tilde{O'}_{ij}}^R
\end{eqnarray}

\subsection{$Z^0 \neu_i^0 \neu_j^0$}

The vertex is \cite[Fig.\ 75]{HaberKane}
\vertex{fig/gff.eps}{$Z^0$}{$\neu_j^0$}{$\neu_i^0$}
  {$\gamma^{\mu} \left( g^L_{Zij} P_L + g^R_{Zij} P_R \right)$}
where
\begin{eqnarray}
  g^L_{Zij} & = & g_{Zij} \varepsilon_i \varepsilon_j \\
  g^R_{Zij} & = & - g_{Zij}^* \\
  g_{Zij} & = & \frac{g}{2 \cos \theta_W} \left(
    - Z_{i3} Z_{j3}^* + Z_{i4} Z_{j4}^* \right)
\end{eqnarray}

\subsection{$\gamma \neu_i^+ \neu_i^+$}

The vertex is \cite[Fig.\ 75]{HaberKane}
\vertex{fig/gff.eps}{$\gamma$}{$\neu_i^+$}{$\neu_i^+$}
  {$\gamma^{\mu} \left( g^L_{\gamma ii} P_L + g^R_{\gamma ii} P_R \right)$}
where
\begin{eqnarray}
  g^L_{\gamma ii} & = & - e \\
  g^R_{\gamma ii} & = & - e
\end{eqnarray}


\section{$\neu$-$q$-$\tilde{q}$ vertices}

Note that if we have the vertex
\vertex{fig/sff.eps}{$\tilde{f}'$}{$\neu_j^+ / \neu_j^0$}{$f$}
  {$g^L_{\tilde{f}' f j} P_L + g^R_{\tilde{f}' f j} P_R$}
then it follows \cite{archange} that the vertex with reversed arrows
is given by
\vertex{fig/sffoutin.eps}{$\tilde{f}'$}{$\neu_j^+ / \neu_j^0$}{$f$}
  {$g^{R*}_{\tilde{f}' f j} P_L + g^{L*}_{\tilde{f}' f j} P_R$}

\subsection{$\neu^+ \tilde{q} q$}

\subsubsection{$\neu_j^+ \tilde{d}_L u$}

The vertex is \cite[Fig.\ 22]{GuHa86}
\vertex{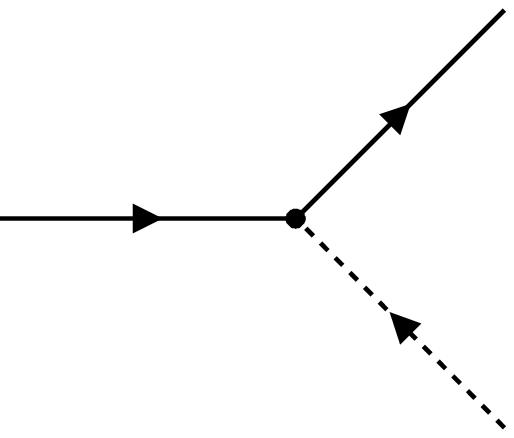}{$\neu_j^+$}{$\tilde{d}_L$}{$u$}
  {$g^L_{\tilde{d}_L u j} P_L + g^R_{\tilde{d}_L u j} P_R$}
where
\begin{eqnarray}
  g^L_{\tilde{d}_Luj} & = & \frac{g m_u V_{j2}^*}
    {\sqrt{2} m_W \sin \beta} \\
  g^R_{\tilde{d}_Luj} & = & - g U_{j1}
\end{eqnarray}

\subsubsection{$\neu_j^+ \tilde{d}_R u$}

The vertex is \cite[Fig.\ 22]{GuHa86}
\vertex{fig/fsfin.eps}{$\neu_j^+$}{$\tilde{d}_R$}{$u$}
  {$g^L_{\tilde{d}_Ruj} P_L + g^R_{\tilde{d}_Ruj} P_R$}
where
\begin{eqnarray}
  g^L_{\tilde{d}_Ruj} & = & 0 \\
  g^R_{\tilde{d}_Ruj} & = & \frac{g m_d U_{j2}}
    {\sqrt{2} m_W \cos \beta}
\end{eqnarray}

\subsubsection{$\neu_j^+ \tilde{u}_L d$}

The vertex is \cite[Fig.\ 22]{GuHa86}
\vertex{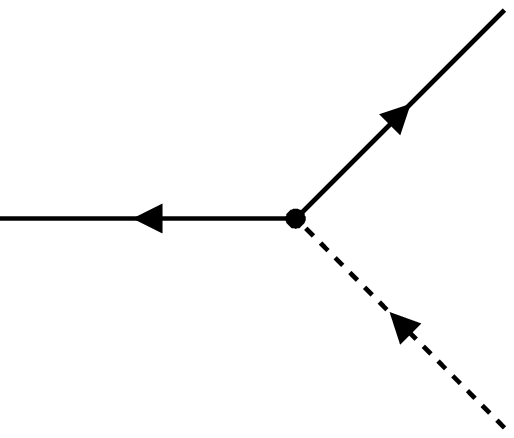}{$\neu_j^+$}{$\tilde{u}_L$}{$d$}
  {$[g^L_{\tilde{u}_Ldj} P_L + g^R_{\tilde{u}_Ldj} P_R] C$}
where
\begin{eqnarray}
  g^L_{\tilde{u}_Ldj} & = & \frac{g m_d U_{j2}^*}
    {\sqrt{2} m_W \cos \beta} \\
  g^R_{\tilde{u}_Ldj} & = & - g V_{j1}
\end{eqnarray}

\subsubsection{$\neu_j^+ \tilde{u}_R d$}

The vertex is \cite[Fig.\ 22]{GuHa86}
\vertex{fig/fsfout.eps}{$\neu_j^+$}{$\tilde{u}_R$}{$d$}
  {$[g^L_{\tilde{u}_Rdj} P_L + g^R_{\tilde{u}_Rdj} P_R] C$}
where
\begin{eqnarray}
  g^L_{\tilde{u}_Rdj} & = & 0 \\
  g^R_{\tilde{u}_Rdj} & = & \frac{g m_u V_{j2}}
    {\sqrt{2} m_W \sin \beta}
\end{eqnarray}

%

%

\subsubsection{$\neu_j^+ d \tilde{u}_L$}

The vertex is \cite[Fig.\ 23]{GuHa86}
\vertex{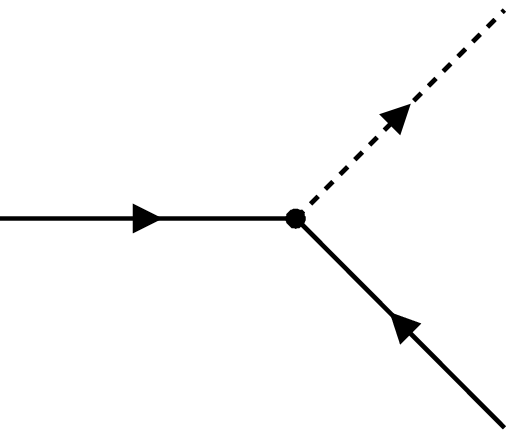}{$\neu_j^+$}{$d$}{$\tilde{u}_L$}
  {$C^{-1}[g^L_{jd\tilde{u}_L} P_L + g^R_{jd\tilde{u}_L} P_R]$}
where
\begin{eqnarray}
  g^L_{jd\tilde{u}_L} & = & g V_{j1}^* \\
  g^R_{jd\tilde{u}_L} & = & - \frac{g m_d U_{j2}}
    {\sqrt{2} m_W \cos \beta}
\end{eqnarray}

\subsubsection{$\neu_j^+ d \tilde{u}_R$}

The vertex is \cite[Fig.\ 23]{GuHa86}
\vertex{fig/ffsin.eps}{$\neu_j^+$}{$d$}{$\tilde{u}_R$}
  {$C^{-1}[g^L_{jd\tilde{u}_R} P_L + g^R_{jd\tilde{u}_R} P_R]$}
where
\begin{eqnarray}
  g^L_{jd\tilde{u}_R} & = & - \frac{g m_u V_{j2}^*}
    {\sqrt{2} m_W \sin \beta} \\
  g^R_{jd\tilde{u}_R} & = & 0
\end{eqnarray}

\subsection{$\neu^0 \tilde{q} q$}

First let us define
\begin{eqnarray}
  Z'_{j1} & = & Z_{j1} \cos \theta_W + Z_{j2} \sin \theta_W \\
  Z'_{j2} & = & -Z_{j1} \sin \theta_W + Z_{j2} \cos \theta_W \\
  Z'_{j3} & = & Z_{j3} \\
  Z'_{j4} & = & Z_{j4}
\end{eqnarray} 


\subsubsection{$\neu_j^0 \tilde{u}_L u$}

The vertex is \cite[Fig.\ 24]{GuHa86}
\vertexrev{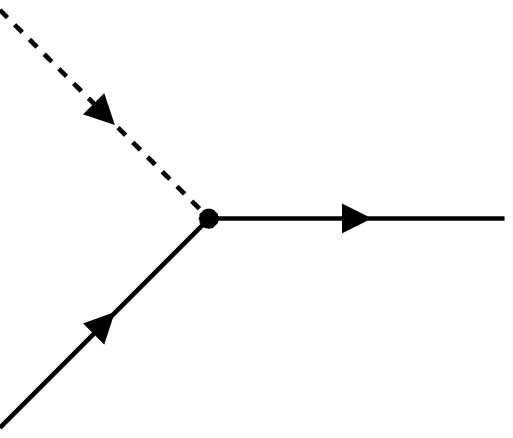}{$\neu_j^0$}{$\tilde{u}_L$}{$u$}
  {$g^L_{\tilde{u}_Luj} P_L + g^R_{\tilde{u}_Luj} P_R$}
where
\begin{eqnarray}
  g^L_{\tilde{u}_Luj} & = & - \frac{g m_u Z_{j4}^*}
    {\sqrt{2} m_W \sin \beta} \varepsilon_j \\
  g^R_{\tilde{u}_Luj} & = & -\sqrt{2} e e_u Z'_{j1}-
    \frac{g \sqrt{2}}{\cos \theta_W} 
    \left(\frac{1}{2}-e_u \sin^2 \theta_W \right)
    Z'_{j2} \nonumber \\
  & = & - \frac{g}{\sqrt{2} \cos \theta_W} 
    \left[(2 e_u-1) \sin \theta_W Z_{j1} + \cos \theta_W Z_{j2}\right] 
\end{eqnarray}

\subsubsection{$\neu_j^0 \tilde{u}_R u$}

The vertex is \cite[Fig.\ 24]{GuHa86}
\vertexrev{fig/fsfrev.eps}{$\neu_j^0$}{$\tilde{u}_R$}{$u$}
  {$g^L_{\tilde{u}_Ruj} P_L + g^R_{\tilde{u}_Ruj} P_R$}
where
\begin{eqnarray}
  g^L_{\tilde{u}_Ruj} & = & \sqrt{2} \left[ 
    e e_u Z^{\prime *}_{j1} - \frac{g e_u \sin^2 \theta_W}
    {\cos \theta_W} Z^{\prime *}_{j2} \right] \varepsilon_j \\
     & = & \frac{\sqrt{2} e_u g \sin \theta_W}{\cos \theta_W} Z^*_{j1} 
    \varepsilon_j \\
  g^R_{\tilde{u}_Ruj} & = & - \frac{g m_u}
    {\sqrt{2} m_W \sin \beta} Z_{j4}
\end{eqnarray}

\subsubsection{$\neu_j^0 \tilde{d}_L d$}

The vertex is \cite[Fig.\ 24]{GuHa86}
\vertexrev{fig/fsfrev.eps}{$\neu_j^0$}{$\tilde{d}_L$}{$d$}
  {$g^L_{\tilde{d}_Ldj} P_L + g^R_{\tilde{d}_Ldj} P_R$}
where
\begin{eqnarray}
  g^L_{\tilde{d}_Ldj} & = & - \frac{g m_d}
    {\sqrt{2} m_W \cos \beta} Z_{j3}^* \varepsilon_j \\
  g^R_{\tilde{d}_Ldj} & = & -\sqrt{2} e e_d Z'_{j1}+ 
    \frac{g \sqrt{2}}{\cos \theta_W} \left(
    \frac{1}{2} + e_d \sin^2 \theta_W \right) Z'_{j2} \\
  & = & - \frac{g}{\sqrt{2} \cos \theta_W} \left[
   (2 e_d + 1) \sin \theta_W Z_{j1} - \cos \theta_W Z_{j2} \right] 
\end{eqnarray}

\subsubsection{$\neu_j^0 \tilde{d}_R d$}

The vertex is \cite[Fig.\ 24]{GuHa86}
\vertexrev{fig/fsfrev.eps}{$\neu_j^0$}{$\tilde{d}_R$}{$d$}
  {$g^L_{\tilde{d}_Rdj} P_L + g^R_{\tilde{d}_Rdj} P_R$}
where
\begin{eqnarray}
  g^L_{\tilde{d}_Rdj} & = & \sqrt{2} \left[
    e e_d Z^{\prime *}_{j1} - \frac{g e_d \sin^2 \theta_W}
    {\cos \theta_W} Z^{\prime *}_{j2} \right] \varepsilon_j \\
  & = & \frac{\sqrt{2} e_d g \sin \theta_W }{\cos \theta_W} 
    Z^*_{j1} \varepsilon_j \\
  g^R_{\tilde{d}_Rdj} & = & - \frac{g m_d}
    {\sqrt{2} m_W \cos \beta} Z_{j3}
\end{eqnarray}


\section{$H$-$q$-$q'$ vertices}

\subsection{$H^0 q \bar{q}$}

These vertices can be written \cite[Fig.\ 7 and 8]{GuHa86}
\vertex{fig/sff.eps}{$H^0$}{$\bar{q}$}{$q$}
  {$g_{Hqq}^L P_L + g_{Hqq}^R P_R$}
where
\begin{eqnarray}
  g_{H_1uu}^L & = & g_{H_1uu}^R = -\frac{g m_u \sin \alpha}
    {2 m_W \sin \beta} \\
  g_{H_1dd}^L & = & g_{H_1dd}^R = - \frac{g m_d \cos \alpha}
    {2 m_W \cos \beta} \\
  g_{H_2uu}^L & = & g_{H_2uu}^R = - \frac{g m_u \cos \alpha}
    {2 m_W \sin \beta} \\
  g_{H_2dd}^L & = & g_{H_2dd}^R = \frac{g m_d \sin \alpha}
    {2 m_W \cos \beta} \\
  g_{H_3uu}^L & = & - g_{H_3uu}^R = - \frac{i g m_u \cot \beta}
    {2 m_W} \\
  g_{H_3dd}^L & = & - g_{H_3dd}^R = - \frac{i g m_d \tan \beta}
    {2 m_W} \\
\end{eqnarray}

\subsection{$H^- du$}

This vertex can be written \cite[Fig.\ 8]{GuHa86}
\vertex{fig/sff.eps}{$H^-$}{$u$}{$d$}
  {$g_{H^-du}^L P_L + g_{H^-du}^R P_R$}
where
\begin{eqnarray}
  g_{H^-du}^L & = & \frac{g}{\sqrt{2} m_W} 
    m_d \tan \beta  \\
  g_{H^-du}^R & = & \frac{g}{\sqrt{2} m_W}  
    m_u \cot \beta 
\end{eqnarray}

\subsection{$H^+ u \bar{d}$}

From the vertex above it follows that \cite[Fig.\ 8]{GuHa86}
\vertex{fig/sff.eps}{$H^+$}{$d$}{$u$}
  {$g_{H^+ud}^L P_L + g_{H^+ud}^R P_R$}
where
\begin{eqnarray}
  g_{H^+ud}^L & = & \frac{g}{\sqrt{2} m_W} 
    m_u \cot \beta  \\
  g_{H^+ud}^R & = & \frac{g}{\sqrt{2} m_W}  
    m_d \tan \beta 
\end{eqnarray}


\section{$V$-$q$-$q'$ vertices}


These vertices can be written \cite[Fig.\ 71]{HaberKane}
\vertex{fig/gff.eps}{$V$}{$q$}{$q$}
  {$\gamma^{\mu} \left[ g_{Vqq}^L P_L + g_{Vqq}^R P_R \right]$}
where
\begin{eqnarray}
  g_{\gamma qq}^L & = & g_{\gamma qq}^R = - e e_q \\
  g_{Wdu}^L & = & - \frac{g}{\sqrt{2}} \\
  g_{Wdu}^R & = & 0 \\
  g_{Wud}^L & = & - \frac{g}{\sqrt{2}} \\
  g_{Wud}^R & = & 0 \\
  g_{Zuu}^L & = & - \frac{g}{2 \cos \theta_W} 
    (1-2e_u \sin^2 \theta_W) \\
  g_{Zuu}^R & = & \frac{g}{\cos \theta_W} 
    e_u \sin^2 \theta_W \\
  g_{Zdd}^L & = & \frac{g}{2 \cos \theta_W}
    (1+2e_d \sin^2 \theta_W) \\
  g_{Zdd}^R & = & \frac{g}{\cos \theta_W}
    e_d \sin^2 \theta_W
\end{eqnarray}


\section{$H$-scalar-scalar vertices}

\subsection{$HHH$}

These vertices can all be written \cite[Fig.\ 9 and 10]{GuHa86}
\vertex{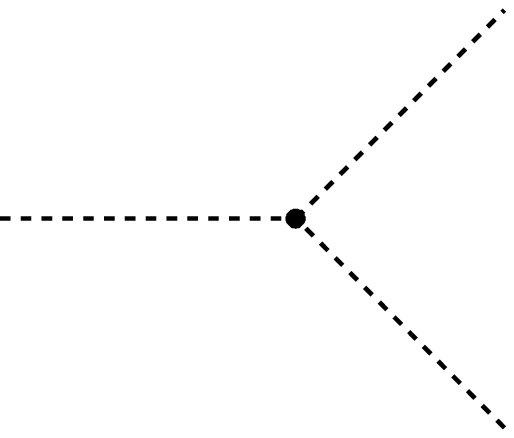}{$H$}{$H$}{$H$}
  {$m_W g_{HHH}$}
where 
\begin{eqnarray}
  g_{H_1H^+H^-} & = & - g \left[ \cos (\beta-\alpha) 
    - \frac{m_Z}{2 m_W \cos \theta_W} \cos 2 \beta
    \cos (\beta+\alpha) \right] \\
  g_{H_2H^+H^-} & = & - g \left[ \sin (\beta-\alpha) 
    + \frac{m_Z}{2 m_W \cos \theta_W} \cos 2 \beta
    \sin (\beta+\alpha) \right] \\
  g_{H_1H_1H_1} & = & - \frac{3 g m_Z}{2 m_W \cos \theta_W}
    \cos 2 \alpha \cos (\beta+\alpha) \\
  g_{H_2H_2H_2} & = & - \frac{3 g m_Z}{2 m_W \cos \theta_W}
    \cos 2 \alpha \sin (\beta+\alpha) \\
  g_{H_1H_2H_2} & = & - \frac{g m_Z}{2 m_W \cos \theta_W}
    \big[ 2 \sin 2 \alpha \sin (\beta+\alpha) \nonumber \\
    & & \mbox{}- \cos 2 \alpha \cos (\beta + \alpha) \big] \\
  g_{H_1H_3H_3} & = & \frac{g m_Z}{2 m_W \cos \theta_W}
    \cos 2 \beta \cos (\beta + \alpha) \\
  g_{H_2H_1H_1} & = & \frac{g m_Z}{2 m_W \cos \theta_W}
    \big[ 2 \sin 2 \alpha \cos (\beta+\alpha) \nonumber \\
    & & \mbox{}+ \cos 2 \alpha \sin (\beta + \alpha) \big] \\
  g_{H_2H_3H_3} & = & - \frac{g m_Z}{2 m_W \cos \theta_W}
    \cos 2 \beta \sin (\beta + \alpha)
\end{eqnarray}

\subsection{$H \tilde{q} \tilde{q}$}

These vertices can all be written \cite[Fig.\ 11--14]{GuHa86}
\vertex{fig/sss.eps}{$H$}{$\tilde{q}$}{$\tilde{q}$}
  {$m_W g_{H\tilde{q} \tilde{q}}$}
where
\begin{eqnarray}
  g_{H^+\tilde{d}_L\tilde{u}_L} & =&  - \frac{g}{\sqrt{2}} \left[
    \sin 2 \beta - \frac{m_d^2 \tan \beta + m_u^2 \cot \beta}
    {m_W^2} \right] \\
  g_{H^+\tilde{d}_R\tilde{u}_R} & = & \frac{g m_u m_d}{\sqrt{2} m_W^2} 
    (\cot \beta + \tan \beta) \\
  g_{H^+\tilde{d}_R\tilde{u}_L} & = & - \frac{g m_d}{\sqrt{2} m_W^2} 
    (\mu + A_d  \tan \beta) \\
  g_{H^+\tilde{d}_L\tilde{u}_R} & = & - \frac{g m_u}{\sqrt{2} m_W^2} 
    (\mu + A_u  \cot \beta) \\
  g_{H_1\tilde{u}_L\tilde{u}_L} & = & - \frac{g m_Z}{m_W \cos \theta_W}
    \left( \frac{1}{2}-e_u \sin^2 \theta_W \right)
    \cos (\alpha+\beta) \nonumber \\
    & & \mbox{} - \frac{g m_u^2}{m_W^2 \sin \beta} \sin \alpha \\
  g_{H_1\tilde{u}_R\tilde{u}_R} & = & - \frac{g m_Z}{m_W \cos \theta_W}
    e_u \sin^2 \theta_W \cos (\alpha+\beta) \nonumber \\
    & & \mbox{} - \frac{g m_u^2}{m_W^2 \sin \beta} \sin \alpha \\
  g_{H_1\tilde{u}_R\tilde{u}_L} & = & - \frac{g m_u}{2 m_W^2 \sin \beta}
    \left[ A_u  \sin \alpha - \mu \cos \alpha \right] \\
  g_{H_1\tilde{d}_L\tilde{d}_L} & = & \frac{g m_Z}{m_W \cos \theta_W}
    \left( \frac{1}{2} + e_d \sin^2 \theta_W \right)
    \cos (\alpha+\beta) \nonumber \\
    & & \mbox{} - \frac{g m_d^2}{m_W^2 \cos \beta} \cos \alpha \\
  g_{H_1\tilde{d}_R\tilde{d}_R} & = & - \frac{g m_Z}{m_W \cos \theta_W}
    e_d \sin^2 \theta_W \cos (\alpha+\beta) \nonumber \\
    & & \mbox{} - \frac{g m_d^2}{m_W^2 \cos \beta} \cos \alpha \\
  g_{H_1\tilde{d}_R\tilde{d}_L} & = & - \frac{g m_d}{2 m_W^2 \cos \beta}
    \left[ A_d  \cos \alpha - \mu \sin \alpha \right] \\
  g_{H_2\tilde{u}_L\tilde{u}_L} & = & \frac{g m_Z}{m_W \cos \theta_W}
    \left( \frac{1}{2}-e_u \sin^2 \theta_W \right)
    \sin (\alpha+\beta) \nonumber \\
    & & \mbox{} - \frac{g m_u^2}{m_W^2 \sin \beta} \cos \alpha \\
  g_{H_2\tilde{u}_R\tilde{u}_R} & = & \frac{g m_Z}{m_W \cos \theta_W}
    e_u \sin^2 \theta_W \sin (\alpha+\beta)
    - \frac{g m_u^2}{m_W^2 \sin \beta} \cos \alpha \\
  g_{H_2\tilde{u}_R\tilde{u}_L} & = & - \frac{g m_u}{2 m_W^2 \sin \beta}
    \left[ A_u  \cos \alpha + \mu \sin \alpha \right] \\
  g_{H_2\tilde{d}_L\tilde{d}_L} & = & - \frac{g m_Z}{m_W \cos \theta_W}
    \left( \frac{1}{2} + e_d \sin^2 \theta_W \right)
    \sin (\alpha+\beta) \nonumber \\
    & & \mbox{} + \frac{g m_d^2}{m_W^2 \cos \beta} \sin \alpha \\
  g_{H_2\tilde{d}_R\tilde{d}_R} & = & \frac{g m_Z}{m_W \cos \theta_W}
    e_d \sin^2 \theta_W \sin (\alpha+\beta)
    + \frac{g m_d^2}{m_W^2 \cos \beta} \sin \alpha \\
  g_{H_2\tilde{d}_R\tilde{d}_L} & = & \frac{g m_d}{2 m_W^2 \cos \beta}
    \left[ A_d  \sin \alpha + \mu \cos \alpha \right] \\
  g_{H_3\tilde{u}_L\tilde{u}_R} & = & - \frac{i g m_u}{2 m_W^2}
    ( A_u \cot \beta + \mu ) \\
  g_{H_3\tilde{d}_L\tilde{d}_R} & = & - \frac{i g m_d}{2 m_W^2}
    ( A_d \tan \beta + \mu ) \\
  g_{H_3\tilde{u}_R\tilde{u}_L} & = & \frac{i g m_u}{2 m_W^2}
    ( A_u \cot \beta + \mu ) \\
  g_{H_3\tilde{d}_R\tilde{d}_L} & = & \frac{i g m_d}{2 m_W^2}
    ( A_d \tan \beta + \mu ) \\
\end{eqnarray}


\section{$V$-scalar-scalar vertices}

\subsection{$VHH$}

These vertices can all be written \cite[Fig.\ 1 and 2]{GuHa86}
\vertexmom{fig/gss.eps}{$V$}{$H_{in}$}{$H_{out}$}
  {$g_{VH_{out}H_{in}} (p_{out}+p_{in})^{\mu}$}{$p_{in}$}{$p_{out}$}
where
\begin{eqnarray}
  g_{WH^+H_1} & = & - \frac{g}{2} \sin (\alpha-\beta) \\
  g_{WH_1H^+} & = & -\frac{g}{2} \sin (\alpha-\beta) \\
  g_{WH^+H_2} & = & - \frac{g}{2} \cos (\alpha-\beta) \\
  g_{WH_2H^+} & = & -\frac{g}{2} \cos (\alpha-\beta) \\
  g_{WH^+H_3} & = & - \frac{i g}{2} \\
  g_{WH_3H^+} & = & \frac{i g}{2} \\
  g_{ZH_3H_1} & = & - \frac{i g \sin (\alpha-\beta)}{2 \cos \theta_W} \\
  g_{ZH_3H_2} & = & - \frac{i g \cos (\alpha-\beta)}{2 \cos \theta_W} \\
  g_{ZH^+H^+} & = & - \frac{g \cos 2 \theta_W}{2 \cos \theta_W} \\
  g_{\gamma H^+H^+} & = & -e 
\end{eqnarray}

\subsection{$V \tilde{q} \tilde{q}$}

These vertices can all be written \cite[Fig.\ 72]{HaberKane}
\vertexmom{fig/gss.eps}{$V$}{$\tilde{q}$}{$\tilde{q'}$}
  {$g_{V\tilde{q}' \tilde{q}} (p_{out}+p_{in})^{\mu}$}{$p_{in}$}{$p_{out}$}
where
\begin{eqnarray}
  g_{\gamma \tilde{q}_L \tilde{q}_L} & = & - e e_q \\
  g_{\gamma \tilde{q}_R \tilde{q}_R} & = & - e e_q \\
  g_{W\tilde{d}_L\tilde{u}_L} & = & - \frac{g}{\sqrt{2}} \\
  g_{Z\tilde{u}_L\tilde{u}_L} & = & \frac{g}{2 \cos \theta_W}
    (-1 + 2e_u \sin^2 \theta_W) \\
  g_{Z\tilde{u}_R\tilde{u}_R} & = & \frac{g}{\cos \theta_W}
    e_u \sin^2 \theta_W \\
  g_{Z\tilde{d}_L\tilde{d}_L} & = & \frac{g}{2 \cos \theta_W}
    (1 + 2e_d \sin^2 \theta_W) \\
  g_{Z\tilde{d}_R\tilde{d}_R} & = & \frac{g}{\cos \theta_W}
    e_d \sin^2 \theta_W
\end{eqnarray}


\section{$H$-$V$-$V$ vertices}

These vertices can all be written \cite[Fig.\ 3]{GuHa86}
\vertex{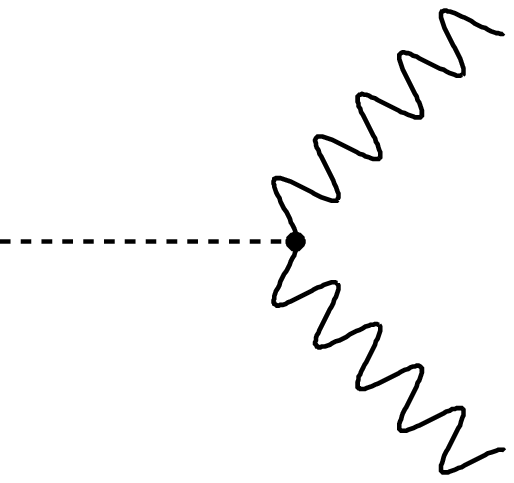}{$H$}{$V$}{$V$}{$m_W g_{HVV} g^{\mu \nu}$}
where
\begin{eqnarray}
  g_{H_1WW} & = & g \cos (\beta-\alpha) \\
  g_{H_2WW} & = & g \sin (\beta-\alpha) \\
  g_{H_1ZZ} & = & \frac{g m_Z}{m_W \cos \theta_W} \cos (\beta-\alpha) \\
  g_{H_2ZZ} & = & \frac{g m_Z}{m_W \cos \theta_W} \sin (\beta-\alpha) \\
\end{eqnarray}


\section{$V$-$V$-$V$ vertices}

These vertices ($Z^0 W^+ W^-$ and $\gamma W^+ W^-$) can all 
be written \cite{Mandl}
\vertexgauge{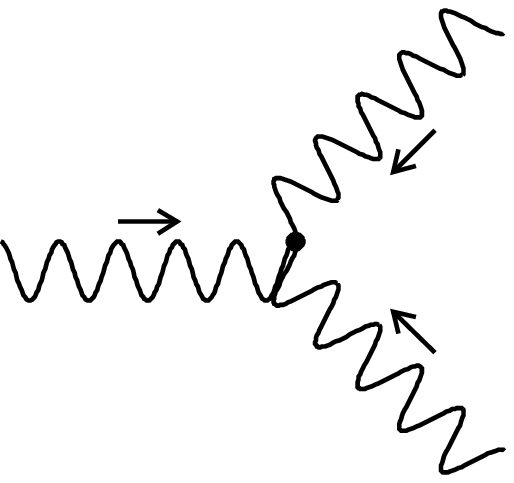}{$V$}{$V$}{$V$}
  {$g_{VVV} [g^{\alpha \beta} (k_1-k_2)^{\gamma} +
    g^{\beta \gamma} (k_2-k_3)^{\alpha}$}{$\mbox{}+g^{\gamma \alpha} (k_3-k_1)^{\beta} ]$}
where
\begin{eqnarray}
  g_{ZW^+W^-} & = & g \cos \theta_W \\
  g_{\gamma W^+W^-} & = & e
\end{eqnarray}


\section{Propagators}
\label{sec:Props}

First define the Mandelstam variables
\begin{eqnarray}
  s & = & (p_1+p_2)^2 = (p_3+p_4)^2 \\
  t & = & (p_1-p_3)^2 = (p_2-p_4)^2 \\
  u & = & (p_1-p_4)^2 = (p_2-p_3)^2.
\end{eqnarray}
We then have the following propagators \cite{Mandl} for internal particles
\begin{eqnarray}
  \mbox{Internal scalar: } & & \Delta_H (q) = 
  - \frac{1}{q^2-m_H^2+i \varepsilon} \\
  \mbox{Internal gauge boson: } & & D_{F \alpha \beta}(q) =
  -\Delta_V (q) \left( g_{\alpha \beta} - \frac{q_{\alpha} 
   q_{\beta}}{m_V^2} \right) \nonumber \\
  & & \mbox{} = 
  \frac{g_{\alpha \beta} - \frac{q_{\alpha} q_{\beta}}{m_V^2}}
  {q^2-m_V^2+i \varepsilon} \\
  \mbox{Internal fermion: } & & S_F(q) =
  - \frac{1}{\not{q}-m_f+i \varepsilon} \nonumber \\
  & & \mbox{} =  
  - \frac{\not{q}+m_f}{q^2-m_f^2+i \varepsilon}
\end{eqnarray}
where as usual $\varepsilon=\Gamma m$ with $\Gamma$ being the width of the
propagating particle and $m$ its mass.
The fermion propagator can in the $t$-channel be written
\begin{equation}
  S_F(q) = [(\not{p_1}-\not{p_3})+m_f] \Delta_t
\end{equation}
where
\begin{equation}
  \Delta_t = - \frac{1}{t-m_f^2+i \varepsilon}
\end{equation}
and in the $u$-channel be written
\begin{equation}
  S_F(q) = - [(\not{p_1}-\not{p_4})+m_f] \Delta_u = 
  [(\not{p_2}-\not{p_3})+m_f] \Delta_u
\end{equation}
where
\begin{equation}
  \Delta_u = - \frac{1}{u-m_f^2+i \varepsilon}
\end{equation}

The $u$-diagram can, when the incoming particles are Majorana fermions
(like the neutralinos), be written in the convenient form
\begin{equation}
  \begin{picture}(9,1.9)(0,0)
  \put(0.0,0.05){$\neu_i^0$}
  \put(0.0,1.7){$\neu_j^0$}
  \put(5.5,0.05){$\neu_i^0$}
  \put(5.5,1.7){$\neu_j^0$}
  \put(4.0,0.9){$=$}
  \put(4.7,0.9){$-$}
  \put(5.1,0.9){$\left( \raisebox{0.0ex}[6.0ex][6.0ex]
    {\makebox[4.0cm]{}} \right)$}
  \put(8.6,0.9){$i \leftrightarrow j$}
  \hspace{0.5cm}\epsfig{file=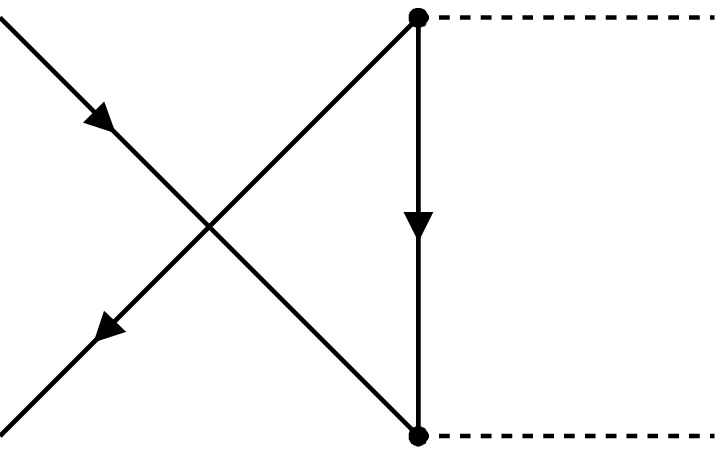, height=1.75cm}
  \hspace{2.7cm}\epsfig{file=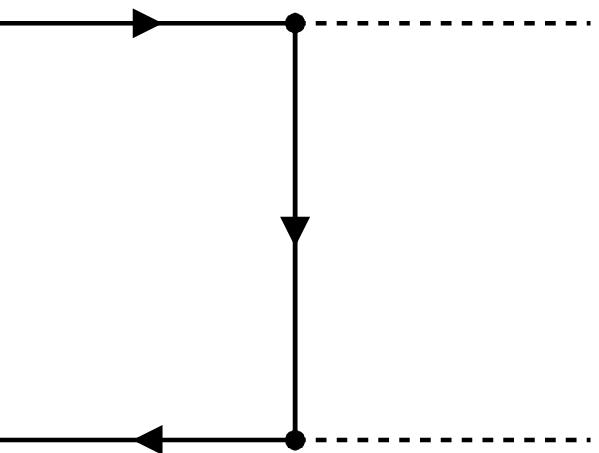, height=1.75cm}
  \end{picture}
\end{equation}




\setlength{\parindent}{\temp}

\cleardoublepage
\def\rightmark{Summary of the Papers}
\addcontentsline{toc}{chapter}{Summary of the Papers}
\def\leftmark{Summary of the Papers}
\vspace*{15mm}
\noindent
{ \Huge\bf Summary of the Papers}

\vspace{10mm}

\begin{list}{-}{\renewcommand{\makelabel}[1]{\makebox[\labelwidth][l]{#1}}}

\item[I.]
  
  The whole chain of processes from the neutralino annihilation
  products in the core of the Sun or Earth to a muon flux at a
  detector is calculated with Monte Carlo simulation techniques.

\item[II.]  

  If a positive signal of WIMP annihilations in the Sun or Earth is
  seen at a neutrino telescope, it is here shown that the mass of the
  WIMP can be inferred from the width of the angular distribution.

\item[III.] 

  The indirect detection rates at neutrino telescopes for neutralino
  annihilation in the Sun or Earth are evaluated and also compared
  with direct neutralino dark matter searches and searches at LEP2.

\item[IV.] 

  The minimal exposures needed to make a discovery of WIMP
  annihilations in the Sun or Earth are calculated for
  different WIMP masses, compositions and signal fluxes.

\item[V.] 

  The relic density of neutralinos is calculated including so called
  coannihilation processes between the lightest neutralino and heavier
  neutralinos and charginos.

\end{list}

\cleardoublepage
\def\rightmark{Acknowledgments}
\def\leftmark{}
\addcontentsline{toc}{chapter}{Acknowledgments}
\vspace*{15mm}
\noindent
{\Huge\bf Acknowledgments}

\vspace{10mm} 

\noindent First of all I would like to thank my supervisor H{\'e}ctor
Rubinstein for his great support and my assistant supervisor Lars
Bergstr{\"o}m who has been most helpful during my studies and for his
careful reading of the manuscript.

I would also like to thank Paolo Gondolo for the many valuable
discussions we have had, for teaching me the helicity amplitude method
for calculating cross sections and for letting me use and develop the
programs he has written.

I also want to thank Gunnar Ingelman for interesting discussions in
the very early stages of this work. My thanks also go to Anders Edin
and Mats Thunman for the discussions we have had.

My colleagues and the personnel at the Department of Theoretical
Physics also deserve to be thanked for making my studies more
pleasant and interesting.

I would also like to thank my dog, Molly, for not letting me write on
my thesis too much by requiring frequent walks and for her deep
insight that she is the most important dark matter in the Universe
(she is a \emph{black} Labrador Retriever).

Finally, my warmest thanks go to my wife, Lisa, for her love and support
and, not to forget, help with the frequent walks with our dog Molly.

\def\rightmark{Bibliography}
\def\leftmark{Bibliography}


\begin{thebibliography}{99}
\addcontentsline{toc}{chapter}{Bibliography}
\frenchspacing
\def\rightmark{Bibliography}
\def\leftmark{Bibliography}

\bibitem{pecvel} 
M.~Strauss and J.~Willick, Phys.\ Rep.\ {\bf 261} (1995) 271;
A.~Dekel, Ann.\ Rev.\ Astron.\ Astrophys.\ {\bf 32} (1994) 319.

\bibitem{KolbTurner}
E.W.~Kolb and M.S.~Turner, \emph{The Early Universe},
Addison-Wesley (1990).

\bibitem{primabund}
See e.g.\ C.~Copi, D.N.~Schramm and M.S.~Turner, Science {\bf 267} 
(1995) 192, Phys.\ Rev.\ Lett.\ {\bf 75} (1995) 3981;
N.~Hata et al., Phys.\ Rev.\ Lett.\ {\bf 75} (1995) 3977.

\bibitem{jkg} G.~Jungman, M.~Kamionkowski and K.~Griest, Phys.\ Rep.\
  {\bf 267} (1996) 195.
  
\bibitem{HaberKane}
H.E.~Haber and G.L.~Kane, Phys.\ Rep.\ {\bf 117} (1985) 75.  

\bibitem{GuHa86}
J.F.~Gunion and H.E.~Haber, Nucl.\ Phys.\ {\bf B272} (1986) 1
[Erratum-ibid.\ {\bf B402} (1993) 567].

\bibitem{WessBagger}
J.~Wess and J.~Bagger, \emph{Supersymmetry and Supergravity}, 2nd 
Ed.\ (Princeton University Press, Princeton, 1992).

\bibitem{HiggsHunter}
J.F.~Gunion, H.E.~Haber, G.~Kane and S.~Dawson, \emph{The Higgs
  Hunter's Guide}, Addison-Wesley (1990).

\bibitem{carena}
M.~Carena, J.R.~Espinosa, M.~Quir{\'o}s and C.E.M.~Wagner, Phys.\
Lett.\ {\bf B355} (1995) 209.

\bibitem{effpot}
M.~Berger, Phys.\ Rev.\ {\bf D41} (1990) 225; H.E.~Haber and R.~Hempfling,
Phys.\ Rev.\ Lett.\ {\bf 66} (1991) 1815; P.H.~Chankowski, S.~Pokorski and
J.~Rosiek, Phys.\ Lett.\ {\bf B281} (1992) 100; J.~Ellis, G.~Ridolfi
and F.~Zwirner, Phys.\ Lett.\ {\bf B257} (1991) 83; Y.~Okada, M.~Yamaguchi and
T.~Yanagida, Phys.\ Lett.\ {\bf B262} (1991) 54;
A.~Brignole, Phys.\ Lett.\ {\bf B281} (1992) 284;
J.~Kodaira, Y.~Yasui and K.~Sasaki, Phys.\ Rev.\ {\bf D50} (1994) 7035.

\bibitem{HiggsLEP2}
E.~Accomando et al., \emph{Higgs Physics at LEP2}, hep-ph/9602250, to
appear in Vol.~1, \emph{Report of the Workshop on Physics at LEP2},
G.~Altarelli, T.~Sj{\"o}strand and F.~Zwirner (eds), CERN 96-01.

\bibitem{NeuLoop1}
M.~Drees, M.M.~Nojiri, D.P.~Roy and Y.~Yamada, hep-ph/9701219.

\bibitem{NeuLoop2}
D.~Pierce and A.~Papadopoulos, Phys.\ Rev.\ {\bf D50} (1994) 565, 
Nucl.\ Phys.\ {\bf B430} (1994) 278; A.B.~Lahanas, K.~Tamvakis and 
N.D.~Tracas, Phys.\ Lett.\ {\bf B324} (1994) 387.

\bibitem{B-functions}
M.~Drees, K.~Hagiwara and A.~Yamada, Phys.\ Rev.\ {\bf D45} (1992) 
1725.

\bibitem{PDG}
R.M.~Barnett et al (Particle Data Group), Phys.\ Rev.\ {\bf D54} (1996) 1.

\bibitem{Mandl}
F.~Mandl and G.~Shaw, {\em Quantum Field Theory}, John Wiley \& Sons, 1984.

\bibitem{ChargLEP2}
See e.g.\ A.S.~Belyaev and A.V.~Gladyshev, hep-ph/9703251.

\bibitem{LEP2}
G.~Cowan (ALEPH Collaboration), talk presented at the special CERN
particle physics results from the LEP run at 172 GeV, 25 February 1997.

\bibitem{CLEO} M.S.~Alam et al.\ (CLEO Collaboration), Phys.\ Rev.\
  Lett. {\bf 71} (1993) 674; Phys.\ Rev.\ Lett. {\bf 74} (1995) 2885.

\bibitem{bsg}
S.~Bertolino, F.~Borzumati, A.~Masiero and G.~Ridolfi, Nucl.\ Phys.\
{\bf B353} (1991) 591.

\bibitem{bsgLarsPaolo}
L.~Bergstr{\"o}m and P.~Gondolo, Astropart.\ Phys.\ {\bf 5} (1996) 263.

\bibitem{GriestSeckel}
K.~Griest and D.~Seckel, Phys.\ Rev.\ {\bf D43} (1991) 3191.

\bibitem{relcalc} Some selected references are: H.~Goldberg, Phys.\ 
  Rev.\ Lett.\ {\bf 50} (1983) 1419; L.M.~Krauss, Nucl.\ Phys.\ {\bf
    B227} (1983) 556; J.~Ellis et al., Nucl.\ Phys.\ {\bf B238} (1984)
  453; K.~Griest, Phys.\ Rev.\ {\bf D38} (1088) 2357 [erratum ibid
  {\bf D39} (1989) 3802]; J.~Scherrer and M.S.~Turner, Phys.\ Rev.\ 
  {\bf D33} (1986) 1585 [erratum ibid {\bf D34} (1986) 3263];
  K.A.~Olive and M.~Srednicki, Phys.\ Lett.\ {\bf B230} (1989) 78,
  Nucl.\ Phys.\ {\bf B355} (1991) 208;
  K.~Griest, M.~Kamionkowski and M.S.~Turner, Phys.\ Rev.\ {\bf D41}
  (1990) 3565; G.B.~Gelmini, P.~Gondolo, and E.~Roulet, Nucl.\ Phys.\ 
  {\bf B351} (1991) 623; A.~Bottino, V.~de~Alfaro, N.~Fornengo,
  G.~Mignola and S.~Scopel, Astropart.\ Phys.\ {\bf 1} (1992) 61;
  A.~Bottino, V.~de~Alfaro, N.~Fornengo, G.~Mignola, M.~Pignone,
  Astropart.\ Phys.\ {\bf 2} (1994) 67; H.~Baer and M.~Brhlik, Phys.\ 
  Rev.\ {\bf D53} (1996) 597. For further references, see G.~Jungman,
  M.~Kamionkowski, and K.~Griest, Phys.\ Rep.\ {\bf 267} (1996) 195.

\bibitem{McDonald}
J.~McDonald, K.A.~Olive and M.~Srednicki, Phys.\ Lett.\ {\bf B283}
(1992) 80.

\bibitem{MizutaYamaguchi}
S.~Mizuta and M.~Yamaguchi, Phys.\ Lett.\ {\bf B298} (1993) 120.

\bibitem{SWO}
M.~Srednicki, R.~Watkins and K.A.~Olive, Nucl.\ Phys.\ {\bf B310} (1988) 
693.

\bibitem{DreesNojiri}
M.~Drees and M.~Nojiri, Phys.\ Rev.\ {\bf D 47} (1993) 376.

\bibitem{GondoloGelmini}
P.~Gondolo and G.~Gelmini, Nucl.\ Phys.\ {\bf B360} (1991) 145.

\bibitem{JEpaper5}
J.~Edsj{\"o} and P.~Gondolo, hep-ph/9704361, submitted to Phys.\ Rev.\
{\bf D}. 

\bibitem{gondoloedsjo}
P.~Gondolo and J.~Edsj{\"o}, in preparation.

\bibitem{reduce}
{\sc Reduce 3.5}. A.C.~Hearn, RAND, 1993.

\bibitem{paolohel}
P.~Gondolo, in preparation.

\bibitem{ChenDreesGunion}
C.H.~Chen, M.~Drees, and J.F.~Gunion, Phys.\ Rev.\ {\bf D55} (1997) 330.

\bibitem{PressSpergel} W.H.~Press and D.N.~Spergel, Astrophys.\ J.\ 
{\bf 296} (1985) 679.
  
\bibitem{SOS} J.~Silk, K.~Olive, and M.~Srednicki, Phys.\ Rev.\ Lett.\
  {\bf 55} (1985) 257.
  
\bibitem{earlysun} 
L.M.~Krauss, K.~Freese, D.N.~Spergel and W.H.~Press, Astrophys.\ 
J.\ {\bf 299} (1985) 1001;
T.~Gaisser, G.~Steigman and S.~Tilav, Phys.\ Rev.\ {\bf D34} (1986) 
2206;
B.A.~Campbell, J.~Ellis, K.~Enqvist, D.V.~Nanopoulos, J.~Hagelin and 
K.A.~Olive, Phys.\ Lett.\ {\bf B173} (1986) 270;
M.~Srednicki, K.~Olive and J.~Silk, Nucl.\ Phys.\ {\bf B279} (1987) 
804;
K.~Griest and D.~Seckel, Nucl.\ Phys.\ {\bf B283} (1987) 681;
J.~Hagelin, K.~Ng and K.~Olive, Phys.\ Lett.\ {\bf B180} (1987) 375;
K.~Ng, K.A.~Olive and M.~Srednicki, Phys.\ Lett.\ {\bf B188} (1987) 
138.

\bibitem{earlyearth} 
K.~Freese, Phys.\ Lett.\ {\bf B167} (1986) 295;
L.~Krauss, M.~Srednicki and F.~Wilczek, Phys.\ Rev.\ {\bf D33} (1986) 
2079;
T.~Gaisser, G.~Steigman and S.~Tilav, Phys.\ Rev.\ {\bf D34} (1986) 
2206.

\bibitem{Baksan} M.M.~Boliev et al., Bull.\ Acad.\ Sci.\ USSR, Phys.\
  Ser.\ {\bf 55} (1991) 126 [Izv.\ Akad.\ Nauk.\ SSSR, Fiz.\ {\bf 55} (1991) 
  748]; M.M.~Boliev et al., in {\it TAUP 95}, proceedings of the
  Workshop, Toledo, Spain, September 17--21, 1995, edited by 
  A.~Morales, J.~Morales and J.A.~Villar, [Nucl.\ Phys.\ (Proc.\ Suppl.)
  {\bf B48} (1996) 83] (North-Holland, Amsterdam, 1996).

\bibitem{Macro} E.~Diehl, Ph.D.~Thesis, U.~of Michigan (1994).
  
\bibitem{Kamiokande} M.~Mori et al.\ (Kamiokande Collaboration), {\rm
    Phys.~Lett.} {\bf B289} (1992) 463; M.~Mori et al.\ (Kamiokande
  Collaboration), {\rm Phys.~Rev.} {\bf D48} (1993) 5505.
  
\bibitem{SuperKamiokande} Y.~Totsuka, in {\it TAUP 95}, proceedings of
  the Workshop, Toledo, Spain, September 17--21, 1995, edited by 
  A.~Morales, J.~Morales and J.A.~Villar, [Nucl.\ Phys.\ (Proc.\ Suppl.)
  {\bf B48} (1996) 547].
  
\bibitem{Amanda} P.O.~Hulth et al (AMANDA Collaboration), to be
  published in Proceedings of the XVII International Conference in
  Neutrino Physics and Astrophysics, Neutrino 96, Helsinki, Finland
  13-19 June 1996, eds.  K.~Enqvist, K.~Huitu and J.~Maalampi
  (World Scientific, Singapore, 1997).
  
\bibitem{Nestor} L.~Resvanis, Europhys.\ News {\bf 23} (1992) 172.
  
\bibitem{GriestSeckel87}
K.~Griest and D.~Seckel, Nucl.\ Phys.\ {\bf B283} (1987) 681.

\bibitem{Gould87} A.~Gould, Astrophys.\ J.\ {\bf 321} (1987) 560.
  
\bibitem{Gouldcapt} A.~Gould, Astrophys.\ J.\ {\bf 321} (1987) 571; 
ibidem {\bf 368} (1991) 610; ibidem {\bf 388} (1992) 338.

\bibitem{rhochi}
J.N.~Bahcall, M.~Schmidt and R.M.~Soneira, Astrophys.\ J.\ {\bf 265} 
(1983) 730;
J.A.R.~Caldwell and J.P.~Ostriker, Astrophys.\ J.\ {\bf 251} (1981) 61;
M.S.~Turner, Phys.\ Rev.\ {\bf D33} (1986) 889;
R.A.~Flores, Phys.\ Lett.\ {\bf B215} (1988) 73;
E.I.~Gates, G.~Gyuk and M.S.~Turner, Astrophys.\ J.\ Lett.\ {\bf 
449} (1995) L123.

\bibitem{JEdiploma} J.~Edsj{\"o}, \emph{Neutrino-induced Muon Fluxes from 
Neutralino Annihilations in the Sun and in the 
Earth}, Diploma Thesis, TSL/ISV-93-0091, ISSN 0284-2769.

\bibitem{JEpaper2} J.~Edsj{\"o} and P.~Gondolo, Phys.\ Lett.\ {\bf
    B357} (1995) 595.

\bibitem{RS} S.~Ritz and D.~Seckel, Nucl.\ Phys.\ {\bf B304} (1988) 877.

\bibitem{JEpaper1}
J.~Edsj{\"o}, Nucl.\ Phys.\ {\bf B} (Proc.\ Suppl.) {\bf 43} (1995) 
265.

\bibitem{neuprod} 

G.F.~Giudice and E.~Roulet, Nucl.\ Phys.\ {\bf B316} (1989) 429;
F.~Halzen, T.~Stelzer and M.~Kamionkowski, Phys.\ Rev.\ {\bf D45} (1992) 4439; 
M.~Drees, G.~Jungman, M.~Kamionkowski and M.M.~Nojiri, Phys.\ Rev.\ {\bf D49}
(1994) 636;
R.~Gandhi, J.L.~Lopez, D.V.~Nanopoulos, K.~Yuan and A.~Zichichi, 
Phys.\ Rev.\ {\bf D49} (1994) 3691;
A.~Bottino, N.~Fornengo, G.~Mignola and L.~Moscoso, Astropart.\ Phys.\
{\bf 3} (1995) 65;
G.~Jungman and M.~Kamionkowski, Phys.\ Rev.\ {\bf D51} (1995) 328;
V.~Berezinsky, A.~Bottino, J.~Ellis, N.~Fornengo, G.~Mignola and
S.~Scopel, Astropart.\ Phys.\ {\bf 5} (1996) 333.

\bibitem{Jackson}
J.D.~Jackson, \emph{Classical Electrodynamics}, 2nd Ed., John Wiley \& 
Sons (1975).

\bibitem{Bahcall}
J.N.~Bahcall and R.K.~Ulrich, Rev.\ Mod.\ Phys.\ {\bf 60} (1988) 297.

\bibitem{Povh}
B.~Povh and J.~H{\"u}fner, Phys.\ Lett.\ {\bf B245} (1990) 653.

\bibitem{Jetset}
T.~Sj\"{o}strand, Comp.\ Phys.\ Comm.\ {\bf 82} (1994) 74;
T.~Sj\"{o}strand, {\em PYTHIA 5.7 and JETSET 7.4. Physics and Manual}, 
CERN-TH.7112/93, hep-ph/9508391 (revised version).

\bibitem{GRV} M.~Gl{\"u}ck, E.~Reya and A.~Vogt, Z.\ Phys.\ {\bf C53}
  (1992) 127.  

\bibitem{CTEQ} 
J.~Botts et al., Phys.\ Lett. {\bf B304} (1993) 159.

\bibitem{JungmanKam}
  G.~Jungman and M.~Kamionkowski, Phys.\ Rev.\ {\bf D51} (1995) 328.

\bibitem{Lohmann}
W.~Lohmann, R.~Kopp and R.~Voss, \emph{Energy loss of muons in the energy 
range 1--10000 GeV}, CERN Report 85-03 (1985).

\bibitem{JEpaper3} L.~Bergstr{\"o}m, J.~Edsj{\"o} and P.~Gondolo,
  Phys.\ Rev.\ {\bf D55} (1997) 1765.

\bibitem{AtmMu}
L.V.~Volkova, Yad.\ Fiz.\ {\bf 31} (1980) 1510 [Sov.\ J.\ Nucl.\
Phys.\ {\bf 31} (1980) 784];
T.K.~Gaisser, T.~Stanev and G.~Barr, Phys.\ Rev.\ {\bf D38} (1988) 85;
G.~Barr, T.K.~Gaisser and T.~Stanev, Phys.\ Rev.\ {\bf D39} (1989)
3532;
P.~Lipari, Astropart.\ Phys.\ {\bf 1} (1993) 195.
M.~Thunman, G.~Ingelman and P.~Gondolo, AStrop.\ Phys.\ {\bf 5} (1996)
309.

\bibitem{honda}
M.~Honda et al., Phys.\ Rev.\ {\bf D52} (1995) 4985.

\bibitem{GaisserStanev}
T.K.~Gaisser and T.~Stanev, Phys.\ Rev.\ {\bf D30} (1984) 985.

\bibitem{SunBgd} D.~Seckel, T.~Stanev and T.K.~Gaisser, Astrophys.\ 
  J.\ {\bf 382} (1991) 652; G.~Ingelman and M.~Thunman, Phys.\ Rev.\ 
  {\bf D54} (1996) 4385.

\bibitem{BottinoMoscoso} A.~Bottino, N.~Fornengo, G.~Mignola and
  L.~Moscoso, Astropart.\ Phys.\ {\bf 3} (1995) 65.

\bibitem{JEpaper4} L.~Bergstr{\"o}m, J.~Edsj{\"o} and 
M.~Kamionkowski, astro-ph/9702037, Astrop.\ Phys., in press.

\bibitem{CMBdettheory}
R.J.~Bond, R.~Crittenden, R.L.~Davis, G.~Efstathiou and P.J.~Steinhardt,
Phys.\ Rev.\ Lett. {\bf 72} (1994) 13;
G.~Jungman, M.~Kamionkowski, A.~Kosowsky and D.N.~Spergel, 
Phys.\ Rev.\ Lett. {\bf 76} (1996) 1007, Phys.\ Rev.\ {\bf D54} (1996)
1332;
J.R.~Bond, G.~Efstathiou and M.~Tegmark, astro-ph/9702100.

\bibitem{CMBdetexp}
C.~Bennett et al., MAP home page, http://map.gsfc.nasa.gov/;
M.~Bersanelli et al., ESA Report D/SCI(96)3, PLANCK home page,
http://astro.estec.esa.nl/SA-general/Projects/Cobras/cobras.html

\bibitem{archange}
Derived from the rules in Fig.~83 in Ref.~\cite{HaberKane} or directly from
the Lagrangian.

\bibitem{paoloprivate}
P.~Gondolo, private communication.

\end{thebibliography}
\end{document}